# Introduction to the Physics of Free Electron Laser and Comparison with Conventional Laser Sources

*G. Dattoli[1], M. Del Franco*, M. Labat*, P.L. Ottaviani[§]*

ENEA, Gruppo Fisica Teorica e Matematica Applicata, Centro Ricerche Frascati, C.P. 65 - 00044 Frascati, Rome (Italy)

* ENEA Guest

[§] INFN Sezione di Bologna

*S. Pagnutti*

ENEA, Gruppo Fisica Teorica e Matematica Applicata, Centro Ricerche Bologna

**Abstract**

This report is a primer introduction on the sources of coherent syncrotron radiation from FELs compared to conventional lasers. Even though the underlying mechanisms are different, FELs can be considered a laser source to all effects. In fact FEL is based on the stimulated emission of virtual photons by a relativistic electron beam, passing through an undulator/wiggler, much like conventional lasers exploit the stimulated emission by an atomic/molecular system on which a population inversion has been realized. FEL operation is illustrated by means of the usual set of parameters (gain, saturation intensity…) used to describe the functioning of conventional lasers.

1) Corresponding author:
G. Dattoli
Tel +39 06 94005421
e-mail Dattoli@frascati.enea.it
Fax +39 06 94005334

# 1. Introduction

This work provides an introduction to the physics of coherent radiation sources using free electron beams, also called Free Electron Lasers (FEL), and makes a comparison with conventional laser sources [1].

The FEL can be fully considered as lasers, even though the mechanisms of operation are different from those of the conventional sources. Indeed, the stimulated emission process is not performed in an atomic or a molecular system, with population inversion, but in a relativistic electron beam passing through the magnetic field of an undulator.

Along this work, we will define several fundamental parameters (such as gain, saturation intensity, etc.) common to both the conventional and the free electron lasers. Thanks to these parameters, we will explain the FEL principles, compare them to the conventional systems, to finally demonstrate why FEL can be fully considered as lasers.

Laser physics started with quantum mechanics, at the beginning of the century, when M. Planck derived the spectral distribution of the "blackbody" radiation. Such a system absorbs all the incident radiation and emits with a spectrum which only depends on its temperature. A perfectly reflective cavity with a little hole can be considered as a blackbody, its radiated energy resulting from the standing wave or resonant modes of this cavity (see Fig. 1).

The distribution law of the blackbody, derived by Planck (1900), is the following:

$$I(\omega)d\omega = \frac{\hbar\omega^3}{\pi^2 c^2\left[\exp\left(\frac{\hbar\omega}{kT}\right)-1\right]}d\omega$$

(1)

$$\hbar = \frac{h}{2\pi},\ \omega = 2\pi\nu,\ \nu = \text{frequency}$$

with the three fundamental constants: $h$ the Planck constant, $k$ the Boltzmann constant and $c$ the light velocity. $I(\omega)d\omega$ is the energy quantity per unit surface, per unit time, emitted in the interval $\omega, \omega + d\omega$. In a very simplified way, Eq. (1) illustrates the equilibrium between a radiation at a given frequency ω and matter at a given temperature *T*. The blackbody distribution as a function of the wavelength (in nm) and of the frequency of the radiationis represented in Fig. 2, together with the Rayleigh-Jeans law, formulated before the Planck law and which reproduced the observations at long wavelengths.

It can be noticed that the Rayleigh-Jeans law diverges as the photon energy increases. It is relevant here to note that Wien had reached the conclusion (1896), through out theoretical considerations on the equipartition of the energy, that for high frequencies (or low temperature), the blackbody distribution could be written as:

(2) $$I(\omega) = c_1\omega^3 exp(-c_2 \frac{\omega}{T}),$$

where $c_{1,2}$ are constants. It is obvious that the Wien law is in fact a particular case of Eq. (1), for which Planck assumes that the changes in energy are not continuous but discrete, after discretizing the phase space with the constant *h*. The quantum concept results from this assumption and was further used by Einstein to explain the photoelectric effect (1905) and by Bohr to justify the stability of the atoms (1913).

The Bohr's theory was making a simple, but efficient, assumption: the electrons in an atomic system are constrained on a stationary orbit, whose energetic level is fixed and could only exchange quanta of radiation with their external environment. We will now see how the hypothesis of a quantified atomic system can naturally lead to the blackbody distribution expression and consequently the considerations that Einstein did in 1917.

We consider an atomic system with two states of energy $E_1$ and $E_2$ as illustrated in Fig. 3, 1 being the lower energy state. The excitation of the system to the higher state 2 (see Fig. 3) can result, with a probability $B_{1,2}$, from the absorption of a photon with an energy of:

(3) $$\varepsilon = \hbar\omega,\ \ \omega = \frac{E_2 - E_1}{\hbar}$$

If the system is in the excited state 2, it can decay to the lower state spontaneously, thanks to the emission of a photon of the same energy as the energy absorbed (see Fig. 4).

Such decay process is analog to the principle of minimum energy in the classical mechanical systems. At this point can be made the following crucial assumption: there is another process of decay, referred as stimulated, for which the transition to the lower state (see Fig. 5) occurs, with probability $B_{2,1}$, in the presence of a photon of energy equal to the energy difference between the two states as defined in Eq. (3).

We consider now a system of *N* atoms, among which $N_1$ are in the lower state and $N_2$ in the higher state ( $N_1 + N_2 = N$ ), in interaction with an electromagnetic field of intensity $I(\omega)$. According to what has been said previously, the equilibrium condition of the atomic system with the radiation is given by:

(4) $$N_1 B_{1,2} I(\omega) = N_2 B_{2,1} I(\omega) + N_2 A_{2,1}$$

where $A_{2,1}$ represents the spontaneous emission probability, independent of $I(\omega)$.

In the conditions of thermodynamical equilibrium at temperature *T*, the number of atoms per unit volume in a given state of energy $E_i$ is described by the Boltzmann distribution law: $N_i \propto e^{-iE_i/KT}$. From this relation, one can derive the ratio $\frac{N_1}{N_2}$ in terms of energy difference and absolute temperature:

(5) $$\frac{N_1}{N_2} = exp\left(\frac{\hbar\omega}{kT}\right)$$

Combining it with Eq. (4), we get:

(6) $$I(\omega) = \frac{A_{2,1}}{B_{1,2}\, exp\left(\frac{\hbar\omega}{kT}\right) - B_{2,1}}$$

Assuming that the absorption and stimulated emission processes are perfectly symmetric, i.e.

that $B_{m,n} = B_{n,m}$, and using the Wien law, the spontaneous emission can be written as:

(7) $$A_{n,m} = \frac{\hbar\omega^3}{\pi^2 c^2} B_{m,n}$$

which corresponds to the Planck distribution law. The most appropriate comment to understand how important are these results may be Einstein's:

***" ..the three hypotesis of absorption and emission of radiation, used here, are not justified by the fact that they lead to the Planck's law. The simplicity of its derivation and the generality of the related considerations and lead to believe that those could provide with the fundamental theoretical elements for the next derivations….".***

## 2. Amplification of the Radiation and Stimulated Emission

According to what we said before, we could ensure that, being given a certain number of atoms, all (or most of them) in the higher level, one can get a substantive amplification process from one single photon causing decay versus the most populated state(see Fig. 6).

The main difference between the two emission processes, spontaneous and stimulated, is that in the case of the spontaneous emission, the atoms decay randomly in time, generating photons uncorrelated in phase and direction, while in the case of the stimulated emission, the atoms decay at the same time, generating coherent photons, with equal phase and direction to those of the incident photon.

Therefore, we consider a medium of $N_1 + N_2 = N$ systems of two states, in interaction with an electromagnetic field with the energy of a single photon $\hbar\omega = E_2 - E_1$. During the propagation in the $z$ direction (see Fig. 7), the variation of the number of photons $n$ of the field is driven by a differential equation deriving from Eq. (4):

(8) $$\frac{dn}{dz} = (N_2 - N_1) b_{1,2} n + a_{2,1} N_2$$

The coefficients of emission and absorption have been redefined to take into account the physical dimensions.

Keeping in mind that the populations remain constant and are independent of the field intensity, the solution of Eq. (8) is:

$$n(z) = n_0 exp(\alpha z) + \beta \frac{exp(\alpha z) - 1}{\alpha} \tag{9}$$

$$\alpha = (N_2 - N_1) b_{2,1}, \quad \beta = a_{2,1} N_2$$

where $n_0$ is the initial number of photons of the field. This solution is simple but of high interest considering the physics. It consists of two parts, one being independent of the initial number of photons, which illustrates the previous statements. We consider the case in which the spontaneous emission term can be neglected. If $N_2 < N_1$, i.e. if there is less atoms in the excited than in the ground state, then $\alpha < 0$ and Eq. (9) becomes:

$$n(z) = n_0 exp(-|\alpha| z) \tag{10}$$

which corresponds to the Beer-Lambert law of absorption and whose coefficient $\alpha$ can be interpreted as a coefficient of linear absorption. The Beer-Lambert law is represented in Fig. 8. Its physical meaning is far from insignificant in spite of its common aspect. Figure 8 shows a beam of intensity $I_0$, partially absorbed passing through a medium of length *l*. With $C$ the concentration of matter in the medium, $\alpha = \varepsilon_\lambda C$, and $\varepsilon_\lambda$ is the mass absorption coefficient $C$ at the wavelength $\lambda$.

When $N_2 = N_1$, the number of photons remains constant. The amplification is possible when $N_2 > N_1$, i.e. when a population inversion has occurred. In this case, the coefficient $\alpha$ is interpreted as the small signal gain coefficient. Still assuming the population inversion, the amplification can be seen as a chain reaction: a photon causes the decay of the excited state, creating a "clone", and following the same process, both lead to the "cloning" of another two photons, and so on. Adding the spontaneous emission, the mechanism can be started spontaneously without the need of an external field. Such mechanism provides the fundamental elements of the amplification and we will see how it can be implemented to realize laser oscillator.

# 3. Laser Oscillator

The radiation amplification, based on the stimulated emission mechanism, requires an active medium in which population inversion occurs, and an environment which can provide a feedback to maintain the system in operation as an oscillator (see Fig. 9).

Without entering into the details of the population inversion mechanism, we can notice that this effect can not be obtained simply by heating the active medium. Indeed, if we consider a two states system, the electrons obey the statistics of Fermi-Dirac and we have:

(11) $$\frac{N_2}{N_1} = \frac{1}{exp\left(\frac{\hbar\omega}{kT}\right)+1}$$

It's obvious that the population inversion process is not the result of a heating. At a temperature close to 0° K, the higher state is depopulated while at high temperature ($T \rightarrow +\infty$) $\frac{N_2}{N_1} = \frac{1}{2}$, i.e. the two states are equally populated. An extensively used mechanism for population inversion is the **optical pumping**: the active medium is slightly more complex than a two states system, which remains a convenient abstraction thus hardly actually feasible.

We now move on to the feedback system. Since the spontaneous emission is essentially isotropic (see Fig. 10), a first direction selection can be operated implementing the medium in an optical cavity, consisting in its most simple configuration of two facing mirrors. The emitted photons result selected according to their direction, increasing the spatial coherence of the radiation at each pass in the cavity.

Being assumed the existence of a gain, one expects the radiation intensity to increase and that at each reflection on the semi-reflective mirror a fraction of this radiation is extracted of the cavity. Since only a part of the radiation is reflected, the evolution of the system will strongly depend on the losses of the cavity and we can expect a saturation mechanism that will drive the system to an equilibrium state. As for other physical systems, the saturation mechanism is non linear and controlled by the radiation intensity itself.

We can now write the rate equations, i.e. the evolution equations of the laser intensity inside the cavity:

(12) $$I_{n+1} = (1-\eta)(G(I_n)+1)I_n$$

or even

(13) $$I_{n+1} - I_n = \left[1-\eta)G(I_n) - \eta\right]I_n$$

$n$ represents the number of round trips inside the optical cavity and $\eta$ the cavity losses (we do not yet distinguish the active and the passive [2] losses, so that $\eta$ are the total losses). In order to take into account the effects of saturation, we consider that the gain $G(I)$ is a decreasing function of the intensity defined as following:

(14) $$G(I) = \frac{g}{1+\frac{I}{I_s}}$$

$I_s$, which is being introduced in a phenomenological way, is a fundamental quantity which represents the saturation intensity. The effects of saturation are then all defined with respect to this reference. One can notice that for $I = I_s$, the gain corresponds to one half of the small signal gain $g$, which we no longer refer to as $\alpha$ to take into account the fact that the active medium is not a simple two sates system. The former equation obviously provides interesting information that we will now discuss.

---

2) The active losses correspond to the transmission of the radiation to the external environment, while the passive losses correspond to the absorption of the radiation by the medium and the cavity mirrors.

Using the following approximation (see Appendix B for more details):

(15) $$I_{n+1} - I_n \cong T_R \frac{dI}{dt}$$

Equation (12) can be rewritten in terms of a differential equation:

(16) $$\frac{dI}{d\tau} = [(1-\eta)G(I) - \eta]I.$$

$\tau$ is a dimensionless time related to the actual time $t$ and to the length of the optical cavity $L_c$ according to: $\tau = \frac{t}{T_R}$, $T_R = \frac{2L_c}{c}$, with $T_R$ the duration of a round trip in the cavity.

Transforming Eq. (16) into:

(17) $$\frac{dX}{g[(1-\eta)\frac{1}{1+X} - r]X}$$
$$X = \frac{I}{I_s}, \quad r = \frac{\eta}{g}$$

it becomes obvious that for $X << 1$, i.e. far from saturation (see Appendix B), the solution evolves exponentially:

(18) $$X = X_0 \exp\{[(1-\eta)g - \eta]\tau\}$$

The equilibrium ($\frac{dX}{d\tau} = 0$) is reached when the intensity in the cavity is:

(19) $$I_E = [\frac{1-\eta}{\eta} g - 1]I_s$$

or even

(20) $$G(I) = \frac{\eta}{1-\eta}$$

It is interesting to note that, as in most cases $\eta << 1$, the extracted power at equilibrium (assuming that the losses are only active losses) is given by:

(21) $$I_{out} = \eta I_E \cong g I_s$$

This relation is an additional evidence of the importance of the saturation intensity, found in the definition of all the quantities of practical interest.

In Figure 11 is presented the evolution of the laser power inside the cavity as a function of the number of round trips.

It is the typical evolution of the logistic function, which is discussed in more details in the next paragraph and in Appendix B.

The intensity first increases exponentially and then slows down because of the saturation mechanism. When the gain equals the losses, the system reaches a stationary state and the power can be extracted off the cavity. Assuming that the losses are active, i.e. only due to the transmission of the second mirror, without additional passive effects such as a heating of the cavity, an expression of the extracted intensity as a function of the saturation intensity and the small signal gain is obtained.

Independently of this result, one may wonder if the extracted power is related to any other reference power which would characterize the efficiency of the system.

Despite not mentioning it before, it is obvious that the population inversion requires a sufficient level of power, which we refer to as pump, that will be further partially transformed into laser power. The ratio between the extracted laser power and the pump power represents the efficiency of the system.

# 4. Laser Oscillators and Optical Cavities

We have seen that the optical cavity is an essential element of an oscillator laser since it confines the electromagnetic radiation, selects the transverse and longitudinal modes, thus creating the spatial and partially temporal coherence of the radiation.

The radiation field inside the optical cavity, whose components have a small angle with respect to the cavity axis, is succesivly reflected by the mirrors. As already seen, this enables to select the components travelling in a direction parallel to the cavity axis. The interfering of the waves in the cavity leads to the formation of stationary waves at given frequencies, depending on the distance between the mirrors. In the case of planar and parallel mirrors separated by a distance *L*, the stationary condition for the wave frequencies is: $\nu_n = n\frac{c}{2L}$, *n* being integer. The various frequencies obtained, varying *n*, are referred as **longitudinal modes** of the cavity. For each longitudinal mode, various transverse sections are possible, which modify the transverse intensity distribution in the plans perpendicular to the optical axis. Those are referred as **transverse modes**.

## *Transverse modes*

In a conceptual point of view, an optical cavity consists of two parallel mirrors (see Fig. 12). The mirrors configuration sets the field shape inside the cavity, as illustrated in Fig. 12. Two flat mirrors, i.e. with an infinite radius of curvature, lead to a constant transverse mode distribution. When the mirrors are confocal, the transverse mode has a parabolic longitudinal profile, which we will discuss more in details later. Additional configurations are shown in Fig. 12.

In Figure 13 are presented the equilibrium conditions of the optical cavity. Indeed, the radii of curvature of the mirrors and the cavity length must satisfy certain conditions to ensure that the optical beam is periodically focused and always confined within the cavity. If not, i.e. if the beam size increases and becomes larger than the mirrors size, the mode does not remain confined and the cavity is said unstable.

Without giving details on the derivation, we provide the criterion of stability:

(22) $$0 < (1-\frac{L}{R_1})(1-\frac{L}{R_2}) < 1$$

Using:

(23) $$g_{1,2} = 1-\frac{L}{R_{1,2}}$$

Equation (22) becomes:

(24) $$0 < g_1 g_2 < 1$$

This expression has a simple geometrical interpretation reported in Fig. 13. The limits of the stability domain are on the branches of the hyperbole defined by $g_1 g_2 = 1$. Of course, the plane-parallel and confocal cavities are stable. We will come back to this issue later.

In the case of a confocal cavity, the transverse modes are the Gauss-Hermite modes and, as previsouly mentioned, the longitudinal profile of the mode is parabolic (see Fig. 14).

The fundamental mode corresponds to a Gaussian distribution defined as:

(25) $$I(r,z) = I_0(\frac{w_0}{w(z)})^2 e^{-\frac{2r^2}{w(z)^2}},$$

where $r$ is the distance to the axis, $w$ is the rms spot size of the transverse mode:

(26) $$w(z) = w_0[1+(\frac{z}{z_R})^{1/2}]$$

In addition, $w_0$ is the beam waist, i.e. the minimum optical beam transverse dimension:

(27) $$w_0 = \sqrt{\frac{\lambda L}{2\pi}}$$

and $z_R$ the Rayleigh length, corresponding to the distance for which $w(z) = \sqrt{2}w_0$:

(28) $$z_R = \frac{\pi w_0^2}{\lambda}$$

For the fundamental mode, $z_R = L/2$ .

The divergence of the mode is related to the former parameters according to:

(29) $$\theta = \frac{\lambda}{\pi w_0}$$

$\theta$ is referred as the angle of diffraction in the far field approximation.

Additional details are given in Appendix A.

## *Longitudinal modes*

For the study of the longitudinal modes, we will refer to Fig. 15. The longitudinal modes are equally separated in frequency by $\Delta f$, which depends on the cavity length according to:

(30) $$\Delta f = \frac{c}{2L}$$

The gain bandwidth of the active medium performs the mode selection and therefore allows the amplification of the modes within this bandwidth only. Those modes oscillate independently with random phase and beating.

A longitudinal mode-locking enables to fix the longitudinal modes phases so that they all oscillate with one given phase [2].

It seems obvious that, in theory, varying the losses and therefore the gain profile would allow to suppress the competition in between modes and choose for instance the one with higher gain, as illustrated in Fig. 16.

But it's also obvious that this is obtained to the detriment of the output laser power. We will now discuss alternative and less drastic methods to suppress the destructive mode competition using the previously mentioned mode-locking (see also Appendix C).

In a mode-locked laser, the output radiation does not fluctuate in a chaotic manner. It consists of a periodic train of pulses (see Fig. 17), with specific duration and temporal separation. We consider an electromagnetic field in an optical cavity consisting of $N_m$ longitudinal modes locked in phase with a $\Delta f$ frequency gap [3]:

(31) $$A(z,t) = E_0 \sum_{m=-\frac{N_m-1}{2}}^{m=\frac{N_m-1}{2}} e^{2\pi i m \Delta f (t-z/c)}$$

For simplicity, we assumed that all the modes have the same amplitude $E_0$. The sum can be replaced using:

(32) $$\sum_{m=0}^{m=q-1} a^m = \frac{a^q - 1}{a-1}$$

This leads to:

(33) $$A(z,t) = E_0 \frac{sin[\pi N_m \Delta f (t-z/c)]}{sin[\pi \Delta f (t-z/c)]}$$

The intensity of the laser is defined as the squared moduleus of the former expression, which gives, at $z$ =0:

(34) $$I(t) = \propto E_0^2 \left[ \frac{\sin[\pi N_m \Delta f \cdot t]}{\sin[\pi \Delta f \cdot t]} \right]^2$$

The result of Eq. (34) is illustrated in Fig. 17. The radiation consists of a train of peaks separated by a distance $T = \frac{1}{\Delta f}$ with a width $\frac{1}{N_m \Delta f}$. The intensity of the peaks is proportional to the square of the number of involved modes $I_p \propto N_m^2 E_0^2$, while the average intensity is proportional to $N_m$, i.e. $I_M \propto N_m E_0^2$. Finally, it is important to notice that in the case of random phases, the output laser intensity corresponds to the average intensity of the mode-locked laser and that the fluctuations have a correlation time equal to the duration of one single pulse of the mode-locked laser (more details are given in Appendix C).

---

3) One can notice that the frequency gap is not always the one given in Eq. (30). It can be larger if some modes are suppresses using for instance an etalon.

Since the former considerations are of high interest for the laser applications and for further discussions, we provide with additional precisions on the radiation pulse train presented in Fig. 18. The train is characterized by an energy per pulse $E_p$, a delay between pulses $T_R$, a pulse duration $\delta_t$ (full width half maximum), (Fig. 19), an average power $P_M$ and a peak power $P_p$ [4)].

All these quantities are linked as it follows:

$$P_p = \frac{E_p}{\delta_t} \tag{35}$$

$$P_M = P_p \frac{\delta_t}{T_R}$$

The ratio $\frac{\delta_t}{T_R}$ corresponds to the duty cycle.

We recall that the full width half maximum width $\delta_t$ is related to the rms width $\sigma$ of a Gaussian according to: $\delta_t = 2\sqrt{\ln(2)} \quad \sigma_t \cong 2.33\sigma_t$.

The mode-locking techniques allow to generate ultrashort pulses. The evolution of the pulse durations $\Delta t$ and pulse spectral widths $\Delta\omega$ achieved over the past few years, in the case of Ti:Sa lasers, is presented in Figs. 20 and 21.

As we'll see later, the ultrashort pulses are used for various applications, as for instance ultrafast spectroscopy and femto-chemistry. To illustrate the first application, an example of pump-probe experiment for the determination of the constants relative to the charge carrier dynamics in the semiconductors is presented in Fig. 22.

As far as femto-chemistry is concerned, ultrashort pulses are used to study molecular dynamics as reported in Fig. 23: a sodium iodide molecule is excited by an ultrashort pulse (a few hundred of fs) to the state $(NaI)^{*}s$ and the complete dynamics which leads the system back to the initial state or to the ionized state is analyzed.

---

4) We recall that the electric field is $E_p = \sqrt{2Z_0 I_p}$, with $I_p = \frac{P_p}{A_{eff}}$ the peak intensity and $A_{eff}$ the effective area of the optical mode, and $Z_0$ is the vacuum impedance.

Before moving on to the description of the various FEL configurations, we will give a definition of the spatial and temporal coherence, that we simply mentioned previously.

## *Temporal coherence*

Two waves are said temporally coherent if they keep a constant difference in phase. The coherence time is the time over which this property remains valid.

In the case of Fig. 24a, the coherence time is infinite, while it is finite in the case of b).

A quantitative definition of the coherence time $\tau_c$ could also be given:

**The temporal coherence is a measurement of the degree of monochromaticity of an electromagnetic wave.**

As a consequence, we can write:

(36) $$\tau_c \Delta f \cong 1$$

In the case of a mode-locked laser, the coherence time is essentially given by the distance in between the radiation packets. The coherence length $l_c$ can be calculated using Eq. (36):

(37) $$L_c \cong \frac{\lambda^2}{\Delta\lambda}$$

## *Spatial coherence*

The spatial coherence of a wave is the measurement of the temporal auto-correlation between two different points of a same transverse plane of the wave (see Fig. 25).

It is obvious that the spatially coherent area of a wave will produce a diffraction figure passing through a slit. This concept can be quantified using the Von Cittert Zernike theorem, which states that the coherence area of a light source is given by:

(38) $$S_c \cong \frac{D^2\lambda^2}{\pi d^2},$$

with $D$ the distance to the source and $d$ its diameter.

The former considerations allowed to give a general idea of the issues relative to conventional lasers. We will now discuss analogous issues relative to free electron lasers.

# 5. Relativistic Kinematics and Dynamics

In the former chapters, we gave the principles of amplifier and oscillator lasers, and we saw that those require an active medium, atomic or molecular, in which population inversion occurs. We will see in this chapter how to realize a laser with an active medium consisting of free relativistic electrons [3,5].

Before coming into details, we recall that the adjective relativistic refers to a particle or a system whose velocity is close to the light velocity, and we provide hereafter with some formula of relativistic kinematic. The total energy $E$ of a particle of mass $m$ moving with a velocity $\upsilon$ is given by:

(39) $$E = m_0\gamma c^2, \gamma = \frac{1}{\sqrt{1-\beta^2}}, \beta = \frac{v}{c}.$$

$\beta$ is the reduced velocity and $\gamma$ the relativistic factor. The total energy corresponds to the sum of the kinetic and mass energy, the mass energy being given by $m_0c^2$. The factor $\gamma$ is a straight measure of the particle energy, and $\gamma$-1 of the kinetic energy. In addition:

(40) $$\beta = \sqrt{1-\frac{1}{\gamma^2}}.$$

In the high energy case, $\gamma >> 1$ and the reduced velocity becomes:

(41) $$\beta \cong 1-\frac{1}{2\gamma^2}$$

In this last case, the particle is said ultra-relativistic. Finally, the relativistic moment is defined as:

(42) $$\vec{p} = m_0 \gamma \vec{v}$$

The second principle of dynamics states that:

(43) $$\frac{d\vec{p}}{dt} = \vec{F}$$

$\vec{F}$ being the external forces applied to the system. For a better understanding of Eq. (43), we describe the movement of a particle, initially steady, submitted to a constant force. From Eq. (43) and Eq. (39), it come, out sthat:

(44) $$v = \frac{at}{\sqrt{1+(\frac{at}{c})^2}}, a = \frac{F}{m_0}.$$

Since in relativistic mechanics, velocity and space are related by the same definition, the curvilinear coordinate of the particle $s$ can be derived from Eq. (44), assuming $s_0$=0:

(45) $$s = \frac{c^2}{a}[\sqrt{1+(\frac{at}{c})^2} - 1].$$

In the case of low velocities, the former equations correspond to the classical Newtonian dynamics equations.

We now consider the main characteristics of the electrons: a charge $|e| \cong 1.6\times10^{-19}$ C and a mass $m_e \cong 0.51$ MeV (1 MeV=$10^6$eV, 1 eV=$1.6\times10^{-19}$ J). The adjective relativistic is used when the kinetic energy of the electrons is of the order of a few MeV. Finally, a charged particle beam is associated to a power equal to the product of its energy with its current:

(46) $$P(MW) = I(A)E(MeV).$$

A 20 MeV beam with a current of 2 A is associated to a a power of 40 MW. Such power is delivered to the beam by the accelerator. From now on, we will refer to LINAC accelerator, i.e. linear accelerators, where the acceleration is performed in radiofrequency cavities. The electron beam power corresponds in the case of free electron lasers, to the pump power in the case of conventional lasers.

# 6. The Free Electron Laser

Free electrons passing through a magnetic field produces flashes or Bremsstrahlung radiation. Synchrotron light sources rely on this process. We consider the case of a magnetic field delivered by an undulator:

(47) $$\vec{B} = B_0(0, sin(\frac{2\pi z}{\lambda_u}), 0)$$

Such field is oscillating in the vertical direction with a peak value $B_0$ and a periodicity $\lambda_u$. It can be obtained with two series of magnets with alternative *N-S* orientation, as illustrated in Fig. 26. The Lorentz force resulting from the undulator field leads to an oscillating trajectory of the electrons in the *x-z* plan. The undulator introduces an essential transverse component in the electron motion, initially exclusively longitudinal.

The electron motion in the undulator is determined by the Lorentz force, which in the absence of electric field is given by (Gaussian units):

(48) $$\frac{d\vec{p}}{dt} = -\frac{e}{c}\vec{v} \times \vec{B}.$$

The former relation illustrates the conservation of the electrons energy and velocity module. Averaging over one undulator period, one can derive the transverse and longitudinal velocities:

$$v_x \cong \frac{cK}{\sqrt{2}\gamma}$$

(49) $$v_z \cong c[1 - \frac{2}{2\gamma^{*2}}]$$

$$\gamma * = \frac{\gamma}{\sqrt{1+\frac{K^2}{2}}}, K = \frac{eB_0\lambda_u}{2\pi m_0 c^2}$$

The *K* parameter is a key parameter in the FEL physics, as we will see. It is here introduced to take into account the effect of the transverse motion on the longitudinal velocity.

We saw in the former paragraphs that a fundamental element for the operation of a laser is the frequency selection, which resulted from a natural characteristic of the atomic and molecular systems, the quantified energy levels. Frequency selection is less obvious in the case of free electron emission studied in a classical environment.

Nevertheless, referring to Fig. 27, we consider the emission process inside the undulator of a radiation at the wavelength $\lambda$. After one undulator period, the emitted radiation, traveling at the light velocity, has slept ahead of the electron beam by a quantity:

(50) $$\delta \cong (c - v_z)\frac{\lambda_u}{c} = \frac{\lambda_u}{2\gamma^2}(1 + K^2/2)$$

Since $\delta$ represents a phase advance of the electromagnetic wave, to obtain constructive interference with the radiation emitted at the next undulator period, it must satisfy:

(51) $$\delta = n\lambda,$$

*n* being an integer. From Equation (50) and Eq. (51) finally comes the definition of the undulator radiation wavelength:

(52) $$\lambda_n = \frac{\lambda_u}{2n\gamma^2}(1 + \frac{K^2}{2})$$

*n*=1 represents the fundamental wavelength, while *n*>1 represent the higher harmonics. In a first step, we just consider the radiation on the fundamental and we try to characterize it, giving an evaluation of its relative bandwidth.

The radiation pulse at the end of the undulator has a structure of a step function with a length $N\delta$, corresponding to a pulse duration $\Delta\tau$ of:

(53) $$\Delta\tau \cong \frac{N\delta}{c}$$

The corresponding spectral distribution is given by the Fourier transform of this pulse (see Fig. 28), and the spectral width $\Delta\omega$ results defined according to the Parseval relation:

(54) $$\Delta\tau\Delta\omega \cong \pi.$$

The relative bandwidth is then given by:

(55) $$\frac{\Delta\omega}{\omega} \cong \frac{2}{2N}$$

Consequently, the spectral intensity profile of the radiation emitted by a single undulator is given by:

(56) $$f(\nu) \propto \left[\frac{\sin(\frac{\nu}{2})}{\frac{\nu}{2}}\right]^2$$

$$\nu = 2\pi N\frac{\omega_0 - \omega}{\omega_0}$$

$f(\nu)$ is illustrated in Fig. 29 and can also be written as:

(57) $$f(\nu) = 2Re[\int_0^1 (1-t)e^{-i\nu t}dt]$$

$\nu$, the so-called detuning, is a very useful parameter, which is used to define the spectral profile with respect to the central reference wavelength. Since $\nu$ is a dimensionless parameter, the width of the profile at half of the maximum intensity is close to 2π.

We explained with simple considerations of kinematics of classical physics how is selected the frequency and defined the bandwidth of the radiation, produced by the free electrons travelling in the magnetic field of an undulator.

Therefore, the undulator is an assembly of magnets, which forces the electron beam to move in the transverse direction and consequently produce radiation at a given frequency within quite a narrow bandwidth.

Such a system belongs to the *insertion devices* family. Before going further, we present in Fig. 30 the evolution over the past few years of those magnetic systems, all dedicated to the production of radiation from ultra relativistic beams. The archetype of an insertion device was a bending magnet, which is characterized by a broad band spectrum. It was followed by the wiggler, which can be considered as a succession of bending magnets, by the undulator previously described and finally by the FEL operated in the self-amplified spontaneous emission configuration which we will discuss in details later. The evolution of the insertion devices can be mainly characterized by two parameters: the bandwidth and the intensity of the produced radiation. We will see in the concluding paragraphs the role played by the improvement of these parameters in the users applications.

In between the undulator and the SASE FEL, has been developed a whole technology that can not be shortly described. It mainly concerns the oscillator FELs, which scheme is given in Fig. 31.

An oscillator FEL is similar regarding its main principles to a conventional oscillator laser. Indeed, we find again a spontaneous emission process from a free electron beam in an optical cavity, the storing of the radiation inside the cavity, and the amplification of this radiation through out a mechanism of stimulated emission. Figure 31 also shows that during the interaction between the electrons and the radiation, the electron beam is being modulated in energy which results in a spatial modulation (bunching) at the radiation wavelength. This modulation enables the production of coherent radiation.

The mechanism of interaction in between the electrons and the radiation of a FEL can be considered as the combination of two competing effects:

***Energy loss from the electron and then amplification of the incident photon, absorption of the stimulating photons with consequent yielding of energy to the electrons*.**

Such mechanism can be understood quite easily using a reference system in which the electrons are non relativistic. In such system, the electrons see two electromagnetic fields: the laser field travelling in the same direction, and the undulator field which proceeds in the opposite direction ("Virtual quanta" model of Weizsacker and Williams, 1934). In Figure 32, we reported the processes of stimulated emission and absorption, similar to the processes of stimulated emission in conventional lasers.

1. the stimulated absorption can be interpreted as a backward scattering of the photons by the laser field. The electrons gain energy, which corresponds to a negative gain;
2. the stimulated emission can be interpreted as a forward scattering of the photons by the undulator field, with a frequency close to the laser frequency. The electrons loose energy, corresponding to a positive gain.

The spontaneous emission process is interpreted using Fig. 32 without stimulating radiation.

If the two effects are considered independent, the gain, i.e. the stimulating field energy increase, can be written as the result of these two effects (stimulated emission and absorption):

(58) $$G(\nu) \propto f(\nu_+) - f(\nu_-)$$

$f(\nu_{+/-})$ refers to the cases illustrated in the Figure, to emission and absorption.

Since $\nu_+ - \nu_- \propto \Delta E_{rec}$, where $\Delta E_{rec}$ represents the energy loss of one single electron via recoiling, the development of Eq. (58) at the lowest order leads to:

(59) $$G(\nu) \propto -\frac{\delta f(\nu)}{\delta \nu}$$

Taking into account that the radiation is emitted by an electron beam at a given current, the gain can be finally expressed as follows:

$$G(\nu, g_0) = -\pi g_0 \frac{\delta f(\nu)}{\delta \nu} = 2\pi g_0 \int_0^1 t(1-t)\sin(\nu t)\,dt$$

$$g_0 = 4\pi \frac{|J|}{I_0} (\frac{N}{\gamma})^3 (\lambda_u K^* f_b)^2$$

(60) $$I_0 = \frac{4\pi}{Z_0 c} \frac{m_e c^3}{e} = 1.7045.10^4 A$$

$$f_b = J_0(\frac{K^{*2}}{2(1+K^{*2})}) - J_1(\frac{K^{*2}}{2(1+K^{*2})})$$

$$K^* = \frac{K}{\sqrt{2}}$$

$g_0$ is the small signal gain coefficient, $I_0$ the Alfven current,

The gain curve is reported in Fig. 33 (one can note that the maximum of the gain does not correspond to the maximum of the spontaneous emission intensity, but located at $\nu_0 \cong$ 2.6). J is the current density of the electron beam and $f_b$ the Bessel factor which takes into account the non perfectly sinusoidal trajectory of the electrons in the linear polarized undulator.

It is worth mentioning that the gain curve is asymmetric. It consists of one part of positive gain and one part of absorption where the electrons, instead of giving energy to the radiation, absorb some energy from the radiation.

Now that the existence of a gain is clarified, we will determine the other typical characteristic of an oscillator FEL: the saturation intensity. We already highlighted that the FEL process consists in a power transfer from the electron beam to the laser. The gain curve is asymmetric but the FEL process is maintained until the cinematic conditions are such that a positive gain is guaranteed and that a sort of saturation is reached.

The energy loss can be deduced from the width of the positive gain region, related to the detuning parameter (see Fig. 33 and Eq. (58)) by:

(61) $$\Delta\nu \cong 4\pi N \frac{\Delta\gamma}{\gamma} \cong 2\pi$$

From this equation comes the expression of the energy variation of the electron beam induced

by the FEL interaction:

(62) $$\frac{\Delta\gamma}{\gamma} \cong \frac{1}{2N}$$

We finally get the expression of the power, transfered from the radiation to the electrons (see Eq. (46)):

(63) $$P_L \cong \frac{P_E}{2N}$$

Recording Eq. (21) and assuming that a similar dynamics is valid in the case of FEL, the saturation intensity can be expressed as:

(64) $$I_S \cong \frac{I_E}{2Ng_0}$$
$$I_E = \frac{|J|E}{e}$$

or in more practical units:

(65) $$I_S[\frac{MW}{cm^2}] = 6.9312.10^2 \frac{1}{2}(\frac{\gamma}{N})^4(\lambda_u[cm]K^* f_b)^{-2}$$

In conclusion, we demonstrated that an oscillator FEL behaves like a conventional oscillator laser and therefore that most of its techniques can be applied to the FEL.

# 7. Oscillator FEL and Mode-Locking

In the former paragraph, we demonstrated the existence of gain and expressed the saturation intensity. In analogy with the conventional laser, we can now write the saturation law of the gain:

(66) $$G(X) \cong \frac{G_M}{F(X)}, \quad X = \frac{I}{I_s}$$
$$G_M = 0.85 g_0, \quad \text{for } g_0 \le 0.2$$

$G_M$ represents the maximum gain calculated at $\nu_0 \cong 2.6$ and:

(67) $$F(X) = 1 + \alpha X + \beta X^2, \alpha + \beta = 1, \alpha = 2(\sqrt{2} - 1), \beta = 3 - 2\sqrt{2}$$

One can notice that in the case of the FEL, the saturation results slightly different from the conventional laser (presence of a quadratic term), but this is just a technical detail which does not modify the physics of the process.

The use of rate equations as defined in Eq .(12) enables to obtain the evolution of the laser power in the cavity.

We recall first that the accumulated power in the cavity is given by the following iteration:

(68) $$I_{r+1} - I_r = [(1-\eta)G(X_r) - \eta]I_r,$$

which is an equation describing the evolution of the intra cavity intensity in terms of *r*, i.e. of the number of round trips in the cavity. $\eta$ in Eq. (68) indicates the total losses of the cavity. As already seen, the signal increases initially in the exponential mode, then the increase slows down and finally stops (or nearly) when the gain equals the losses. The equilibrium power in the cavity is obtained from the condition $I_{r+1} = I_r$ , which implies:

(69) $$G(X_e) = \frac{\eta}{1-\eta}$$

Together with Eq. (66) and (67), this leads to the expression of the equilibrium intensity in the cavity:

(70) $$I_e = (\sqrt{2} + 1)(\sqrt{\frac{1-\eta}{\eta} G_M} - 1)I_S$$

The solution of Eq. (68), (69) and (70) can be written as (see Appendix B):

(71) $$I_{r+1} = I_0 \frac{[(1-\eta)(G_M + 1)]^r}{1 + \frac{I_0}{I_e}\left\{[(1-\eta)(G_M + 1)]^r - 1\right\}}$$

$I_0$ is the initial radiation intensity, due for instance to the spontaneous emission.

An example of evolution of the laser signal in the cavity is presented in Fig. 34. The evolution is typically sigmoidal, as in the case of the conventional lasers.

We will come back to this issue later.

We now discuss the mechanism of mode-locking in the case of oscillator FEL. Before coming into details, we try to understand the number of longitudinal modes in the case of the FEL process.

From the gain curve, whose width is given by:

(72) $$(\Delta f)_{FEL} = \frac{c}{2N\lambda}$$

can be retrieved the number of coupled modes:

(73) $$N_m = \frac{(\Delta f)_{FEL}}{(\Delta f)_c} = \frac{L}{N\lambda}$$

It is worth noticing that such value is inversely proportional to the slippage $N\lambda$. The physical meaning of this relation will be discussed later.

In the absence of a specific mechanism of coupling of the longitudinal modes, the oscillator FEL modes also oscillate independently producing a fluctuating output radiation. But in the case of oscillator FEL, there is a natural coupling mechanism, resulting from the electron packets themselves, delivered by the accelerator with a pulse train structure of finite duration.

In Figure 35 is illustrated the structure of an electron beam delivered by a Radio-Frequency accelerator, which consists of a series of macro pulses themselves consisting of a train of micro pulses which we assume to be Gaussian:

(74) $$f(z) = \frac{1}{\sqrt{2\pi}\sigma_z} e^{-\frac{z^2}{2\sigma_z^2}}.$$

The current of a micro pulse corresponds to the peak current, while the current of the macro pulse corresponds to the average current. The case is similar to the one discussed in the case of conventional laser pulses.

The gain process is determined by micro pulses and is therefore highly dependent on the electron beam finite distribution.

As clearly illustrated in Fig. 36, if the FEL process is determined by a pulsed electron beam structure, each packet of electrons will provide a packet of radiation. This is the basis of the mode-locking mechanism, discussed previously.

In order to understand the mechanism, the electron beam pulse must be considered as a gain filter. This can be easily done considering the Fourier transform of the electron beam in the frequency domain of the FEL (see also Fig. 37):

$$\tilde{f}(\nu,\mu_c) = \frac{1}{\sqrt{2\pi}}\int_{-\infty}^{\infty} f(z)e^{-i(k-k_0)z}dz$$

(75)
$$= \frac{1}{\sqrt{2\pi}}\int_{-\infty}^{\infty} f(z)e^{-i\frac{\nu}{N\lambda}z}dz \propto \frac{1}{\sqrt{2\pi}\mu_c}e^{-\frac{\nu^2}{2\mu_c^2}}$$
$$\mu_c = \frac{\Delta}{\sigma_z};\Delta = N\lambda$$

Δ is referred as the **slippage length**. The Fourier transform of *f(z)* is Gaussian too and in the following, we will use the normalized form:

(76)
$$\tilde{f}(\nu,\mu_c) = \frac{1}{\sqrt{2\pi}\,\mu_c}e^{-\frac{\nu^2}{2\mu_c^2}}$$

The filter, representing the electron beam, is therefore regulated by the quantity $\mu_c$, referred as **longitudinal coupling coefficient**.

This quantity has a all-purpose physical meaning, which will be discussed in detail later. Here, we just highlight that when $\mu_c$ increases, the number of coupled modes increases, and that for a continuous beam ($\sigma_z \to \infty$) the operation becomes essentially single mode. The distribution derived from the former equation becomes a Dirac peak.

The gain "filtered" by the electron beam can be written in terms of a convolution:

(77) $$G(\nu,\mu_c) = \int_{-\infty}^{\infty} G(\nu + y)\tilde{f}((\nu,\mu_c)dy = 2\pi g_0 \int_0^1 t(1-t)\sin(\nu t)e^{-(\mu_c t)^{2/2}}dt$$

It appears from this relation that for positive coupling coefficient values, the gain is decreased, simply because there are more modes oscillating in phase, and according to Eq. (75) that those coupled modes are within the rms band:

(78) $$\sigma_\nu \cong \mu_c.$$

The former considerations provide qualitative indications only. More quantitative considerations will be given later.

Before concluding this paragraph, we try to explain why the optical pulse width of an oscillator FEL is related to the electron beam width. The following reasoning is simple but efficient.

It is obvious that in the first steps of the process, the radiation emitted by the electrons is essentially incoherent and the coherence develops only after a certain number of interactions between the optical wave and the electron beam in the cavity. The maximum coherence is reached when the spatial and spectral distribution are correlated. Such correlation is guaranteed by the Schwartz identity. During the initial steps, the optical pulse length and its spectral width are approximately equal to the electron beam pulse length and the spectral width of the gain curve. Going further in time, i.e. higher in the number of round trips, both quantities decrease. We can write:

(79) $$\sigma_r \cong \frac{\sigma_z}{n}$$ $$\Delta\omega = \frac{\omega_0}{2nN}$$

$\sigma_r$ is the rms width of the radiation packet.

The coherence of the laser field is reached when:

(80) $$\frac{\sigma_r}{c}\Delta\omega \cong \pi$$

which is obtained for

(81) $$n^* \cong \frac{1}{\sqrt{\mu_c}}$$

We can now derive an approximate laser pulse duration:

(82) $$(\sigma_r)^* \cong \sqrt{\Delta.\sigma_z}$$

Up to now, the considerations have been qualitative but sufficient for our needs. In the next paragraphs, we will present more rigorous considerations, referring to well-defined experimental cases.

# 8. FEL Oscillator Dynamics and Harmonics Generation

The mechanism that we described previously deals with a sort of active mode-locking and the detailed comparison with what occurs in the case of conventional lasers is out of the scope of this work. But it is worth reconsidering more rigorously some of the notions seen before and to introduce the **FEL equations**.

The dynamics of an electron beam which propagates in the $z$ direction in the magnetic field $\vec{B}_u$ of an undulator (see Eq. (46)) and interacts with a planar electromagnetic wave including subharmonic content, is described by the Lorentz equations coupled to the Maxwell equation (Gaussian units):

$$m_e \frac{d}{dt}(\gamma\vec{v}) = -e\left(\vec{E}_s + \frac{1}{c}\vec{v}\times\vec{B}\right)$$

$$\frac{d\gamma}{dt} = -\frac{e}{m_e c^2}\vec{E}_s \cdot \vec{v}$$

$$\left(\nabla^2 - \frac{1}{c^2}\frac{\partial^2}{\partial t^2}\right)\vec{A} = -\frac{4\pi}{c}\vec{J}$$

$$\vec{E}_s = -\frac{1}{c}\frac{\partial}{\partial t}\vec{A}$$

In the 1D approximation and in the case of a linear polarized undulator, we assume:

(83)
$$\begin{aligned}
&\vec{B} = \vec{B}_s + \vec{B}_u \\
&\vec{B}_u = [0, B_0 sin(k_u z), 0] \\
&\tilde{E}_s = \left[\sum_n E_{s,n}\cos(\psi_n), 0, 0\right] \quad n \ \ is \ \ odd \\
&\vec{B}_s = [0, \sum_n E_{s,n} cos(\psi_n), 0] \\
&\vec{k}_u = \frac{2\pi}{\lambda_u} \\
&\psi_n = k_{s,n} z - \omega_n t + \phi_n
\end{aligned}$$

with $k_{s,n} = \frac{2\pi}{\lambda_{s,n}}$, $\lambda_{s,n} = \frac{\lambda_s}{n}$, $\omega_n = ck_{s,n}$ respectively the wave number, the wavelength and the frequency of the harmonics of the laser field (the index $n$=1 for the fundamental will be omitted in the following).

In the approximation of a continuous electron beam ($\lambda_s << \sigma_z$ and current intensity $J$ constant), averaging over one undulator period, using the paraxial approximation for the Maxwell equation, for slowly varying field amplitude (γ nearly constant along the undulator, i.e. in the small signal gain conditions), one obtains the **Colson equations** (pendulum approximation):

(84)
$$\begin{aligned}
&\frac{d^2}{d\tau^2}\zeta = \sum_n |a_n| cos(\psi_n), \psi_n = n\varsigma + \phi_n \\
&\frac{da_n}{d\tau} = -j_n \left\langle e^{-in\varsigma}\right\rangle, j_n = 2\pi g_{0,n}
\end{aligned}$$

$$\tau = \frac{ct}{L_u} \cong \frac{z}{L_u}$$

(85)

$$\varsigma = (k_u + k_s)\int_0^t \bar{v}_z(t')dt' - \omega t \quad \textit{dimensionless electron phase}$$

$\bar{v}_z$ *average velocity over one undulator period*

$\langle\ \rangle$, *indicates average over electrons*

The dimensionless amplitudes $a_n$ are defined as:

(86)

$$a_n = |a_n| e^{i\phi_n}$$

$$|a_n| = \frac{k_s}{\gamma^4} K^*(N\lambda_u)^2(1+K^{*2}) f_{b,n} e_{s,n}$$

(87)

$$e_{s,n} = \frac{eE_{s,n}}{\sqrt{2} m_e c^2} \quad \textit{normalized amplitude}$$

$$f_{b,n}(\xi) = (-1)^{\frac{n-1}{2}}[J_{\frac{n-1}{2}}(n\xi) - J_{\frac{n+1}{2}}(n\xi)]$$

$$g_{0,n} = g_0\left(\frac{f_{b,n}}{f_{b,1}}\right)^2$$

with $g_0$, $K^*$ and $\xi$ defined in Eq. (60). We note that $|a|^2 = 8\pi^2 \frac{I}{I_s}$. The evolution equation of the amplitude of the field can be written as:

(88)

$$\frac{da}{d\tau} = i\pi g_0 \int_0^\tau \tau' a(\tau - \tau') e^{-i\nu\tau'} d\tau'$$

Such equations take into account some aspects of the FEL dynamics, such as variation of the field phase and intensity. In addition, Eq. (88) includes the effects of high gain, i.e. the corrections associated to the fact that a high gain can vary the field significantly during the interaction.

Using the following definition of the gain:

(89)

$$G = \frac{|a|^2 - |a_0|^2}{|a_0|^2}$$

the antisymmetric curve can be derived from Eq. (88). This curve was previsouly obtained using the small signal gain approximation which assumes $g_0 << 1$ and $a(\tau - \tau') \cong a(\tau)$.

Such approximation is rather satisfactory for small signal gain coefficient values lower than $30\%$. For higher values, the maximum gain is written as:

(90) $$G_M = G(g_0) \cong 0.85 g_0 + 0.192 g_0^2 + 4.23\times 10^{-3} g_0^3, g_0 \le 10$$

and

(91) $$G_M \cong \frac{1}{9}[1 + \frac{1}{\sqrt{3}(\pi g_0)^{1/3}}]\exp[\sqrt{3}(\pi g_0)^{1/3}], g_0 \ge 10$$

Equation (88) considers an ideal electron beam, i.e. with all electrons at the same energy. In reality, the energy distribution of the electron beam can be assumed Gaussian:

(92) $$f(\varepsilon) = \frac{1}{\sqrt{2\pi}\sigma_\varepsilon} e^{-\frac{\varepsilon^2}{2\sigma_\varepsilon^2}}$$
$$\varepsilon = \frac{\gamma - \gamma_0}{\gamma_0}$$

where $\sigma_\varepsilon$ is the electron beam energy spread. To the energy distribution corresponds an analog distribution in the frequency domain of the FEL gain:

(93) $$\tilde{f}(\nu, \mu_\varepsilon) = \frac{1}{\sqrt{2\pi}(\pi\mu_\varepsilon)} e^{-\frac{\nu^2}{2(\pi\mu_\varepsilon)^2}}$$

(94) $$\mu_\varepsilon = 4N\sigma_\varepsilon$$

Following the same procedure of convolution, as in the case of the longitudinal modes distribution, Eq. (88) is modified as following:

(95) $$\frac{\partial a}{\partial \tau} = i\pi g_0 \int_0^\tau \tau' a(\tau - \tau') e^{-i\nu\tau'} e^{-(\frac{\pi\mu_\varepsilon \tau'}{2})^2} d\tau'$$

The reduction of the gain due to the energy distribution is similar to the reduction of the gain due to the inhomogeneous broadening in conventional lasers and the parameter $\mu_\varepsilon$ is one of the inhomogeneous broadening parameter of the FEL. The reduction of the gain due to such effect can be quantified according to:

(96) $$G \cong \frac{G_M}{1+1.7\mu_\varepsilon^2}, \ \mu_\varepsilon \le 1$$

A further source phenomenon of inhomogeneous broadening is due to the fact that the electron beam at the undulator entrance has non zero transverse dimensions. Assuming uncorrelated transverse components, the transverse distribution of the beam in $x, y$ plans can be written as following:

(97)
$$F(x,x',y,y') = W(x,x')W(y,y') \quad \eta' = \frac{1}{c}\frac{d\eta}{dt}$$
$$\mathrm{W}(\eta,\eta') = \frac{1}{2\pi\varepsilon_\eta} exp\left[-\frac{1}{2\varepsilon_\eta}(\beta_\eta\eta'^2 + 2\alpha_\eta\eta\eta' + \gamma_\eta\eta^2)\right]$$

(98)
$$\varepsilon_\eta \ \ emittance \ \ in \ \ the \ \ plan(\eta,\eta')$$
$$\alpha_\eta,\beta_\eta,\gamma_\eta \ \ Twiss \ \ coefficients \ \ with \ \ \beta_\eta\gamma_\eta - \alpha_\eta^2 = 1$$

We will not treat the effect of reduction of the gain due to the emittance. More details are given in Appendix A.

The convolution of the gain with the electron packet for the coupling study of the modes is correct in an heuristic point of view, but does not take into account the fact that the photon packet, having a higher velocity that the electrons, slips over the electron packet (see Fig. 38).

Therefore, Equation (95) should be modified according to:

(99) $$\frac{da(z,\tau)}{d\tau} = i\pi g_0(z+\Delta\tau)\int_0^\tau \tau' a(z+\Delta\tau',\tau-\tau')e^{-i\nu\tau'} e^{-(\frac{\pi\mu_\varepsilon\tau'}{2})} d\tau'$$

$g_0(z+\Delta\cdot\tau)$ is the small signal gain coefficient which takes into account the electrons distribution $f(z)$, Eq. (74) and the slippage.

The former equation is sufficiently general to describe the dynamics of the pulse propagation with the effects of a high gain.

The coupling parameter $\mu_c$ has an additional physical meaning which is worth mentioning: it measures the overlapping of the electron beam with the photon beam. A very short electron beam remains superimposed to the photon beam for a very short time, which drives another reduction of the gain.

The reduction of the gain due to the multi mode coupling and spatial overlap can be quantified as following [5]:

(100) $$G \cong \frac{G_M}{1+\frac{1}{3}\mu_c}$$

An important aspect of the FEL dynamics is the non linear harmonic generation, resulting from the bunching mechanism, which itself is a consequence of the energy modulation mechanism (see Fig. 39). When the electrons are microbunched at a wavelength which corresponds to a submultiple of the fundamental wavelength, coherent emission is produced on higher order harmonics, through out the mechanism of non linear harmonic generation.

The intra cavity evolution of the fundamental and of the higher order harmonics is presented in Fig. 40. The dynamics of the process is understood as follows: the fundamental increases, until it reaches a sufficiently high power level to induce a bunching, that can lead to non linear harmonic generation. The saturation mechanism comes from the combined effects of the energy loss and energy spread increase of theelectron beam due to the same interaction.

The maximum harmonic power is related to the maximum of the fundamental power according to [6]:

---

5) Both Eqs. (97) and (100) are valid in the case of small gain ( $g_0 < 0.3$). For more general hypothesis, the expressions are slightly more complicated and will not be reported here for simplicity.

6) Note that in the linear polarized undulators, radiation is emitted on-axis only on odd order harmonics (3,5,7..).

(101) $$P_n = \frac{1}{2}\frac{\sqrt{n}}{n^3}\sqrt{\frac{n-1}{2}}g_{0,n}\frac{P_E}{4N}$$

where $g_{0,n}$ is the small signal gain coefficient of the harmonic $n$. Note that the coefficient $\frac{P_E}{4N}$ is the maximum power $P_L$ achievable on the fundamental. The higher order harmonics, opposite to the fundamental, are not stored in the cavity. Therefore the powers presented in Fig. 40 are relative to the harmonic generation in one single round trip.

The non linear harmonic generation mechanism is similar to the frequency-mixing mechanism in the case of non linear optics. Eq. (101) can be understood as the non linear response of the medium (the electron beam) to the laser electric field, so that it can then be written in terms of amplitude of the harmonics electric field:

(102) $$E_n = \chi_n E_L$$ $$\chi_n = \sqrt{\frac{g_{0,n}}{2}\frac{\sqrt{n}}{n^3}\sqrt{\frac{n-1}{2}}}$$

Before concluding this paragraph, we will describe the profile of the pulses in a FEL oscillator and the method which can be implemented to shape the pulses simply using the cavity parameters.

We saw that the interaction between the radiation and the electron beam is quite complex, in particular the mechanism that gives rise to the gain and to the FEL dynamics. The process has a dispersive nature: it involves a phase and intensity variation of electromagnetic fields. The dynamics of the optical packet can be characterized by an index refraction corresponding to a group velocity which can be defined as:

(103) $$v_g = \frac{c}{n_g}$$

The former relation can be interpreted as the fact that the centroid velocity of the optical pulse is lower than the light velocity because of the gain itself. More precisely, one can note that such effect is the exact analog of what occurs in materials, i.e. the phase velocity of an electromagnetic wave is reduced because the electric field creates a perturbation in the material charges which results proportional to the permittivity of the medium.

The medium charges, essentially electrons, oscillate with a given phase delay with respect to the perturbative wave, and emit radiation at the same frequency but with a phase delay. The macroscopic sum of the contribution of each single charge is a wave at the same frequency but with a shorter wavelength than that of the perturbative field, with consequent variations of the group velocity. The group refraction index is related to the refraction index by the relation:

(104) $$n_g = n_r(1 - \frac{\lambda}{n}\frac{\partial}{\partial\lambda}n_r)$$

$n_r$ is the medium refraction index and $\lambda$ the wavelength in vacuum.

Without coming into derivation details, we can note that the effect of *breaking* of the FEL optical pulse, also referred as *lethargy*, can be reproduced quantitatively in terms of a complex refraction index related to the complex gain function.

One of the practical consequences of such effect is (see Fig. 41), if the overlap between the electrons and laser beams is to be conserved at each round trip, that the cavity must be tuned with respect to the reference length defined as twice the distance in between the electron packets.

The FEL gain and the output power are then functions of this new parameter $\delta L$:

***the variation of the cavity length with respect to the synchronism condition for an empty cavity***.

This parameter modifies other important quantities such as the profile of the packet, as we will discuss in the next paragraph.

Introducing $\delta L$ ($\delta L > 0$), the expression of the round trip period becomes:

(105) $$T_R = 2\frac{L_1}{c} + (\frac{L_u}{v} + \frac{L_u}{c}) + 2\frac{L_2 - \delta L}{c}$$

Note that the optical pulse interacts with the electrons only during the forward part of the round trip (mirror 1 to 2). In the backward part of the round trip (from 2 to 1), there is no FEL interaction. The synchronism condition is obtained assuming that:

(106) $$T_R = T_E = 2\frac{L_c}{c}, L_c = L_1 + L_u + L_2$$

A relation between the group refraction index and cavity tuning can be deduced from the synchronism condition:

(107) $$n_g = 2\frac{\delta L^*}{L_u} + 1.$$

$\delta L^*$ indicates the mismatch which compensates the effect of lethargy.

# 9. Pulse Generation in Oscillator FEL

We saw in the former paragraphs that the oscillator FEL dynamics, operated with electron pulses, leads to various problems related to the coupling of the longitudinal modes and to the desynchronization of the cavity. Therefore, one expects that the gain and the output power depend on the parameters $\mu_c$ and $\delta L$.

As far as $\mu_c$ is concerned, we clarified that its role is twofold and that it is involved in the FEL process via active mode-locking and via kinematical overlap. The effect of such quantity on the gain has been given in Eq. (100). Including the cavity mismatch then modifies the gain as following (valid for $g_0 \leq 0.3$):

(108) $$G(\theta,\mu_c) = G_M \frac{\theta}{\theta_s}\left[1 - \ln\left(\frac{\theta}{\theta_s}\gamma(\mu_c)\right)\right]$$
$$G_m = 0.85 g_0$$

$$\gamma(\mu_c) = 1 + \frac{\mu_c}{3}, \theta_s \cong 0.456, \theta = \frac{4\delta L}{g_0 \Delta}$$

The saturation mechanism described in the case of oscillator FEL with continuous electron beams is still valid in the case of FEL oscillator with pulsed electron beams. The saturated gain is then defined as:

(109)

$$G(X,\theta,\mu_c) = G_M \frac{\theta}{\theta_s}\left[1 - \ln\left(\frac{\theta}{\theta^*} F(X)\right)\right]$$

$$\theta^* = \theta_s(\gamma(\mu_c))^{-1} \quad \textit{desynchronism parameter}$$

where the saturation function $F(X)$ has the form given in Eq.(66), with $X = \frac{I}{I_s^*}$ and $I_s$ given by:

(110)
$$I_s^* = I_s.[1 - exp(-\frac{3}{2}(\frac{g_0}{\sqrt{\mu_c}})^{1/2})]$$

The equilibrium power can be obtained from Eq. (69):

(111)
$$X_e(\theta, \mu_c) \cong (\sqrt{2}+1)[\sqrt{\frac{\theta_s}{\theta\gamma(\mu_c)}} exp[\frac{1}{2}(1 - \frac{\eta}{1-\eta}\frac{\theta_s}{G_M \theta}] - 1]X_s$$

In Figure 42, we reported the net gain and intensity at equilibrium as a function of the cavity mismatch. The net gain of the oscillator, defined as $\frac{I_{n+1} - I_n}{I_n}$, in the exponential growth region is given by $G_{net} = (1-\eta)G_M - \eta$.

In Figure 43 is presented a comparison between the theoretical equilibrium intensity and the numerical simulation.

As a comment to the previous figures, we note that:

- the maximum of the gain curve as a function of $\theta$ is given in Eq. (100) and is obtained for $\theta = \theta^*$. The corresponding value of the mismatch is:

(112) $$\delta L^* = \frac{\theta_s}{4}\frac{g_0\Delta}{\gamma(\mu_c)}$$

Together with Eq. (107), we obtain the expression of the refraction index:

(113) $$n_g = \frac{\theta_s}{2}\frac{\Delta}{L_u}\frac{g_0}{\gamma(\mu_c)} \cong 0.27\frac{\Delta}{L_u}G_M(\mu_c)+1$$
$$G_M(\mu_c) = \frac{0.85 g_0}{\gamma(\mu_c)}$$

- the maximum of the intra cavity power is not obtained for $\vartheta = \vartheta^*$. The origin of such a difference can be simply explained as follows. When the power increases, the gain reduces and $n_g$ also reduces. Therefore, moving towards saturation, the effect of lethargy becomes less relevant and a lower $\delta L$ is required to optimize the intra cavity power.

The former considerations are valid in general, but the validity range of Eqs. (108-113) is limited to the case of small signal gain.

In the case of high gain, the physical process are basically similar, but the formula are more complicated and will not be reported here for shortness.

$\delta L$ is highly important as far as the control of the laser pulse shape is concerned. This is illustrated in Fig. 44, where we reported the power and the shape of the pulses for various mismatching conditions. The laser and electron pulse at the entrance of the undulator are also presented in the Figure, in the saturated regime. Assuming a cavity shortened more than what is necessary to compensate the effect of lethargy, the radiation pulse is located, at each entrance in the undulator, at the head of the electron beam. Of course, the radiation pulse is located at the rear in the case of a too long cavity.

We now consider what occurs when the photon beam is located slightly at the rear of the electron beam (top right hand side of Fig. 44). The radiation overlaps the electrons all along the undulator and produces a short pulse with a sort of side-band which increases after the main peak in a region where the energy spread is not yet high enough to produce saturation effects. In the case of stronger displacements, the dynamics is even more complicated and the multi-peak structure more pronounced. More details will be given later. If the cavity is significantly shortened, the electron pulse tends to loose part of the radiation which naturally decreases with the cavity losses.

We now detail the role played by the pulses dynamics in the process of non linear harmonic generation.

As we have seen before in the case of a shortening of the cavity length, the radiation pulse tends to be located, at the undulator entrance, at the head of the electron beam. This results in an energy modulation and bunching with deterioration of the local energy distribution. The combination of these effects with the effect of slippage leads to what is shown in Fig. 45 and Fig. 46, i.e. the production of short sub-harmonic pulses.

As shown in Fig. 46, the harmonic pulse width tends to shorten when it gets closer to saturation. Indeed, the electron beam has been deteriorated more significantly where the main interaction with the fundamental occurred.

# 10. FEL Oscillator in the VUV Region

The FEL oscillators covered a large part of the electromagnetic spectrum (see Fig. 47). The ultra-violet region, below 200 nm is covered by FEL based on high energy accelerators such as storage rings. In this paragraph, we discuss the possible operation of a FEL in the VUV region, using a high energy linear accelerator (LINAC) to produce radiation around 120 nm on the fundamental and down to a few tens of nm thanks to the coherent harmonic generation mechanism.

It is obvious that the available spectral range of operation of oscillator FEL is limited by the disponibility of high reflectivity mirrors, i.e. limited to wavelengths longer than 120 nm. Assuming high reflectivity optics down to this limit, one can hope to reach coherent radiation, via harmonic generation, down to 24 nm.

Before presenting a specific example of oscillator FEL at this fundamental wavelength, we consider the problem of the extraction of the power from an optical cavity.

As we have seen before, the intra cavity power depends in the gain and in the losses which generally consist of one active (transmission) and one passive (absorption) component. The total losses, introduced in Eq. (70), are then expressed as following:

(114) $$\eta = \eta_A + \eta_P$$

The output power is related to the losses according to:

(115) $$I_{out} = \eta(\sqrt{2}+1)(\sqrt{\frac{1-\eta}{\eta}G_M} - 1)I_S$$

The power which can be extracted (i.e. transmitted) depends on the active losses:

(116) $$I_T = \eta_A(\sqrt{2}+1)(\sqrt{\frac{1-(\eta_A+\eta_P)}{\eta_A+\eta_P}G_M} - 1)I_S$$

The extraction can be optimized by maximizing the transmitted power with the losses, i.e. solving the following equation (Fig. 48):

(117) $$\frac{\partial}{\partial \eta_A} I_T = 0$$

from which we have:

(118) $$\eta_A = \frac{1}{2(1+r)}[1-(\frac{1}{1+G_M})^{1/2}]; r = \frac{\eta_P}{\eta_A}$$

For small $G_M$ values, the optimal losses and output intensity are given by:

$$\eta_A^* \cong \frac{1}{4}\frac{G_M}{1+r}$$

(119)

$$I_0^* \cong 0.5 G_M I_S$$

In the case of the small signal gain signal and according to Eqs. (63) and (65):

$$I_0^* \cong \frac{0.85}{4N} I_E \tag{120}$$

We now move on to a specific example of oscillator FEL. We consider an oscillator FEL operated with the parameters given in Table I and a 3 m long confocal optical cavity.

In this case, the operation wavelength is 120 nm and the gain is quite large in order to compensate high total cavity losses. In Figure 49 is presented the evolution of the intra cavity intensity evolution on the fundamental and first harmonics in the case of $20\%$ total cavity losses.

In Figure 50, we reported the evolution of the laser pulse after various numbers of round trips, while in Figs. 51-52 are shown the harmonic pulses. Finally, in Figs. 53, 54, 55 and 56 are shown the normalized pulse profiles of the harmonics for various cavity lengths (we have denoted by δ the ratio $\delta = \frac{\delta L}{\lambda}$, negative values of δ correspond to an increase of the cavity length)

We conclude looking at the results reported in Table II, which confirm the possibility of the operation of a FEL oscillator at a fundamental wavelength around 120 nm and delivering coherent radiation in the VUV region thanks to harmonic generation.

The example that we presented is at the limit of the technology as far as mirrors are concerned. Those may not even support the heating induced by the intra cavity power. Further development of the high reflectivity optics would enable to reach even shorter wavelengths.

# 11. FEL Operating in the High Gain Regime

In the former paragraph, we understood that one of the limiting factors of the FEL oscillator are the mirrors which restrict the spectral range to wavelengths longer than a few hundreds of nm. Therefore, the soft X-ray region seems out of the range of operation of the FEL oscillator.

But it is possible to implement high gain mechanism which enables to reach saturation in one single pass in the undulator. One possible scheme is given in Fig. 57. A high peak current electron beam is injected in a long undulator, and amplifies the signal emitted in the first periods of the undulator. Such FEL configuration is referred as Self Amplified Spontaneous Emission (SASE) FEL.

The main steps of the process are:

- Spontaneous emission
- Energy modulation
- Longitudinal bunching of the electrons over a distance of the order of the spontaneous radiation wavelength
- Coherent emission
- Saturation

In order to understand these steps, we go back to the integral form of the FEL equations (see Eq. (84) and (86)), which can be rewritten as follows:

(121)
$$e^{i\nu\tau}\frac{da}{d\tau} = i\pi g_0 \int_0^\tau (\tau-\xi) e^{i\nu\tau} a(\xi) d\xi$$
$$e^{i\nu\tau}\frac{da}{d\tau} = i\pi g_0 \hat{D}_\tau^{-2}[a(\tau)e^{i\nu\tau}$$

$D_\tau^{-2}$ indicates an integration (or a negative derivative) repeated twice. We also used the following Cauchy equality for the integrations:

(122)
$$\hat{D}_x^{-n} f(x) = \frac{1}{(n-1)!}\int_0^x (x-\xi)^{n-1} f(\xi) d\xi$$
$$\hat{D}_x^{-n} f(x) = \int_0^x dx_1 \ldots \int_0^{x_n-1} dx_n f(x_n)$$

Deriving both parts of Eq. (121) with respect to the temporal variables, the former integral equation can be transformed into an ordinary third order equation:

$$(\hat{D}_\tau^3 + 2i\nu \hat{D}_\tau^2 - \nu^2 \hat{D}_\tau) a(\tau) = i\pi g_0 a(\tau)$$

(123) $$\hat{D}_n^n = \frac{d^n}{d\tau^n}$$

$$a\big|_{\tau=0} = a_0, \hat{D}_\tau a\big|_{\tau=0} = 0, \hat{D}_\tau^2 a\big|_{\tau=0} = 0$$

The solution of this equation is obtained with a standard method and can be written as following:

(124)
$$a(\tau) = \frac{a_0}{3(\nu+p+q)} e^{-\frac{2}{3}i\nu\tau} \Big\{(-\nu+p+q)e^{-\frac{1}{3}(p+q)\tau} +$$
$$+2(2\nu+p+q)e^{\frac{i}{6}(p+q)\tau}\left[\cosh\left(\frac{\sqrt{3}}{6}(p-q)\tau\right)+i\frac{\sqrt{3\nu}}{p-q}\sinh\left(\frac{\sqrt{3}}{6}(p-q)\right)\tau\right]\Big\}$$
$$p = \left[\frac{1}{2}(r+\sqrt{d}\right]^{1/3}, \quad q = \left[\frac{1}{2}(r-\sqrt{d})\right]^{1/3}$$
$$r = 27\pi g_0 - 2\nu^3, \quad d = 27\pi g_0\left[27\pi g_0 - 4\nu^3\right]$$

This solution is not so easy to handle in the analytical point of view nor easy to interpret in the physical point of view. A more transparent form can be obtained with $\nu$ =0 in Eq. (123). This is justified since we are considering solutions with a high gain, and that since in this case, maximum gain is reached for small values of the detuning parameter. In the $\nu$ =0 case, Eq. (123) becomes much more simple:

(125) $$\hat{D}_\tau^3 a(\tau) = i\pi g_0 a(\tau)$$

The general solution is a linear combination of the three roots which is given by:

(126) $$a(\tau) = \sum_{j=0}^{2} a_j R_j(\tau) = \sum_{j=0}^{2} a_j e^{\delta_j \tau}$$

$\delta_j$ are the three complex roots of $(i\pi g_0)^{1/3}$. The solution is then finally given by:

(127) $$a(\tau) \propto e^{(\pi g_0)^{\frac{1}{3}}\frac{\sqrt{3}}{2}\tau}$$

In addition, using $\tau = \frac{z}{N\lambda_u}$, the former relation becomes:

(128) $$a(z) \propto e^{\frac{z}{2L_G}}$$

where

(129)
$$L_G = \frac{\lambda_u}{4\pi\sqrt{3}\rho}$$
$$\rho = \frac{1}{4\pi}\left(\frac{\pi g_0}{N^3}\right)^{\frac{1}{3}}$$

$L_G$ is referred as the gain length, while $\rho$ is referred as the Pierce parameter and constitutes one of the fundamental parameters of the SASE FEL. Finally, including the three roots in the expression, the evolution of the power in the small signal gain regime is given by:

(130)
$$|a(z)| = A(z)|a_0|^2$$
$$A(z) = \frac{1}{9}(3 + 2cosh(\frac{z}{L_{g,1}}) + 4cos(\frac{\sqrt{3}}{2}\frac{z}{L_{g,1}})cosh(\frac{z}{2L_{g,1}}))$$

This relation enables to describe the initial zone of non exponential growth. In Figure 58 is presented a comparison between the approximated solution of Eq. (128), Eq. (130) and a 1D simulation.

The former equation only describes the intensity increase of the laser, not its saturation. The physical mechanism which determines the saturation is not different from the one which has been previously discussed in the case of the oscillators. To simplify the treatment, we assume that the evolution is just the one relative to the exponential growth, so that we can write the differential equation relative to the growth process as following:

(131)
$$\frac{d}{dz}P(z) = \frac{P(z)}{L_g}(1 - \frac{P(z)}{P_F})$$
$$P(0) = P_0$$

A quadratic non linearity has been added in order to take into account the effects of saturation.

$P_F$ indicates the final power in the saturation regime, that we will specify later. The solution of the former equation can be obtained easily, using the transformation:

(132) $$\frac{d}{dz}T(z) = -\frac{1}{L_{g,1}}\left[T(z) - \frac{1}{P_{F,1}}\right],\ T(z) = \frac{1}{P_1(z)}$$

and can be written:

(133) $$P(z) = \frac{P_0}{9}\frac{e^{z/L_g}}{1 + \frac{P_0}{P_F}(e^{z/L_g} - 1)}$$

In the more general case, the exponential is substituted by $A(z)$:

(134) $$P(z) = P_0\frac{A(z)}{1 + \frac{P_0}{P_F}(A(z) - 1)}$$

The saturation power still remains to be defined. We already noted that when the gain coefficient increases, non linear elements tend to modify the maximum of the gain curve and its antisymmetric shape (Fig. 59).

The values of the small signal gain coefficient considered in the figure range from 0.1 to 10. The values that we should consider to define a typical high gain SASE FEL are reported in Fig. 60, reproducible with a Gaussian distribution:

(135) $$G(\omega) = \frac{1}{\sqrt{2\pi}\rho}e^{-\frac{(\frac{\omega-\omega_0}{\omega_0})^2}{2\rho^2}}.$$

The relative width of the gain curve results propotionnal to the $\rho$ parameter. According to what has been discussed in the former paragraphs, one expects the ratio between the FEL power and the electron beam power to be as well proportional to $\rho$ .A more accurate analysis demonstrated that this is true, though with an additional numerical factor [6]:

(136) $$P_F \cong \sqrt{2}\rho P_E$$

Before concluding this paragraph, we specify other fundamental quantities of the SASE-FEL regime.

In the case of the oscillators, the required time to reach saturation can be estimated as the required time, i.e. number of round trips, for the gain to decrease to the losses level. This can be deduced from Eq. (71) imposing the condition $I_{r+1} \cong I_e$:

(137) $$r^* \cong \frac{ln(I_e/I_0)}{ln((1-\eta)G_M+1)}$$

which corresponds in the time domain to:

(138) $$t^* \cong \frac{2L_c}{c} r^*$$

In the case of the SASE regime, we no longer deal with a time but with a length of saturation, imposing $P(z) \cong P_F$ in Eq. (133). This leads to:

(139) $$Z_F \cong \ln\left(\frac{9P_F}{P_0}\right) L_g$$

In the high gain regime:

(140) $$N_F \cong \frac{1}{\rho}$$

To get some numerical examples, we can note that if one limits the undulator length to a few tens of meters, given that the undulator period is of the order of a few centimeters and that $P_F \cong 10^8 P_0$, the value of the $\rho$ results around $10^{-3}$.

# 12. SASE FEL and Coherence

In the former paragraphs, we saw that the longitudinal coherence in FEL oscillators operated with short electron bunches, is guaranteed by the mode-locking mechanism.

In the case of the FEL operated in the SASE regime, since there is no optical cavity, we can no longer talk about longitudinal modes, strictly speaking.

Nevertheless, we can repeat the same argumentation on the filter properties of the electron beam to understand if something analog to mode-locking can be defined for the cases of FEL operated in the SASE regime.

As we already saw in the case of the oscillators, the effect of mode-locking is guaranteed in the interaction region where the electrons see a field with constant phase. This region is limited to one slippage length. We repeat the same procedure of Fourier transform which enabled to obtain Eq. (75). Using for the frequency domain the variable $\xi = \frac{\omega - \omega_0}{\omega_0}$ and changing the variable for the space domain according to $\tau = \frac{z}{c}$, $\Delta$ being the slippage length, we obtain the integral:

(141) $$\tilde{f}(\xi) \propto \frac{1}{\sqrt{2\pi}} \int_{-\infty}^{\infty} f(\tau) e^{-i\xi\tau} d\tau$$

from which:

(142) $$\tilde{f}(\xi) = \frac{1}{\sqrt{2\pi}\,\tilde{\mu}_c} e^{-\frac{\xi^2}{2\tilde{\mu}_c^2}} \qquad \tilde{\mu}_c = \frac{\lambda}{\rho\sigma_z}$$

We used the approximation $\rho \cong \frac{1}{N}$. The former relation ensures that a coherence length $l_c$ exists and that it is around $\lambda/\rho$ (a more correct definition will be given later). If the electron beam length is about the coherence length, there would not be any problem of longitudinal coherence because all the modes inside the gain curve would be naturally coupled. Since in practical cases $\lambda/\rho << \sigma_z$ and since there is no clear definition of the longitudinal modes, we should define "macro modes", i.e. a sort of extension of the longitudinal modes. The number of "macro modes" is given by:

(143) $$M_L \cong \frac{\sigma_z}{l_c}$$

Such macro modes introduced in the FEL Physics at the end of the 70's by Dattoli and Renieri are referred as ***supermodes*** (see [4] and references therein).

Each of these modes have a spatial distribution given by:

(144) $$m(z) \cong \frac{1}{\sqrt{2\pi}l_c} e^{-\frac{z^2}{2l_c^2}}, \qquad l_c = \frac{\lambda}{4\pi\sqrt{3}\rho}$$

The relative Fourier transform $\tilde{m}(\xi)$ provides with the frequency distribution. The total spectrum is given by:

(145) $$S(\xi) = \sum_{n=1}^{M_L} \tilde{m}_n(\xi)$$

The phase of each component is totally random. The spatial distribution of the optical pulse is given by the Fourier transform of the spectrum and the results is shown in Fig. 61.

Each randomly phased supermode makes one optical pulse which results in a series of spikes separated by a fixed distance: the coherence length (see Fig. 62).

The evolution of each supermode is nearly independent from the others. There is only the slippage which enables to create a coherence zone (of the same order of the coherence length) and produces a sort of smoothing of the chaos, while the field increases along the undulator (see Fig. 63).

Dealing with a partially chaotic phenomenon, we should also define a probability distribution of energy of these spikes [7]:

(146)

$$P(E) = \frac{M^M}{(M-1)!} x^{M-1} e^{-Mx}$$

$$x = \frac{E}{\langle E \rangle}$$

Such distributions are shown in Fig. 64. it is obvious that, when the number of supermodes increases, the distribution narrows and (as it will be presented in Appendix D) the average quadratic deviation is:

(147) $$\sigma_E \cong \frac{1}{\sqrt{M}}$$

Clarified the notion of longitudinal coherence, we now move on to the definition of the transverse coherence.

As previously, the problem of the definition of a transverse coherence results from the fact that there is no more optical cavity, which prevent us from defining strictly speaking transverse modes.

We should note that in the high gain case, we will no longer be able to talk about free modes but guided modes, because of the strong distortion in the propagation caused by the interaction itself. An example of such behavior is given in Fig. 65. The Figure also illustrates the gain guiding mechanism, which becomes relevant as soon as the Rayleigh length becomes longer than the gain length. It is obvious that even in this case the $\rho$ parameter plays a fundamental role. Indeed, the condition for the guided modes can be written as follows:

(148) $$\rho >> \frac{\lambda_u \lambda}{4\pi^2 \sqrt{3} w_0^2}$$

Without optical cavity, the waist is essentially related to the transverse section of the electronic beam. We will clarify later, in the next paragraphs and the Appendix A the meaning of such relation.

# 13. FEL in the SASE Regime and Example of Application

Before giving a specific example of a SASE FEL, we note that the effects of inhomogeneous widening and relative reduction of the gain can be treated using the same procedure as the one given in the case of a FEL oscillator. In the case of energy spread, the parameter which controls the importance of the effects is the analog of the $\mu_\varepsilon$ coefficient introduced in Eq. (93):

(149) $$\tilde{\mu}_\varepsilon = \frac{2\sigma_\varepsilon}{\rho}$$

The effect of this parameter on the gain length is given by:

(150) $$\tilde{L}_g = \chi L_g$$
$$\chi = 1 + 0.185\frac{\sqrt{3}}{2}\tilde{\mu}_\varepsilon^2$$

The analysis of a SASE FEL system can then be done calculating quantities such as the saturation length and saturation power. One possible combination of the various parameters is given in Fig. 66, where we reported the saturation length, the power and the wavelength of a SASE FEL.

Several SASE FEL have been proposed as fourth generation light sources. The location of these projects is reported in Fig. 67. These FEL will deliver radiation in the extreme ultraviolet and X-ray ranges, with a brightness at least 10 orders of magnitudes times larger than the one presently available on the synchrotron light sources. The evolution of the brightness in time is represented in Fig. 68. The so-called brightness is defined by ***Number of photons per second per source unit area, per unit solid angle, per 0.1% bandwidth*** $(photons/sec/mm^2/mrad^2/0.1\%\Delta\omega)$

The distribution of the SASE FEL projects around the world is illustrated in Fig. 67. The evolution of the brightness in time is given in Fig. 68.

The 10 orders of magnitude jump offered by the SASE FEL results from:

1. While conventional synchrotron light sources use the simple principle of spontaneous emission, SASE FEL are based on stimulated emission which guarantees a photon gain of $G = \frac{1}{9} e^{\frac{L_u}{\lambda_u} 4\pi \sqrt{3}\rho}$, with $L_u = N\lambda_u$ the undulator length. Assuming $N\rho \cong 1$, one gets that with respect to the conventional sources, the photon flux is higher by roughly eight orders of magnitude.
2. The conventional sources use electron beams accelerated in a storage ring, while SASE FEL use an electron beam accelerated in a LINAC (linear accelerator) of high energy. The transverse dimensions of the emitting source (the electron beam) is therefore 100 times smaller in the case of the SASE FEL that in the case of the synchrotron sources. According to what has been said in (1) and given that the brightness is inversely proportional to the transverse section of the beam, the magnification factor reaches ten orders of magnitude.

An other interesting aspect of the SASE FEL is the pulse duration of the radiation: as illustrated in Fig. 69, SASE FEL can produce ultra-short radiation pulses.

The shortness of the X-ray pulse is related to the duration of the current pulse, which generated it. The electron beam length can be understood as follows. Assuming a maximum charge of 1 nC ($10^{-9}$ C) and given that a SASE FEL needs currents of at least 1 kA to obtain a reasonable saturation length, it turns obvious that the duration of the electron beam should be smaller than a few ps.

In conclusion, we found that a SASE FEL can provide radiation in the X-ray range with duration of the order of the ps with extremely high brightness. Such characteristics make the SASE FEL sources of high interest for various types of applications listed below.

1. **Temporal resolution of chemical reactions**
2. **Atomic scale resolution and relative technology for single molecule imaging**
3. **Diagnostics for cancer detection at premature state**
4. **Study of diseases induced by Prion** ….

A scheme of a pump probe type experiment is provided in Fig. 70 and 71. The experiment is meant to resolve the temporal scale of chemical reaction, i.e. in less formal terms, to monitor a chemical reaction while it occurs. The Figure shows a X-ray pulse, used to excite the system, while a second laser is used to monitor the evolution of the system at various times.

A further important possibility offered by FEL operated in the X-ray domain is illustrated in Fig. 72. It deals with the study of a myoglobine molecule, whichis  a protein present in the muscles and which presides over the accumulation of oxygen and its transformation into energy. One of the unsolved problems is indeed the mechanism which enables the oxygen to get inside a molecule. Laue diffraction and methods using a few hundred fs laser pulse could certainly provide with information on such mechanisms.

In Figures 73 and 74, we give a more precise idea of the large range of possibilities offered by SASE FEL in various biological domains.

In Figure 75, we reported the possible regions of interest of spectroscopy for various short wavelength sources including FEL.

We finally cite an application in the plasma physics domain. A SASE FEL operating in the X-ray range could be used to create hot plasmas such as those present in the giant stars (see Fig. 76).

In Figure 77, is reported a scheme of hypothetic SASE FEL facility. Various beam lines are considered, providing light with various characteristics (polarization, wavelength, brightness, ...) for various users. Such variety of characteristics is enabled by the use of various types of undulator.

# Appendix A

## Optical cavity and transverse modes

In this appendix, we consider the problem of the distribution of the transverse modes in optical cavities and we analyze in parallel the problems relative to the optical functions of the electrons which, within certain limits, can be discussed in analogy with the Gaussian optics. All along this work, we described a Gaussian type beam inside the confocal cavity and we characterized it in terms of Rayleigh length, spot size and divergence (see section 4).

The electric field from which the distribution is derived is the solution of the Helm holtz equation in the paraxial approximation and is given by (using cylindrical coordinate) [1,2]:

(A.1) $$E(r,z) = E_0 e^{-\frac{r^2}{w(z)^2}} e^{-ikz-ik\frac{r^2}{2R(z)}+i\zeta(z)}$$

with $k$ the wave vector, $w(z)$ the function defined in Eq. (26), $R(z)$ the radius of curvature of the wavefront and $\zeta$ the phase of the wave.

The evolution of the electric field to the intensity of the mode is obtained via the following relation:

(A.2) $$I(r,z) = \frac{|E(z)|^2}{2Z_0} = I_0\left(\frac{w_0}{w(z)}\right)^2 e^{-\frac{2r^2}{w(z)^2}}$$
$$Z_0 = 377\Omega_{vacuum\ \ impedance}$$

The intensity distribution is therefore a Gaussian with an rms width given by:

(A.3) $$\sigma(z) = \frac{w(z)}{2}$$

The intensity can also be expressed as a function of the divergence angle $\theta$ which verifies $\theta = \frac{r}{z}$. For large $z$ and $r < w_0$, we assume $\theta \approx \frac{r}{z}$ and the intensity becomes:

(A.4)

$$I(\theta, z) \propto \frac{1}{\sqrt{2\pi}\sigma_\theta} e^{-\frac{\theta^2}{2\sigma_\theta^2}}$$

$$\sigma_\theta = \sqrt{\frac{\lambda}{2\pi L}}$$

The quality of a transverse mode can be quantified using the trpoduct between the rms angular with spatial widths. In our case, we have:

(A.5)

$$\sigma(0)\sigma(\theta) = \frac{\lambda}{4\pi}$$

This relation defines a mode limitated only by diffraction. The right hand term corresponds to the surface in the phase space ( $r$ , $\theta$ ) of the radiation.

More generally, taking into account a possible correlation between the spatial and the angular components, the distribution in the phase space of the optical Gaussian mode can be written as following (see Eq. (97):

(A.6)

$$I(r,\theta) = \frac{1}{2n\sum} e^{-\frac{\beta\theta^2 + \gamma r^2 + 2ar\theta}{2\Sigma}}$$

The coefficients $\beta$ , $\gamma$ are related to the section and the divergence of the beam according to:

(A.7)

$$\sigma_r = \sqrt{\beta\Sigma}$$

$$\sigma_\theta = \sqrt{\gamma\Sigma}$$

The coefficient $\alpha$ measures the correlation between the angular and the spatial components. For $\alpha$ =0:

(A.8)

$$\Sigma = \frac{\lambda}{4\pi}$$

At waist, were $\alpha$ =0, the coefficients $\beta$ , $\gamma$ are easily identified as:

$$\beta_0 = \frac{L}{2}$$

(A.9)

$$\gamma_0 = \frac{L}{2}$$

The electronic distribution in the phase space can be treated with an analog method (see section 8, Eq. (97)). The surface in the phase space corresponds to a key quantity referred as the emittance of the electron beam $\varepsilon$ (mm.mrad). The emittance is a measure of the quality of the electron beam.

A low emittance electron beam enables a better focusing and a lower divergence: the beam is a point like source.

In Figure. 78, we reported a geometrical representation of the phase space of the electron beam (or of the radiation) together with the role played by the various coefficients. The ellipse surface is related to the emittance $\varepsilon$ of the electron beam (or to the parameter $\Sigma$ in the case of the radiation).

In the case of the FEL operated in the SASE regime at very short wavelength (VUV-X), the parameters of brightness discussed in the final section are obtained when assuming that the phase space of the electron beam and of the radiation can be superimposed (see Fig. 79), i.e. assuming that:

(A.10)

$$\varepsilon = \frac{\lambda}{4\pi}$$

Such relation is relevant as far as emittance values of the order of $\varepsilon < 3\frac{\lambda}{4\pi}$ are considered.

The relation Eq. (A.10) is highly important and defines the characteristics of the electron beam at source point. The emittance of the beam is a quantity which is conserved during the transport and therefore, taking into account the effects of therelativistic contraction, the pre-accelerated electron beam should have an emittance, referred as normalized emittance:

(A.11)

$$\varepsilon_n \cong \gamma\varepsilon \cong \gamma\frac{\lambda}{4\pi}$$

The operation of a SASE FEL at 1 nm with an electron beam of 1 GeV consequently requires an electron beam of normalized emittance around 0.2 mm.mrad.

Before concluding this paragraph, we note that the electron beam can be guided and focused using specific magnetic systems such as dipoles and quadrupoles.

The scheme of a quadrupole is given in Fig. 80. It behaves as a lens, focusing (or de-focusing) in one of the transverse direction. Opposite to the standard lenses, quadrupoles focusing in one direction, are defocussing in the other direction.

Therefore, the transport of a charged beam requires the use of quadrupoles and drift spaces (sections without field) adjusted as shown in Figs. 81 and 82: ordered in the so-called FODO configuration.

In Figure 82 are presented the strength lines of the magnetic field: it appears obvisou that electron are successively focused in the magnets.

As in the case of the standard lenses, a focal length can be defined:

(A.12) $$f^{-1} = kd$$

with $d$ the lens width and $k$ the strength of the quadrupole:

(A.13) $$k[m^{-2}] \cong 0.3\frac{g[T/m]}{p[GeV/c]};T = tesla = 10^4 gauss$$

$g$ indicates the field gradient and $p$ the particle moment.

The dipoles deflect the electron beam, as shown in Fig. 83.

From the conceptual point of view, we deal with a system which produces a field in the orthogonal direction with respect to the motion axis of the particles. An example of dipole scheme is presented in Fig. 84.

# Appendix B

## Solution of the logistic equations

In the former paragraphs, we saw that the problem of the evolution of the laser systems is regulated by sigmoidal curves, also referred as logistic. Such curves cen be derivated from non linear differential equations of the first order [6].

We recall that it is possible to transform a differential equation which describes the evolution of the intensity in the cavity of an oscillator, in a differential equation as long as the higher orders of the derivative (i.e. $T_R^m(\frac{d}{dt})^m, m \geq 1$ can be neglected):

(B.1) $$\frac{dI}{d\tau} = [(1-\eta)G(I) - \eta]I$$

We try now to find a solution to this equation.

Considering the case of the FEL (see section 7) and writing everything in terms of the dimensionless variable *X*, one obtains:

(B.2) $$\frac{dX}{d\tau} = \left[1-\eta)\frac{G_M}{1+\alpha X+\beta X^2} - \eta\right]X$$

Separating the variables, one finds:

(B.3) $$\frac{1+\alpha X+\beta X^2}{[1-\frac{\eta}{G_N}(\alpha X+\beta X^2)]X}dX = G_N d\tau,$$
$$G_N = (1-\eta)G_M - \eta$$

After integration, the following fundamental is obtained:

(B.4) $$\phi(X) = G_N \tau$$

This equation can not be easily inverted to obtain the explicit form of $X(\tau)$.

As a simple approximation, we consider the case of Eq. (B.2) where the terms relative to saturation are developed while all the contributions in $X^m, m \geq 3$ are neglected. This leads to:

(B.5) $$\frac{d}{d\tau} X = \left[G_N - \eta' \alpha X\right] X;\ \eta' = G_N + \eta \ \frac{d}{dX} X = [G_N - \eta' \alpha X] X; \eta' = G_N + \eta$$

Proceeding as in the case of Eqs. (133) and (134), one gets the solution of the former equation:

(B.6) $$X = X_0 \frac{e^{G_N \tau}}{1 + \frac{X_0}{X_F}(e^{G_N \tau} - 1)}; X_F = \frac{G_N}{\eta' \alpha}$$

The solution of the more complex case (Eq. (B.2) is found numerically. Nevertheless, we could verify that the functional structure remains the same, apart from the fact that $X_F$ is substituted by the result of the relation $(1 - \eta) G(X_F) = \eta$.

Finally, to obtain the solution of the finite difference equations, we assumed that the fuinctionnal dependence is again the same and the exponential function was substituted by the round trip step increase function, i.e.:

(B.7) $$e^{G_n \tau} \rightarrow (1 + G_n)^r$$

This results being the solution of the equation $X_{r+1} - X_r = G_N X_r$ (see Eq. (71)).

# Appendix C

## Active mode-locking

We have stressed that the mechanism of coupling of the longitudinal modes in a FEL oscillator is essentially a mechanism of active mode-locking (AML) (see section 4). We try now to be more precise, showing what are the real analogies and considering the case of AML induced by the modulation of the cavity losses, which we assume to be of the following type [1,2]:

(C.1) $$q(t) = M(1 - cos(\omega_m t)); \omega_m = \frac{2\pi}{T_R}$$

where $T_R = \frac{2L_c}{c}$ is the round trip period in the cavity.

The equations which drive the evolution of the signal in the cavity can then be written, for $\eta << 1$, as:

(C.2) $$T_R \frac{\partial}{\partial T} A(\tau, T) = (\hat{g}(\tau) - \eta - q(\tau)) A(\tau, T)$$

where $T$ is the time in the cavity, $\hat{g}(\tau)$ is the gain as a function of time and $A(\tau, T)$ is the amplitude of the optical pulse in the time domain, i.e. the Fourier transform of the spectral amplitude.

Using the convolution theorem:

(C.3) $$\hat{g}(\tau) A(\tau, T) = \frac{1}{\sqrt{2\pi}} \int_{-\infty}^{+\infty} g(\omega) A(\omega, T) e^{-i\omega\tau} d\omega$$

and assuming that the gain has a Lorentzian shape, i.e.:

(C.4) $$g(\omega) = \frac{g}{1 + \frac{\omega^2}{\Omega^2}}$$

we obtain:

(C.5) $$\hat{g}(\tau)A(\tau,T)=\frac{g}{\sqrt{2\pi}}\int_{-\infty}^{+\infty}\frac{A(\omega,T)}{1+\frac{\omega^2}{\Omega^2}}e^{-i\omega\tau}d\omega$$

Using the approximation:

(C.6) $$g(\omega)\cong g(1-(\frac{\omega}{\Omega})^2)$$

one can write [7]:

(C.7) $$\hat{g}(\tau)A(\tau,T)=(g+\Delta\frac{\partial^2}{\partial\tau^2})A(\tau,T);\Delta=\frac{g}{\Omega^2}$$

Approximating further the modulation function of the losses to the lower order in $\tau$, one gets from Eq. (C.2):

(C.8) $$T_R\frac{\partial}{\partial T}A(\tau,T)=(g+\Delta\frac{\partial^2}{\partial\tau^2}-\eta-R\tau^2)A(\tau,T);R=M\frac{\omega_M^2}{2}$$

The solution of Eq. (C.8) can be written as:

(C.9) $$A(\tau,T)=e^{\hat{G}\frac{T}{T_r}}A(\tau,0);\hat{G}=g+\Delta\frac{\partial^2}{\partial\tau^2}-\eta-R\tau^2$$

where $\hat{G}$ represents the gain operator which is essentially an harmonic oscillator operator. Therefore, using the eigen function of the harmonic oscillator, we have:

(C.10)
$$A(\tau,T)=A_n(\tau)e^{\lambda_n\frac{T}{T_R}}$$
$$A(\tau,T)=\sqrt{\frac{1}{2^n\sqrt{\pi}n!\sigma_\tau}}H_n(\frac{\tau}{\sigma_\tau})e^{-\frac{\tau^2}{2\sigma_\tau^2}}$$

---

7) Note that $\int_{-\infty}^{+\infty}\omega^2A(\omega,T)e^{-i\omega\pi}d\omega=-\frac{\partial^2}{\partial\tau^2}A(\tau,T)$

(C.11) $$\sigma_\tau = (\frac{\Delta}{R})^{1/4}$$

The eigen values are given by:

(C.12) $$\lambda_n = g - \eta - 2R\sigma_\tau^2(n + \frac{1}{2})$$

They determine the increase at each round trip of each eigen function.

The same analysis can be repeated in the case of the FEL and the same results are obtained. We leave this derivation to the reader as a useful exercise.

# Appendix D

## The Gamma distribution

The gamma distribution, used to characterize th energy distribution of the SASE FEL "**spikes**", belongs to the two parameters distribution family. In general, such distribution are defined as follows:

(D.1) $$\Phi(x;\mu,\delta) = \frac{x^{\mu-1}}{\delta^{\mu}}\frac{e^{-\frac{x}{\delta}}}{\Gamma(\mu)}$$

where $\delta$ is referred as the scale parameter and $\mu$ as the form parameter. Besides:

(D.2) $$\Gamma(\mu) = \int_0^{\infty} e^{-t} t^{\mu-1} dt$$

which reduces to the factorial of $\mu$ when $\mu$ is an integer ($\Gamma(\mu+1) = \mu!$).

The average and the quadratic width are easily obtained from the former relation. Indeed:

(D.3) $$\begin{aligned} \langle x^m \rangle &= \int_0^{\infty} x^m \Phi(x;\mu,\delta) dx \\ &= \frac{1}{\Gamma(\mu)}\int_0^{\infty} \frac{e^{-\frac{x}{\delta}}}{\delta^{\mu}} x^{m+\mu+1} dx \\ &= \frac{\delta^m}{\Gamma(\mu)}\Gamma(\mu+m) \end{aligned}$$

For $m$=1, recording that $\Gamma(\mu+1) = \mu\Gamma(\mu)$, we obtain:

(D.4) $$\langle x \rangle = \mu\delta$$

The variance is given by:

(D.5) $$\sigma^2 = \langle x^2 \rangle - \langle x \rangle^2 = \mu\delta^2$$

# Appendix E

## The linear accelerator

In the former chapters, we discussed the various components of a FEL system but we gave little details on the electron beam, which, in general, can be generated in any accelerator as long as the final required electron beam characteristics are guaranteed.

Presently, the linear accelerator is the more common solution for short wavelengths FEL, although accelerating machine such as microtron were used for the operation of far infrared and/or TeraHertz FEL.

A conceptual scheme of a LINAC is given in Fig. 85. The scheme refers to a structure for travelling waves consisting of a radio-frequency system, which delivers the electric field in the accelerating cavities. This field drives the electron beam to the desired energy after a given distance. The high frequency systems usually a klystron which, as we will see, delivers an electromagnetic field of high power with a wavelength of the order of a few micrometers.

Inside the cavity, the electrons "see" a constant electromagnetic field given by:

(E.1) $$E_z = E_0 cos(\phi)$$

where $E_0$ is the field on crest of the wave and $\phi$ the relative phase between the wave and the electrons (see Fig. 86). Therefore, the acceleration of the electrons is ensured only if the electrons and the radiation have the same phase velocity[8].

---

8 When the phase velocity of the electromagnetic mode is higher than $c$, it is necessary to introduce inside the cavity some elements which produce a breking. The simplest way is to introduce irises inside the cavity, i.e. disks with a hole in the center

The aim of such accelerating structure is to accelerate the electrons with maximum efficiency, i.e. so that the power delivered by the radio frequency system $P_{RF}$ is as much as possible transmitted to the electron beam, and as little as possible dissipated inside the cavities. We already know that the electron beam power is related to the current and to the energy according to:

(E.2) $$P_e[MW] = I[A]E[MeV]$$

Using Ohm law, one gets the energy dissipated in the cavity:

(E.3) $$P_c[MW] = \frac{(E[MeV])^2}{R[\Omega]}$$

where *R* is the resistance of the structure, referred as Shunt resistance and defined according to:

(E.4) $$R = R_s l$$

*l* is the accelerating structure length, related to the acceleration gradient (i.e. acceleration per unit length) as follows:

(E.5) $$E_z = \sqrt{R_s P_{RF}}$$

Therefore, in the power balance we should take into account both the power, used to modify the load of the cavity, and the power acquired by the electrons:

(E.6) $$P_{RF} = P_c + P_e = \frac{E^2}{R_s l} + IE$$

*I* is the current relative to the time duration of the radio frequency pulse. The whole system can be represented as an electric circuit (see Fig. 87).

An important quantity is the factor of merit *Q* of the cavity, defined as the ratio between the accumulated and dissipated power along one cycle. Using the analogy with the electric circuit, *Q* can be related to the filling time of the cavity according to:

(E.7) $$T_r = 2\tau \frac{Q}{\omega} = \frac{l}{v_g}$$

where $\tau$ is the attenuation factor of the structure and $\omega$ is the RF field pulsation. The last equality is obvious and indicates that the electrons must be in presence of the field all along their travel in the cavities.

We come back to Eq. (E.7) and try to find how to combine the various terms to obtain some practical quantities. We recall first that the current of the beam appearing in Eq. (E.6), which we will call average current, refers to the klystron pulse duration. But as we already know, inside each macro pulse can be several micro pulses with shorter duration and a current which we will call peak current. Peak and average currents are related as follows:

(E.8) $$I = \delta I_p$$

$\delta$ is the duty cycle, i.e. the ratio between the duration of a micro pulse and the distance in between them (see Fig. 17).

Assuming that an accelerating structure of roughly 30 m and taking into account the typical values of the Shunt impedance: between 50 and 60 $M\Omega$/m, we get what is shown in Fig. 88, where we reported the maximum achievable energy from an electron beam of 600 A peak current and $\delta$=$10^{-4}$.

We now try to understand what is a klystron. A general scheme is given in Fig. 89.

The klystron is a converter of electrons kinetic energy to electromagnetic radiation. In this sense, it not so different from a FEL. In Fig. 89, a continuous electron beam (<500 kV, 500 A) is extracted, through out thermionic emission, from the cathode and is injected in the first cavity, which is excited by a field at a given frequency. The electrons in the cavity are accelerated and decelerated depending on the phase they found ; the energy modulation is then converted in density modulation so that the bunched electrons emit radiation in the second cavity. This radiation is at the same frequency as the one of the bunching radiation.

The electrons, which produced the radiation, are gathered at the end of the line in a collector. The klystron is driven by a modulator which provides voltages of the order of MV, within a duration $\tau_m$ of the order of a few microseconds. The pulse can then be reproduced with a repetition frequency $\nu_{rep}$. Finally, the average klystron power is defined as:

(E.9) $$P_m = \nu_{rep} \tau_m P_r P_{RF}$$

We do not come into further details which are beyond the scope of this Appendix.

Before concluding, we just wanted to remind the operating bands of a radio frequency system: they are summarized in Table E1

**RIASSUNTO**

In questo articolo si discute la Fisica delle sorgenti coerenti, denominate Free Electron Lasers (FEL), che utilizzano fasci di elettroni liberi per produrre radiazione di tipo laser. Tali dispositivi, a differenza delle sorgenti convenzionali, sono basate sul processo di emissione stimolata da parte di fasci di particelle cariche, generalmente elettroni relativistici, in moto in strutture magnetiche tipo ondulatore. Non essendo la radiazione emessa associata a sistemi atomici o molecolari, in cui sia stata realizzata una inversione di popolazione, la Fisica dei FEL e dei dispositivi convenzionali può apparire affatto diversa. Nel seguito si dimostrerà che esiste una forte analogia e vedremo come, a dispetto delle apparenti differenze, concetti quali guadagno, intensità di saturazione……..permettono di stabilire una completa analogia tra FEL e sorgenti laser convenzionali.

# TABLE I

- $E = 500$ MeV
- $I = 600\ A$
- $\sigma_e \approx 5\cdot 10^{-4}$, $\varepsilon_{x,y} \approx 2$ mm×mrad
- $\sigma\tau \approx 100$ fs
- $t_u = 0.056$ m,
- $K = 2.5$
- $N = 50$
- $\mu_c = 0.2$

# TABLE II

| | | |
|---|---|---|
| $n = 1,$ | $\lambda \cong 120$ nm, | $I_1 \cong 2.3 10^6 \dfrac{MW}{cm^2}$ |
| $n = 3,$ | $\lambda \cong 40$ nm, | $I_3 \cong 10^4 \dfrac{MW}{cm^2},\ \delta = 1$ |
| $n = 5,$ | $\lambda \cong 24$ nm, | $I_5 \cong 2 \times 10^3 \dfrac{MW}{cm^2}$ |

**TABLE E1**

| Definition | Frequency Interval |
|---|---|
| L-Band | 1 to 2 GHz |
| S-band | 2 to 4 GHz (Linac convenzionali) |
| C-band | 4 to 8 GHz ( SCSS) |
| X-band | 8 to 12 GHz (KEK-X, SLAC-X) |

## FIGURE CAPTIONS

Fig. 1 - Mode distributions in a hot cavity

Fig. 2 - Energy distribution (arb.unit) of the blackbody compared to the Rayleigh-Jeans law (correct at long wavelength) as a function of a) the wavelength (nm) and b) the frequency of the radiation

Fig. 3 - Absorption process of a photon and consequent state excitation

Fig. 4 - Spontaneous emission process

Fig. 5 - Stimulated emission process

Fig. 6 - Stimulated emission

Fig. 7 - Increment of the radiation

Fig. 8 - Illustration of the Beer-Lambert law

Fig. 9 - Components of an oscillator laser: 1) Optical active medium, 2) Energy delivered to the optical medium, 3) Mirror, 4) Semi-reflective mirror, 5) Output laser beam

Fig. 10 - Selection of the transverse mode in the optical cavity

Fig. 11 - Laser intensity (arb. unit) vs round trip number for $g$=15% and $h$=2%

Fig. 12 - Examples of optical cavities

Fig. 13 - Optical cavities and related stability diagrams

Fig. 14 - Gauss-Hermite mode and main reference quantities

Fig. 15 - Cavity gain and losses profile, longitudinal modes, and frequency profile of the output laser

Fig. 16 - Gain and loss spectra, longitudinal mode locations, and laser output for single mode laser operation

Fig. 17 - Gain and loss spectra, longitudinal mode locations, and laser output for single mode laser operation(a) Mode-locked laser output with constant mode phase. (b) Laser output with randomly phased modes

Fig. 18 - Periodic pulse train

Fig. 19 Definition of the full width half maximum (FWHM)

Fig. 20 - Pulse width of different laser systems by year. The shortest duration, 4-5 fs, has been achieved with a 800 nm laser. 1 ps corresponds to 300 μm, 1 fs=300 nm and 1 as=0.3 Å

Fig. 21 - Pulse width of Ti:Sa lasers by year. Notice that $\delta\tau\Delta\omega \cong$ constant

Fig. 22 - Example of pump-probe experiment

Fig. 23 - Ground and excited state of the *NaI* molecule

Fig. 24 - Temporal coherence between two waves. a) The phase difference is constant in time, corresponding to an infinite coherence time. b) The two waves are initially in phase, and tend to dephase with time, corresponding to a finite coherence time

Fig. 25 - Examples of wavefronts with a) infinite and b) finite coherence length

Fig. 26 - Electron motion in an undulator

Fig. 27 - Phase advance of the radiation with respect to the electrons

Fig. 28 - Step function and corresponding Fourier transform

Fig. 29 - Function *f(ν)*

Fig. 30 - Evolution of the insertion devices

Fig. 31 - Oscillator FEL

Fig. 32 - Stimulated emission and absorption processes, similar to the stimulated emission process in the case of conventional lasers. $\omega_1 = c|k_1|=$ laser frequency ; $\omega_2 = c|k_2|=$ undulator radiation frequency; $\varepsilon_1, p_1=$ initial energy and moment of the electrons; $\varepsilon_2, p_2=$ final energy and moment of the electrons. a) Backward stimulated scattering (negative gain). b) Forward stimulated scattering (positive gain)

Fig. 33 - FEL gain curve

Fig. 34 - Example of FEL signal evolution in the cavity up to saturation. Comparison between simulation (continuous line) and analytic calculation (dotted line)

Fig. 35 - Structure of an electron beam delivered by a Radio-Frequency accelerator

Fig. 36 - Formation of a laser pulse from an electron beam pulse

Fig. 37 - Fourier transform of a Gaussian beam

Fig. 38 - Slippage mechanism

Fig. 39 -Bunching mechanism induced by the FEL interaction. Bunching is performed at the resonant wavelength and on higher harmonics, allowing radiation production on the fundamental and higher harmonics

Fig. 40 - Example of intra cavity evolution of the fundamental and of the high order harmonics (third and fifth). The Figure also reports the energy spread increase. Continuous lines: numerical calculation. Dashed line: analytical approximation

Fig. 41 - Cavity tuning and synchronism condition.

Fig. 42 - Net gain and intra cavity intensity at equilibrium ($\mu_c$=0.5, $\eta$=0.06, $g_0$=0.3). Net gain corresponds to $G_{net} = (1-\eta)G_M - \eta$

Fig. 43 - Intra cavity intensity at equilibrium calculated using Eq.(110) and comparison with the numerical simulation ($\mu_c$=1, $\eta$=0.06, $g_0$=0.3)

Fig. 44 - Intra cavity power as a function of the cavity mismatch and pulse shape

Fig. 45 - Evolution of the third harmonic after various number of round trips. The harmonic pulse width decreases when the FEL approaches saturation. $\lambda$=1 $\mu$m, $\mu_c$=0.3, $\delta L$=-$\lambda$

Fig. 46 - Third harmonic at saturation. $\lambda$=150 $\mu$m, $\mu_c$=0.038, $\delta L$=-25$\lambda$

Fig. 47 - Energy as a function of the wavelength and operating regions of the FEL

Fig. 48 - Power density as a function of the active losses

Fig. 49 - Growth of the fundamental and generation of non linear harmonics to the third and fifth orders, $\delta L = -5\lambda$

Fig. 50 - Evolution of the fundamental along round trips with $\delta$=-5

Fig. 51 - Evolution of the third harmonic along round trips with $\delta$ =-5

Fig. 52 - Evolution of the fifth harmonic along round trips with $\delta$ =-5

Fig. 53 - Normalized pulse profile. $\delta$ =-5. Main harmonic: $P_{max}$=$2.50\times10^{6}\, MW/cm^{2}$, $\sigma_{em}$=0.39$\sigma_{z}$. Third harmonic: $P_{max}$=$5.05\times10^{3}\, MW/cm^{2}$, $\sigma_{em}$=0.24$\sigma_{z}$

Fig. 54 - Normalized pulse profile. $\delta$ =15. Main harmonic: $P_{max}$=$1.35\times10^{6}\, MW/cm^{2}$, $\sigma_{em}$=0.63$\sigma_{z}$. Third harmonic: $P_{max}$=$2.97\times10^{3}\, MW/cm^{2}$, $\sigma_{em}$=0.33$\sigma_{z}$

Fig. 55 - Normalized pulse profile. $\delta$ =-5. Main harmonic: $P_{max}$=$2.50\times10^{6}\, MW/cm^{2}$, $\sigma_{em}$=0.39$\sigma_{z}$. Fifth harmonic: $P_{max}$=$6.23\times10^{2}\, MW/cm^{2}$, $\sigma_{em}$=0.27$\sigma_{z}$

Fig. 56- Normalized pulse profile. $\delta$ =15. Main harmonic: $P_{max}$=$1.35\times10^{6}\, MW/cm^{2}$, $\sigma_{em}$=0.63$\sigma_{z}$. Fifth harmonic: $P_{max}$=$2.79\times10^{2}\, MW/cm^{2}$, $\sigma_{em}$=0.35$\sigma_{z}$

Fig. 57 - SASE FEL process. Microscopic image of the bunching mechanism and laser signal increase along the undulator

Fig. 58 - Comparison between exact and approximated solution

Fig. 59 - Gain curve for different values of the gain coefficient: $g_0$=0.1, 1, 10

Fig. 60 - Shape of the high gain curve (continuous line) and Gaussian approximation (segmented line)

Fig. 61 - Temporal and spectral distribution of the SASE FEL

Fig. 62 - Coherence length and spikes distribution

Fig. 63 - Evolution of the power while signal increase and fluctuation smoothing

Fig. 64 - Energy distribution of the FEL spikes as a function of $x=\frac{E}{\langle E\rangle}$, for various values of $M$. Continuous line: $M$ =10, dotted line: $M$ =20 and segmented line: $M$ =50

Fig. 65 - a) Gaussian mode and b) guided mode

Fig. 66 - Saturation length, power and wavelength as a function of the $K$ parameter for a SASE FEL operating with a 2.8 cm period undulator and a 1 GeV energy, $5.10^{-4}$

energy spread and 1 kA current electron beam

Fig. 67 - SASE FEL projects over the world

Fig. 68 - Evolution of the brilliance of the X-ray conventional sources over the years and perspectives offered by FEL sources

Fig. 69 - Average brightness vs pulse duration of the SASE FEL sources

Fig. 70 - Pump probe experiment

Fig. 71 - Fs dynamics recorded with a pump probe experiment

Fig. 72. Diffraction image obtained with X-rays and 3-D image of a myoglobine molecule

Fig. 73 - Structure of a molecule studied with a 2.7 Å resolution. An X-FEL would allow a higher resolution and the study of the protein dynamics

Fig. 74 - X-FEL sources and structural biology

Fig. 75 - Region of interest for spectroscopy and X-FEL sources

Fig. 76 - Interaction of an intense X-ray laser beam and formation of plasma with a golden target layer

Fig. 77 - Layout of a X-FEL source

Fig. 78 - Twiss coefficient and phase space

Fig. 79 - Surface in the phase space of the radiation (green) and of the electrons

Fig. 80 - Quadrupolar lens and relative strength lines

Fig. 81 - Scheme of a FODO transport system, also referred as strong focusing system

Fig. 82 - Transport of the electrons in a FODO cell and relative transverse focusing

Fig. 83 - Electron beam trajectory in a bending magnet (continuous line) and after the deviation ($B$ =11.1 KG, $\gamma$ =200/0.511, $d$ =30 cm, deviation angle = 434 mrad)

Fig. 84 - Scheme of the geometry and of the reference axis of a dipole magnet

Fig. 85 - Scheme of a LINAC with traveling waves

Fig. 86 - Mechanism of acceleration in a LINAC with traveling waves

Fig. 87 - Coupling between the klystron and the accelerating cavity and equivalent electric

circuit

Fig. 88 - Maximum energy for a 30 m accelerating structure, $I$ =0.6 mA and two different values of RF power: continuous line corresponds to 200 MW and dotted line to 250 MW

Fig. 89 - Operation scheme of a klystron

**TABLE CAPTIONS**

Table I – Example of FEL parameters

Table II – FEL parameters for coherent harmonic radiation in the extreme UV region

Table E1 - Frequency bands and existing LINAC

Radiation modes in a hot cavity provide a test of quantum theory

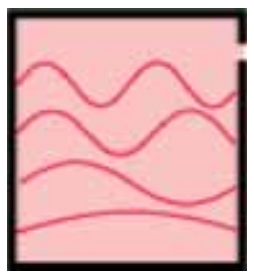

| | Modes per unit frequency per unit volume | Probability of occupiyng modes | Average energy per mode |
|---|---|---|---|
| CLASSICAL | $\frac{8\pi\nu^2}{c^3}$ | Equal for all modes | kT |
| QUANTUM | $\frac{8\pi\nu^2}{c^3}$ | Quantized modes: require hn energy to exciteupper modes, less probable | $\frac{h\nu}{e^{\frac{h\nu}{kT}}-1}$ |

Fig. 1

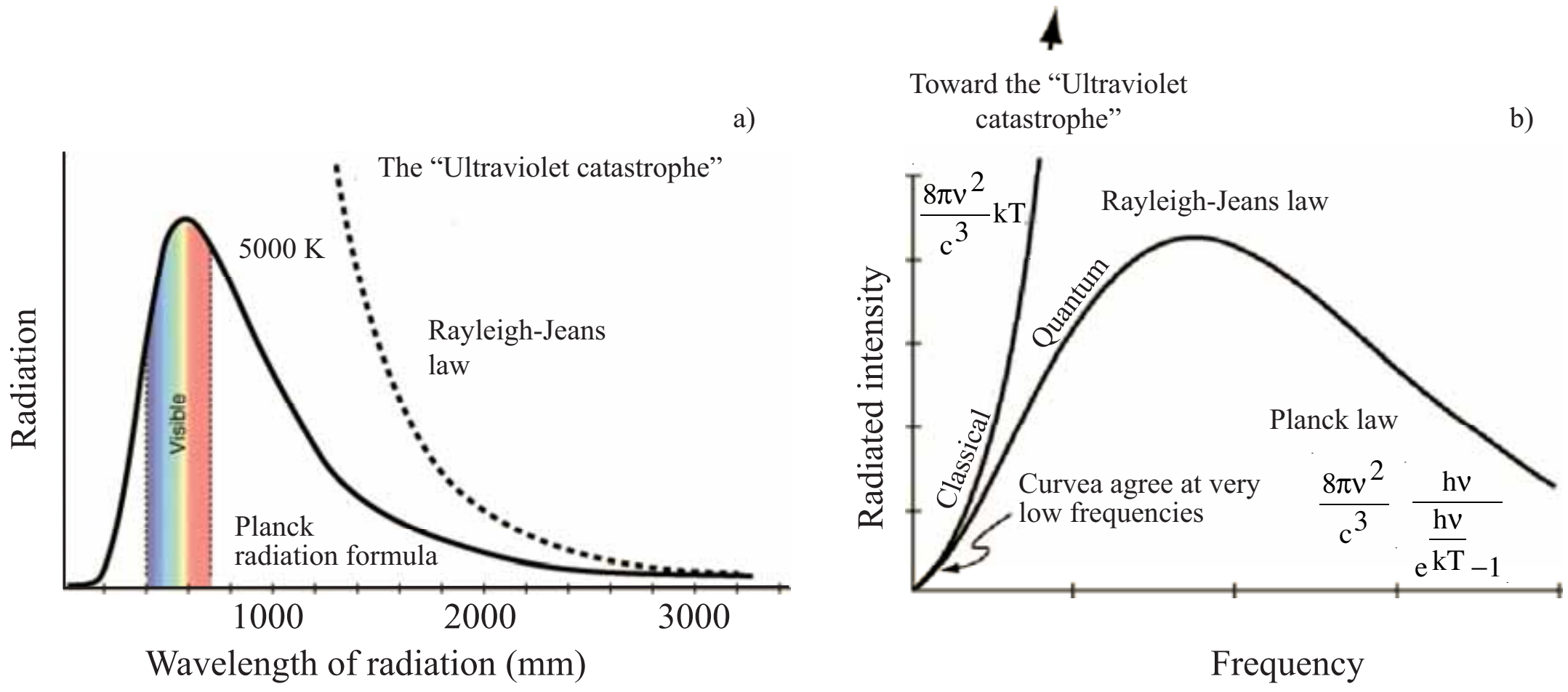

Fig. 2

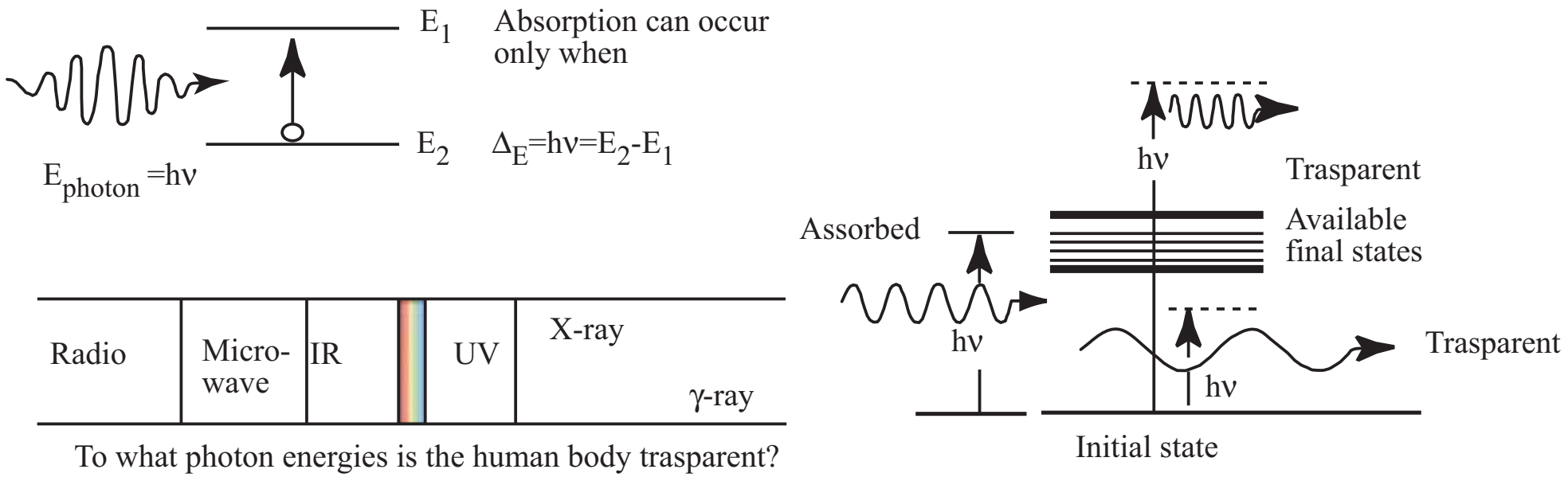

Fig. 3

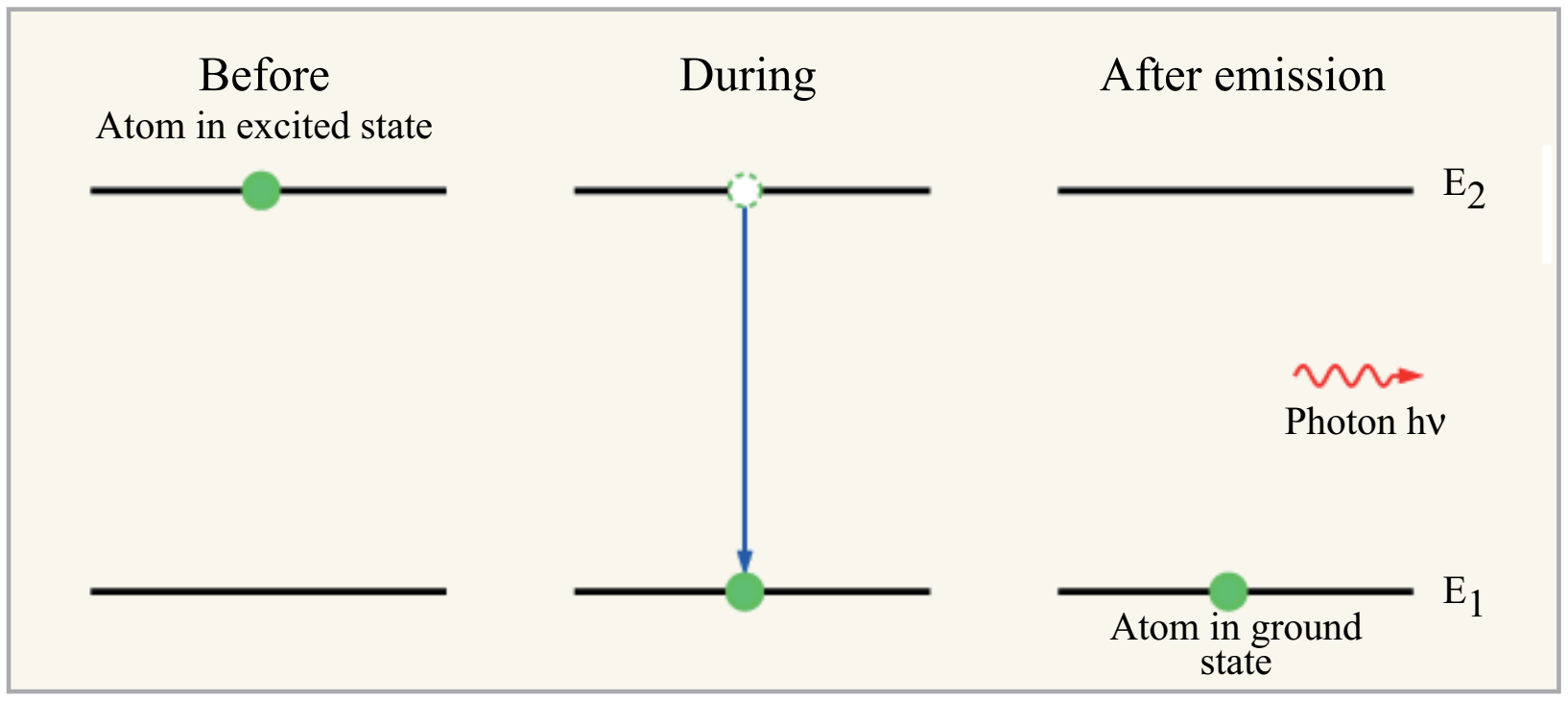

Fig. 4

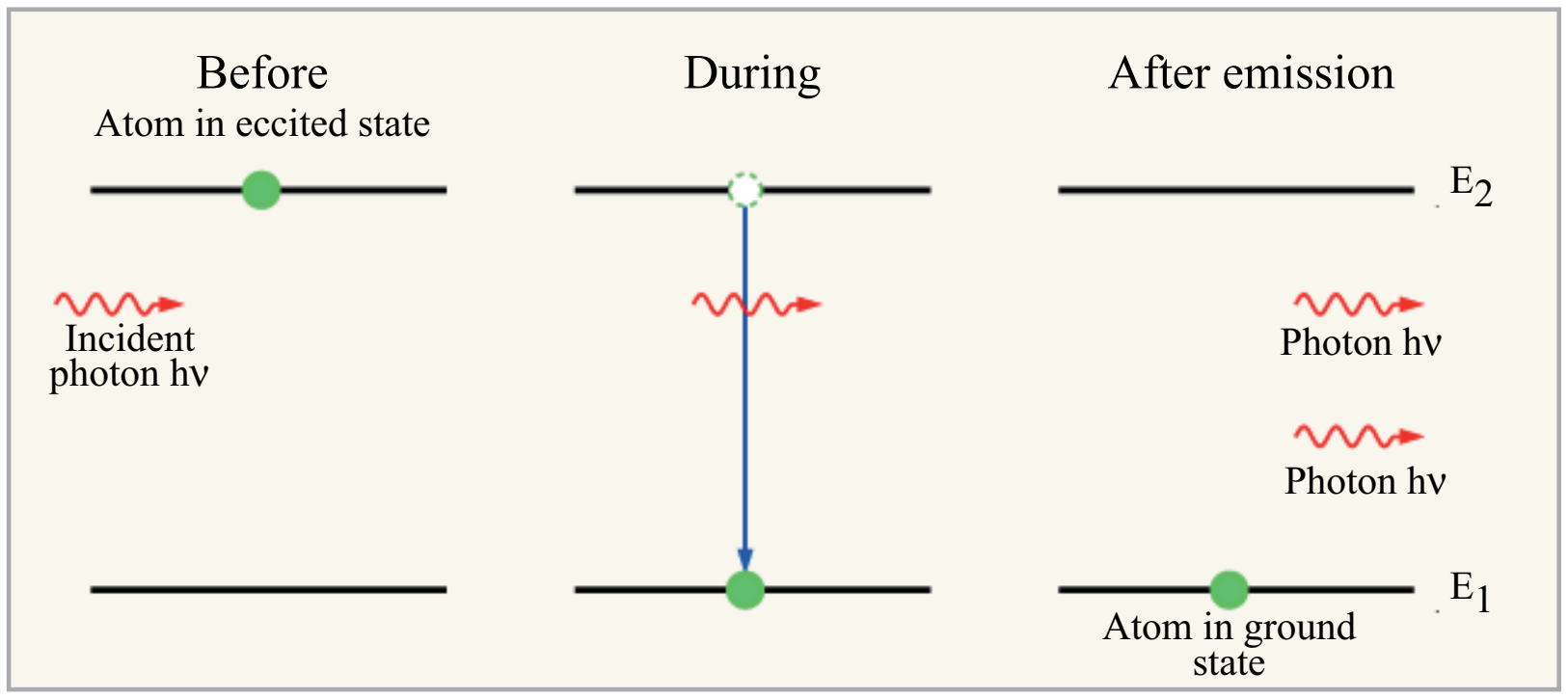

fig. 5

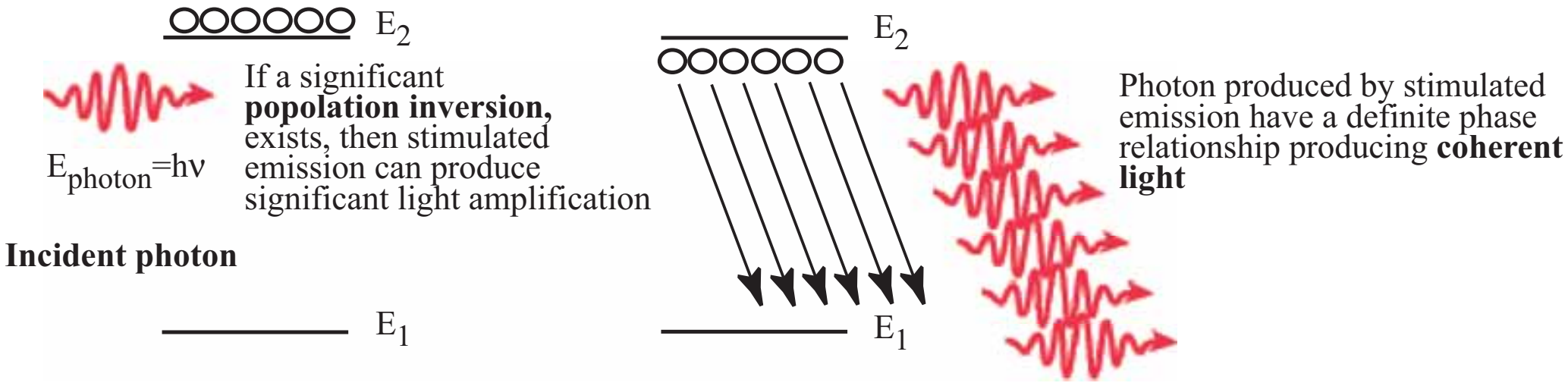

Fig. 6

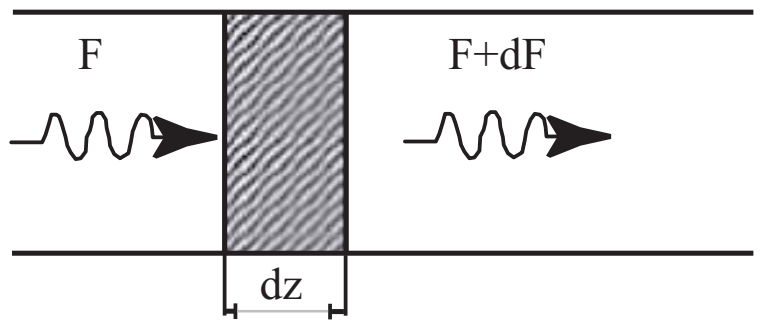

Fig. 7

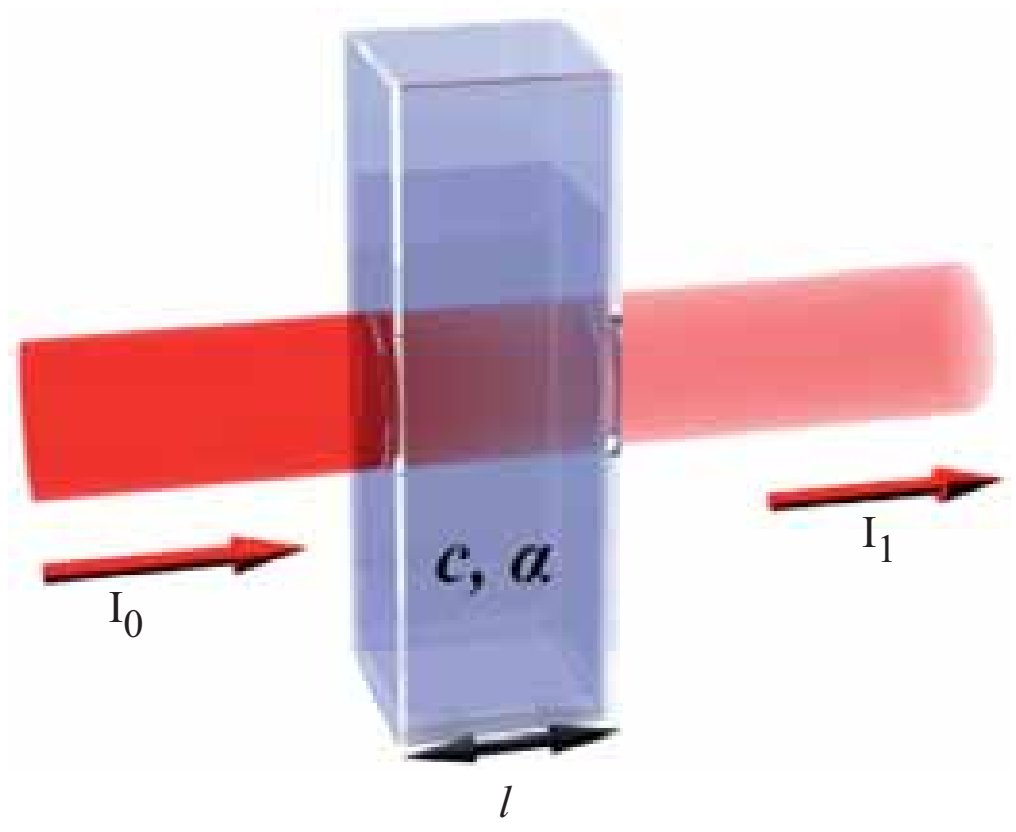

Fig. 8

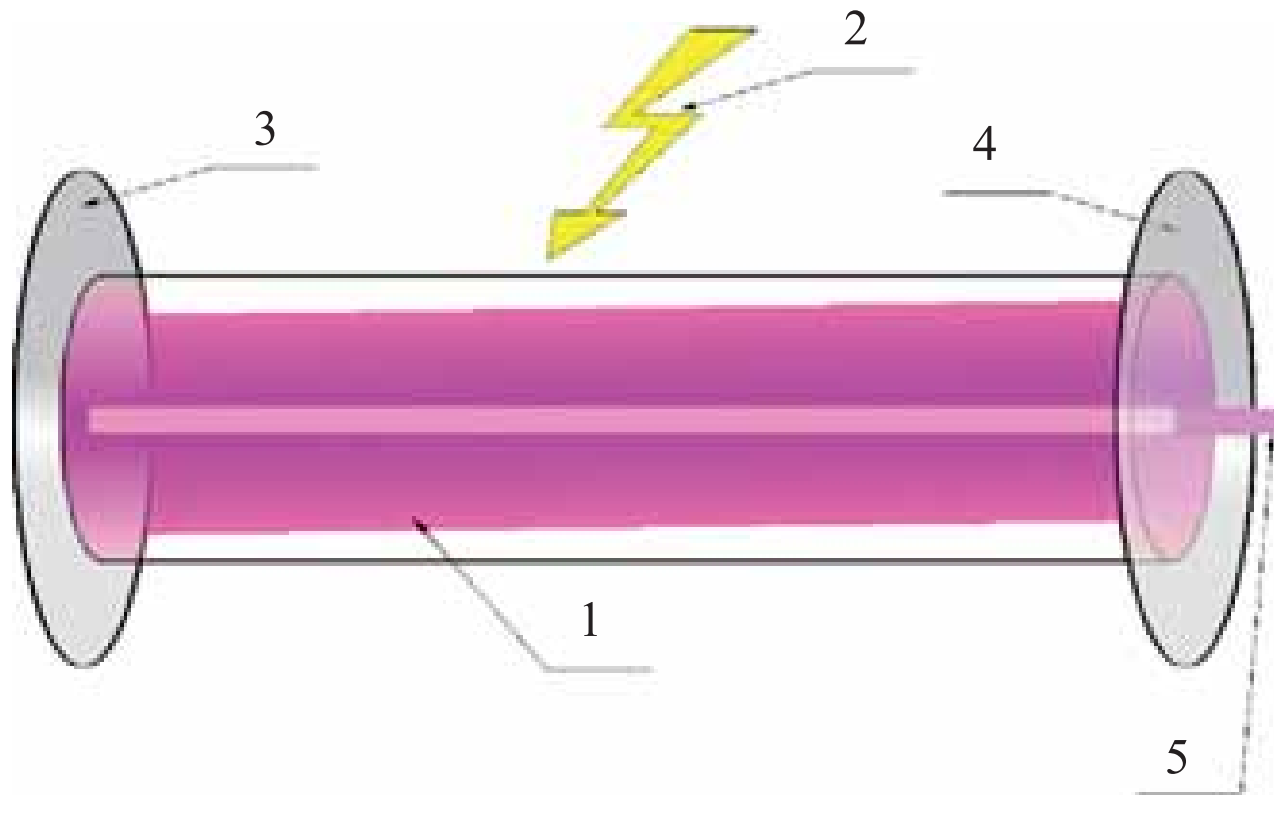

Fig. 9

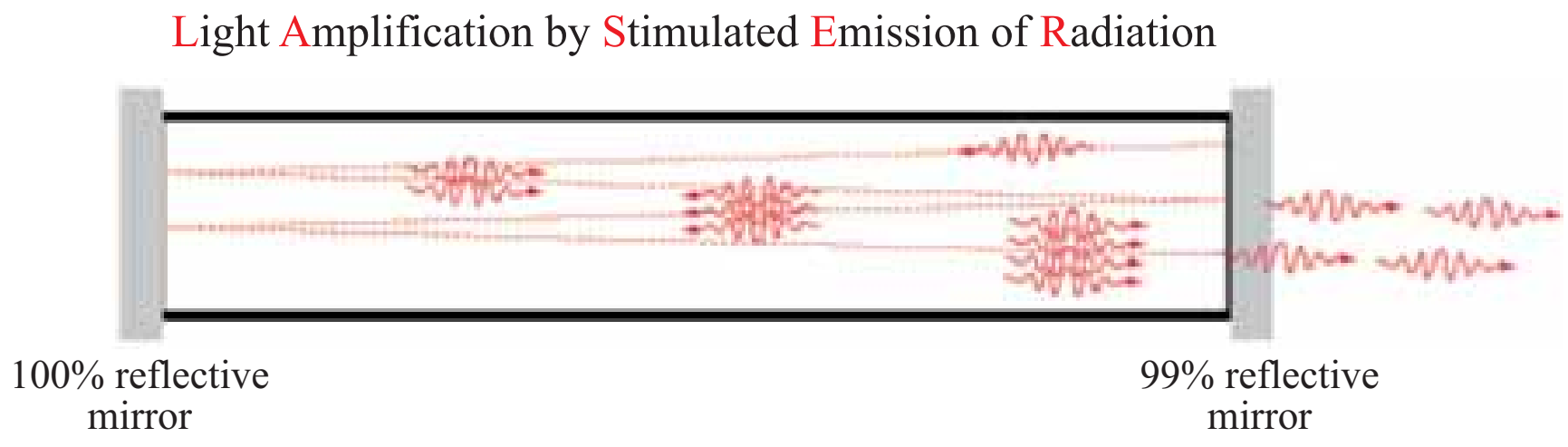

Fig. 10

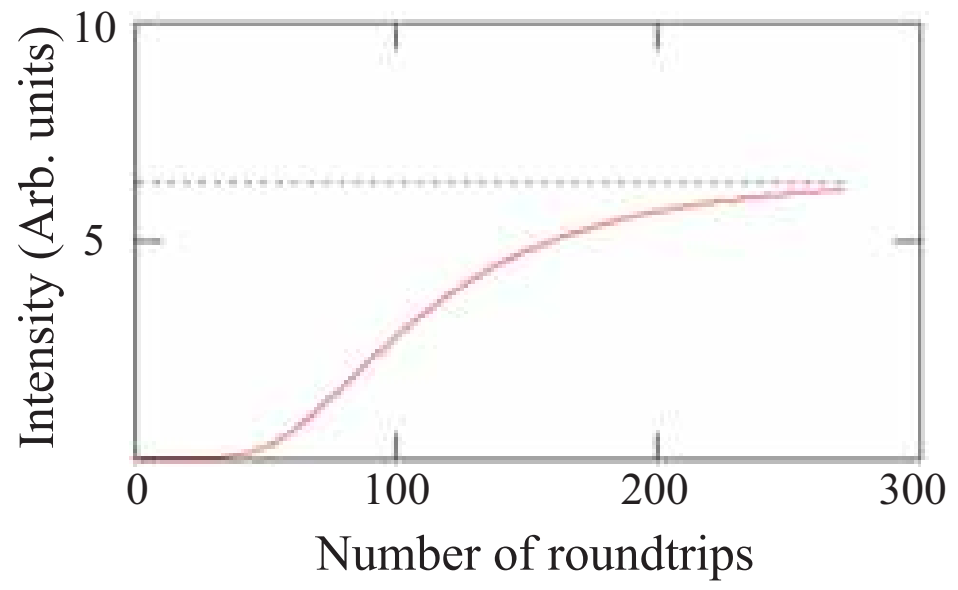

Fig. 11

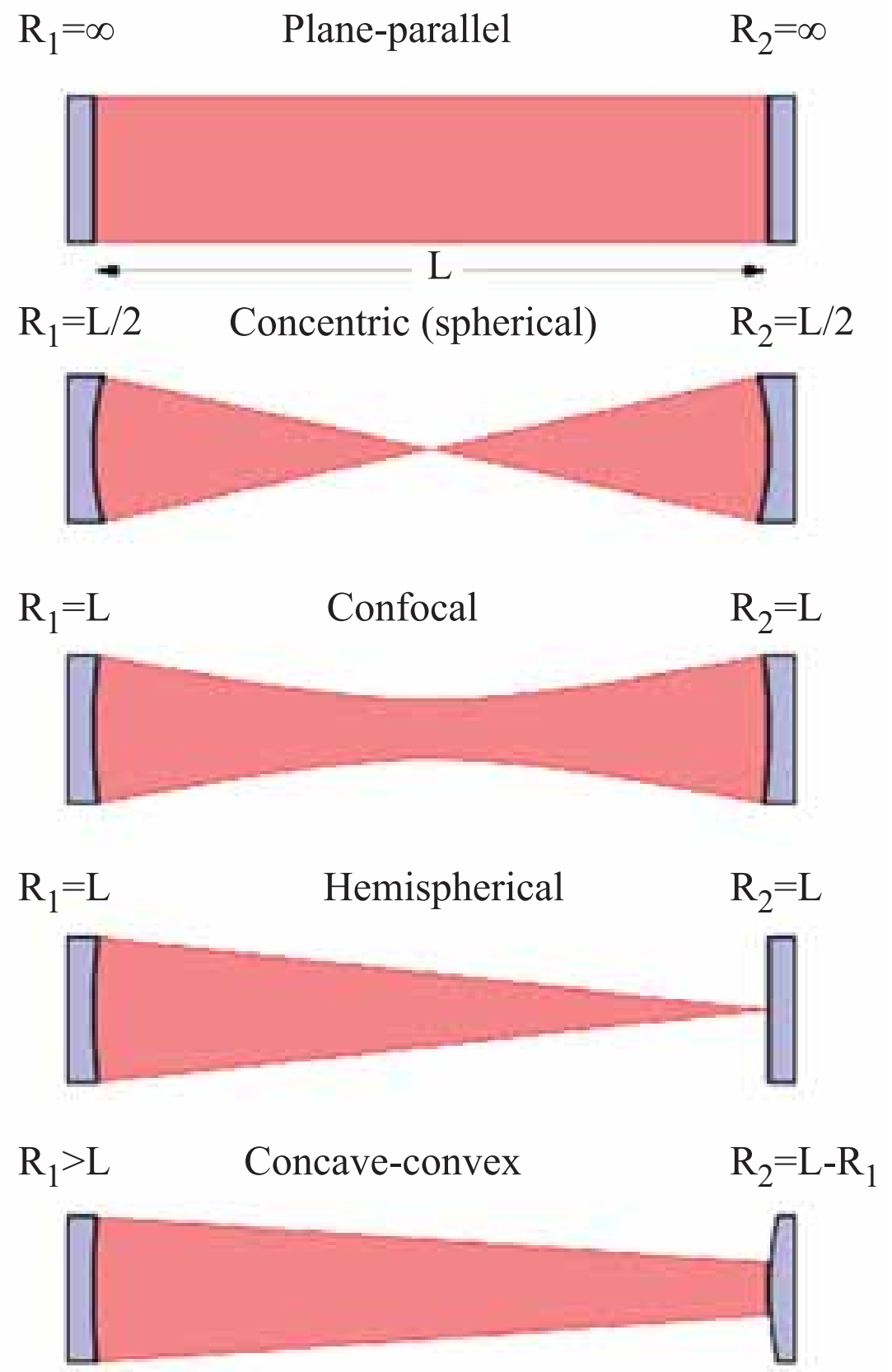

Fig. 12

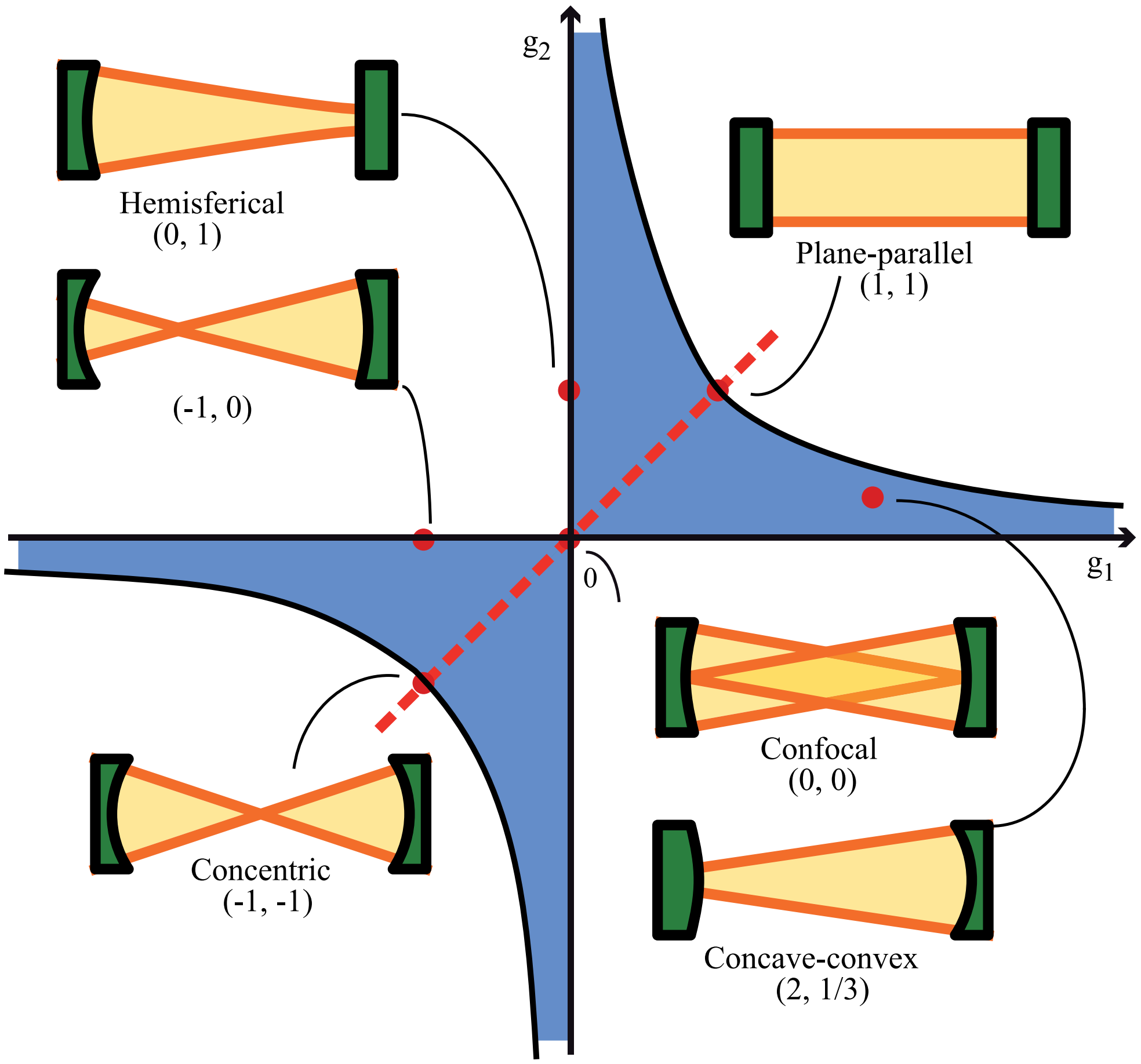

Fig. 13

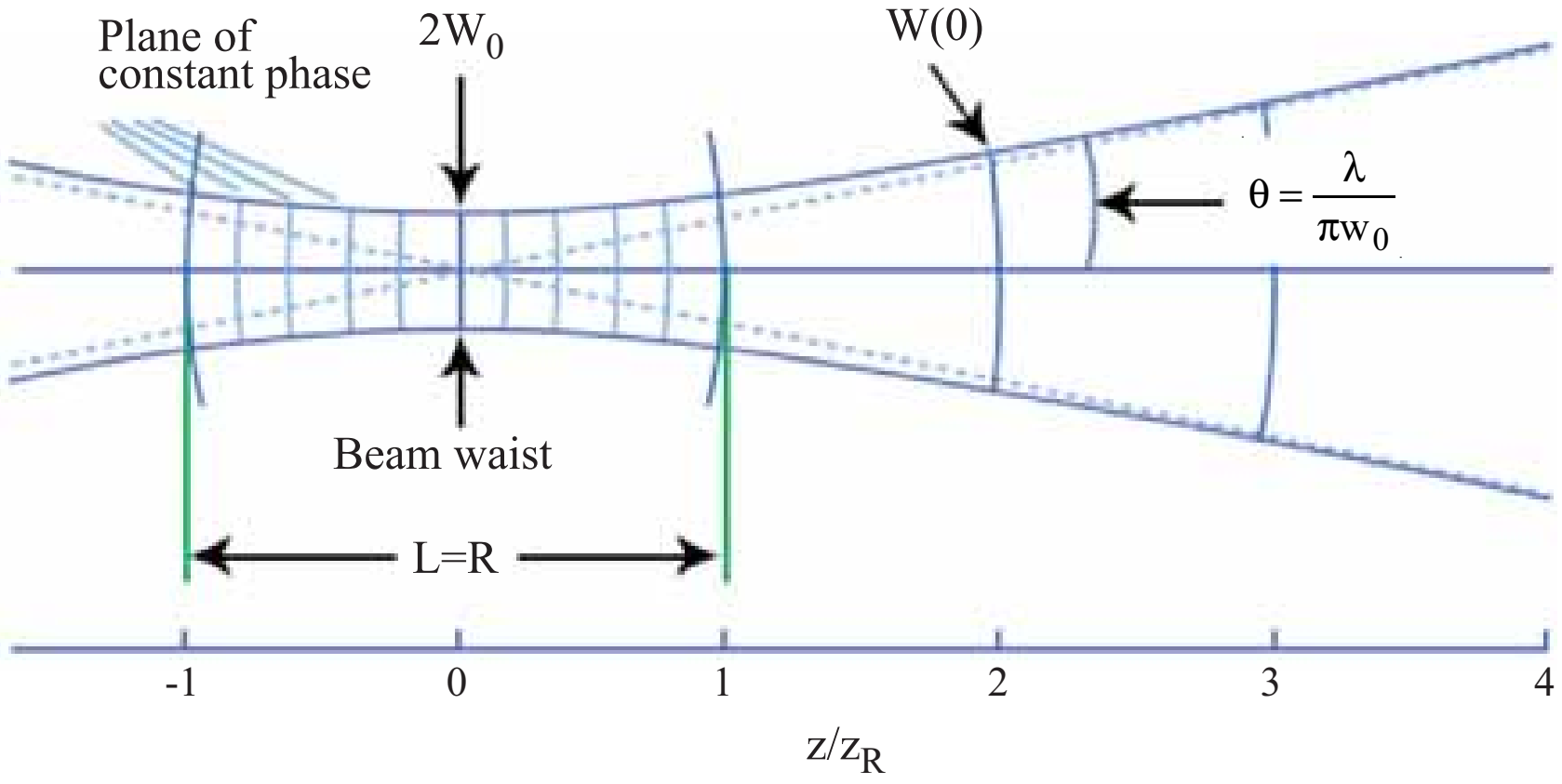

Fig. 14

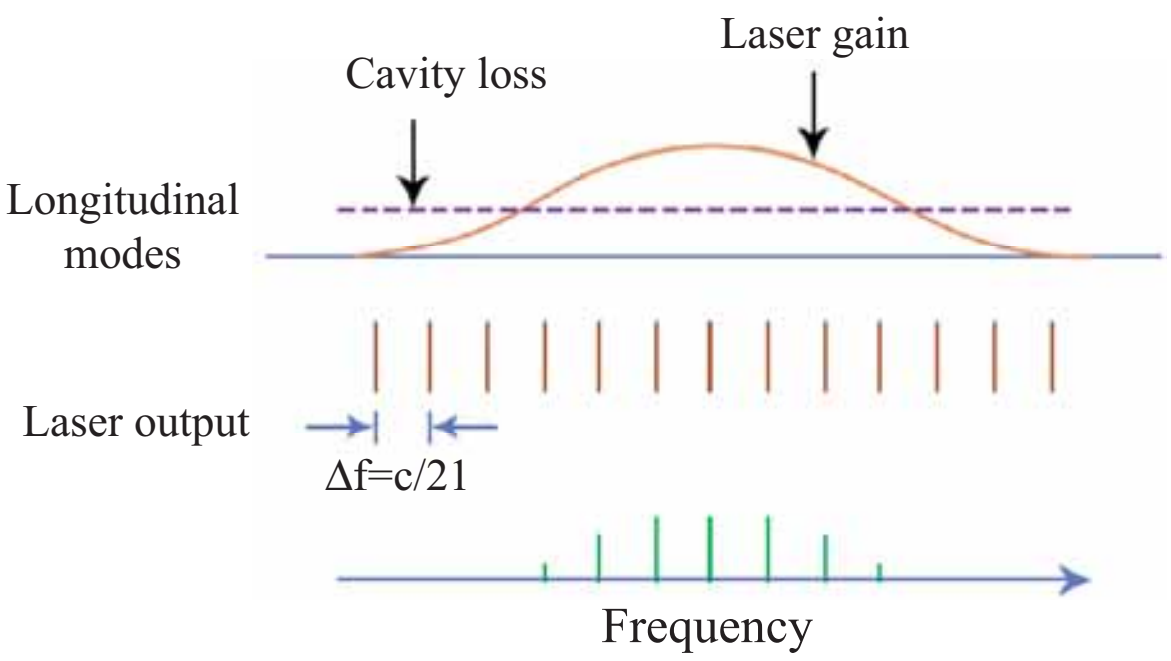

Fig. 15

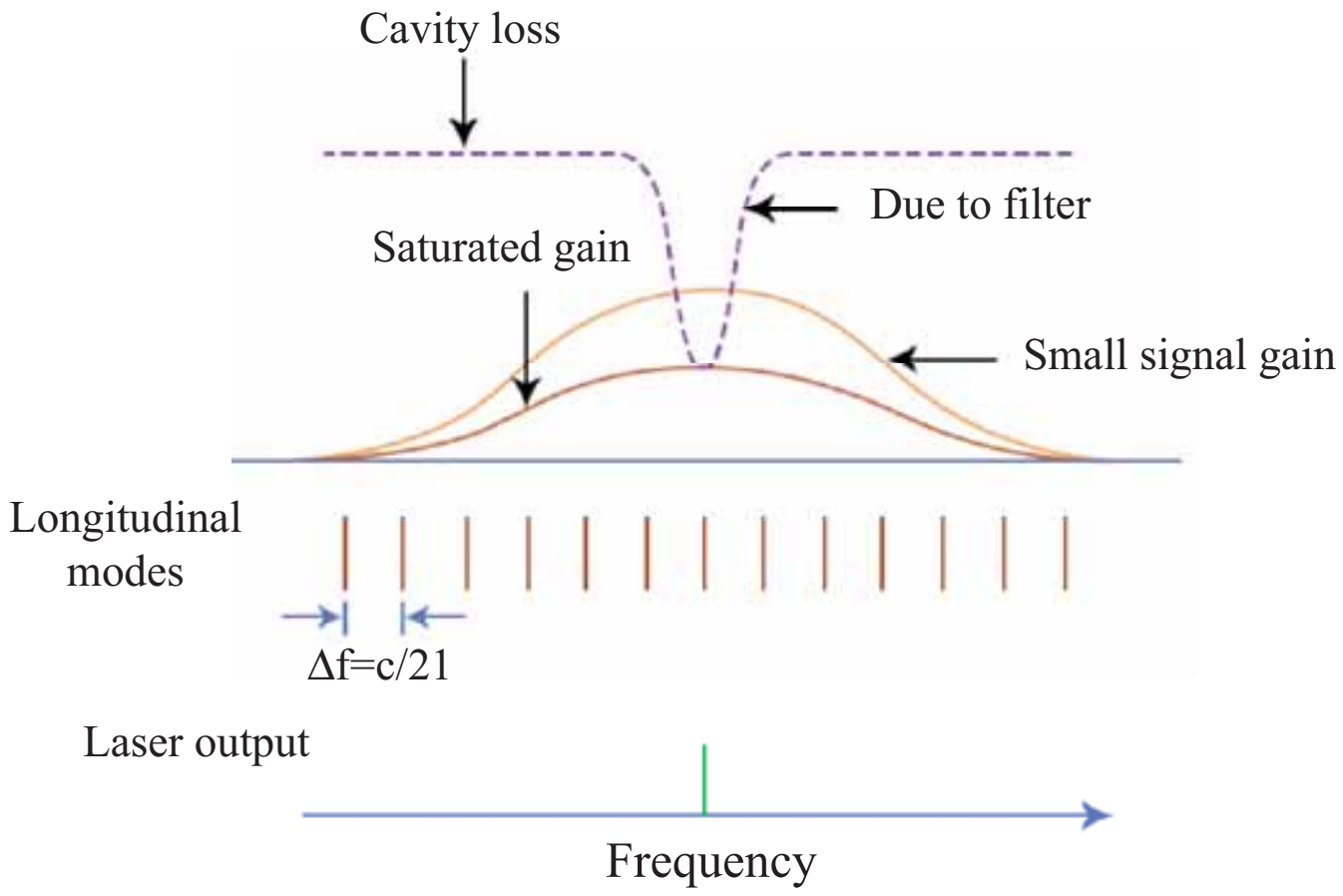

Fig. 16

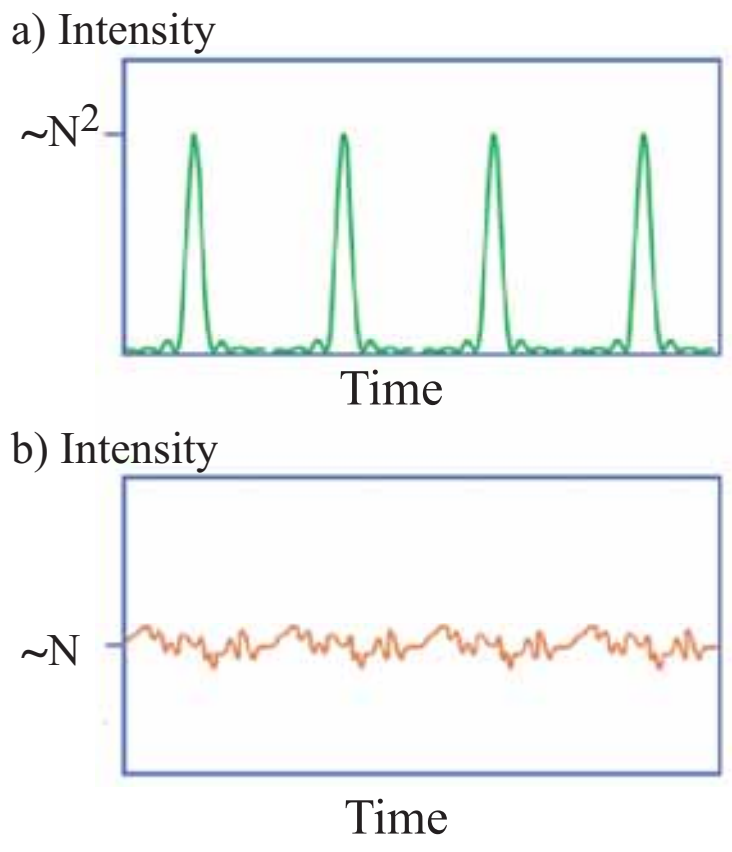

Fig. 17

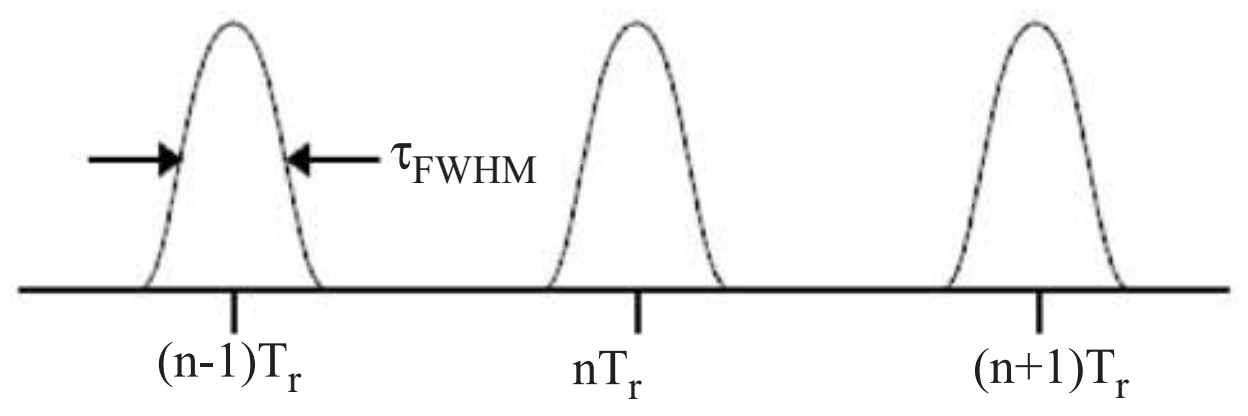

Fig. 18

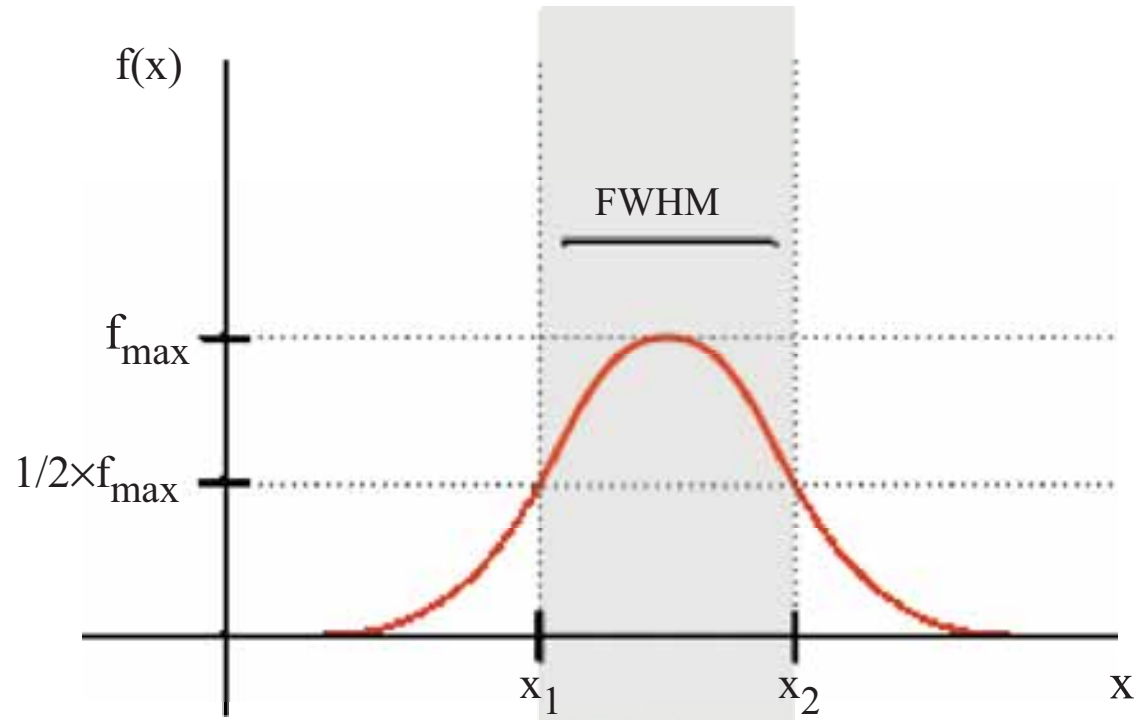

Fig. 19

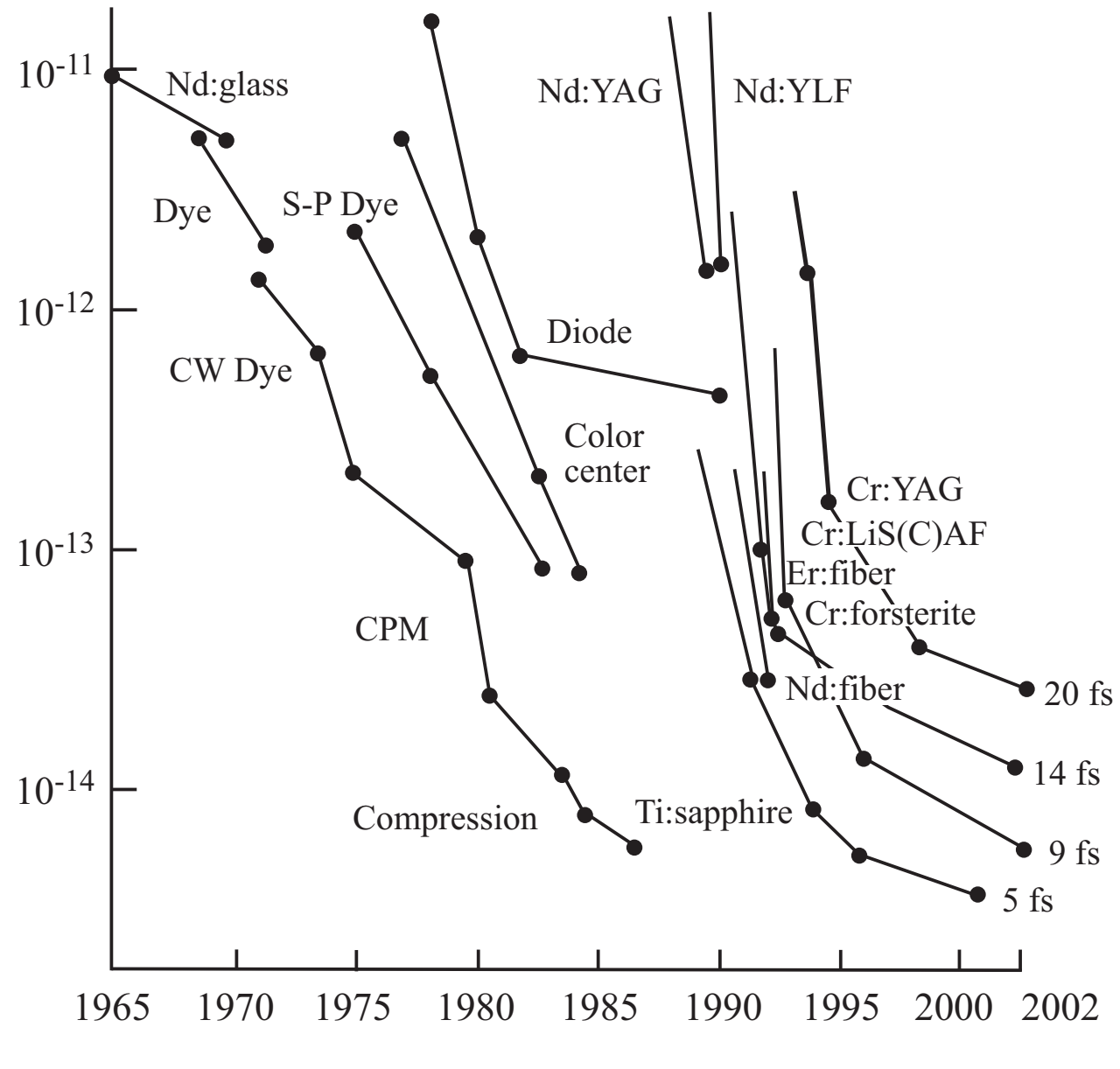

Fig. 20

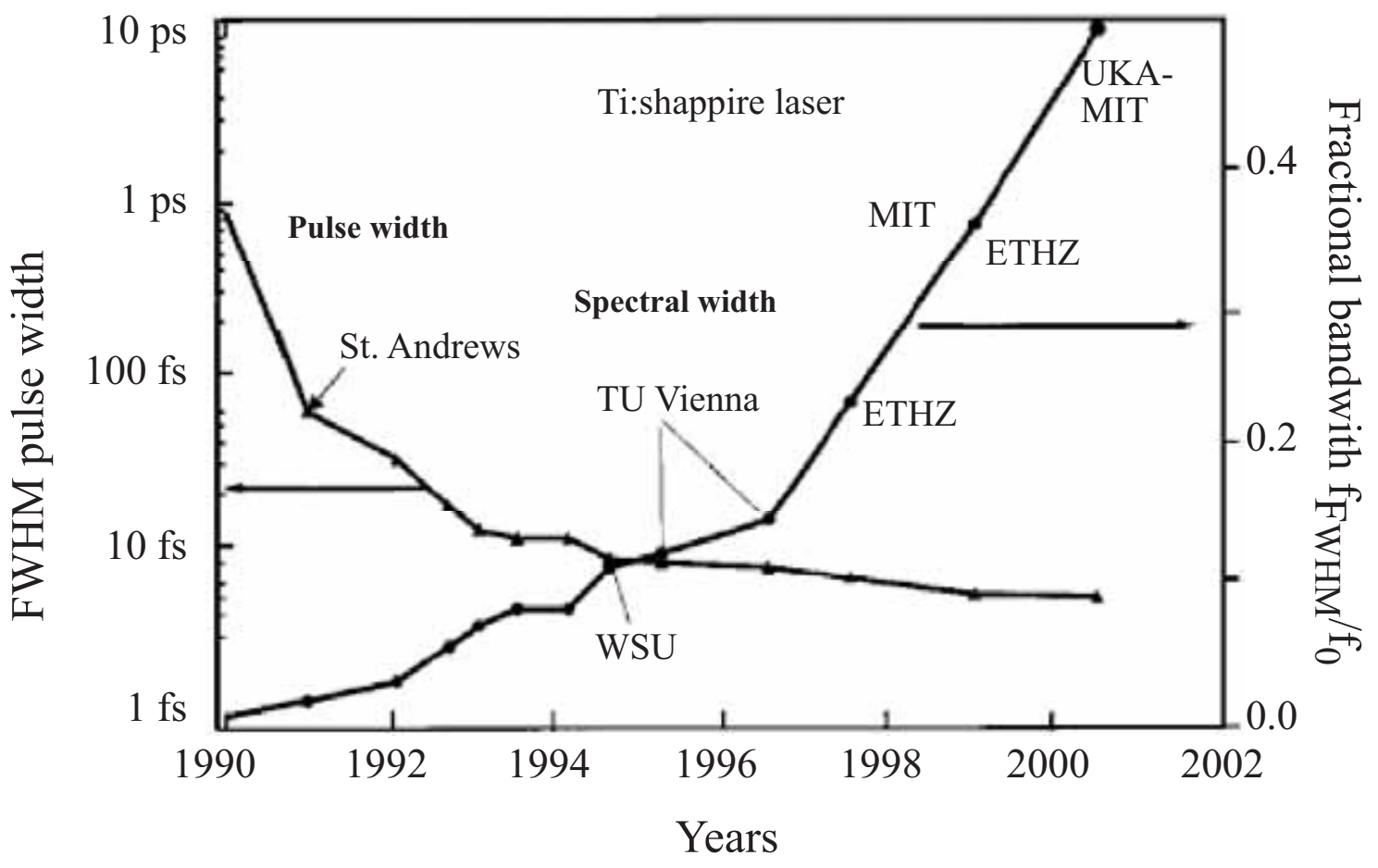

Fig. 21

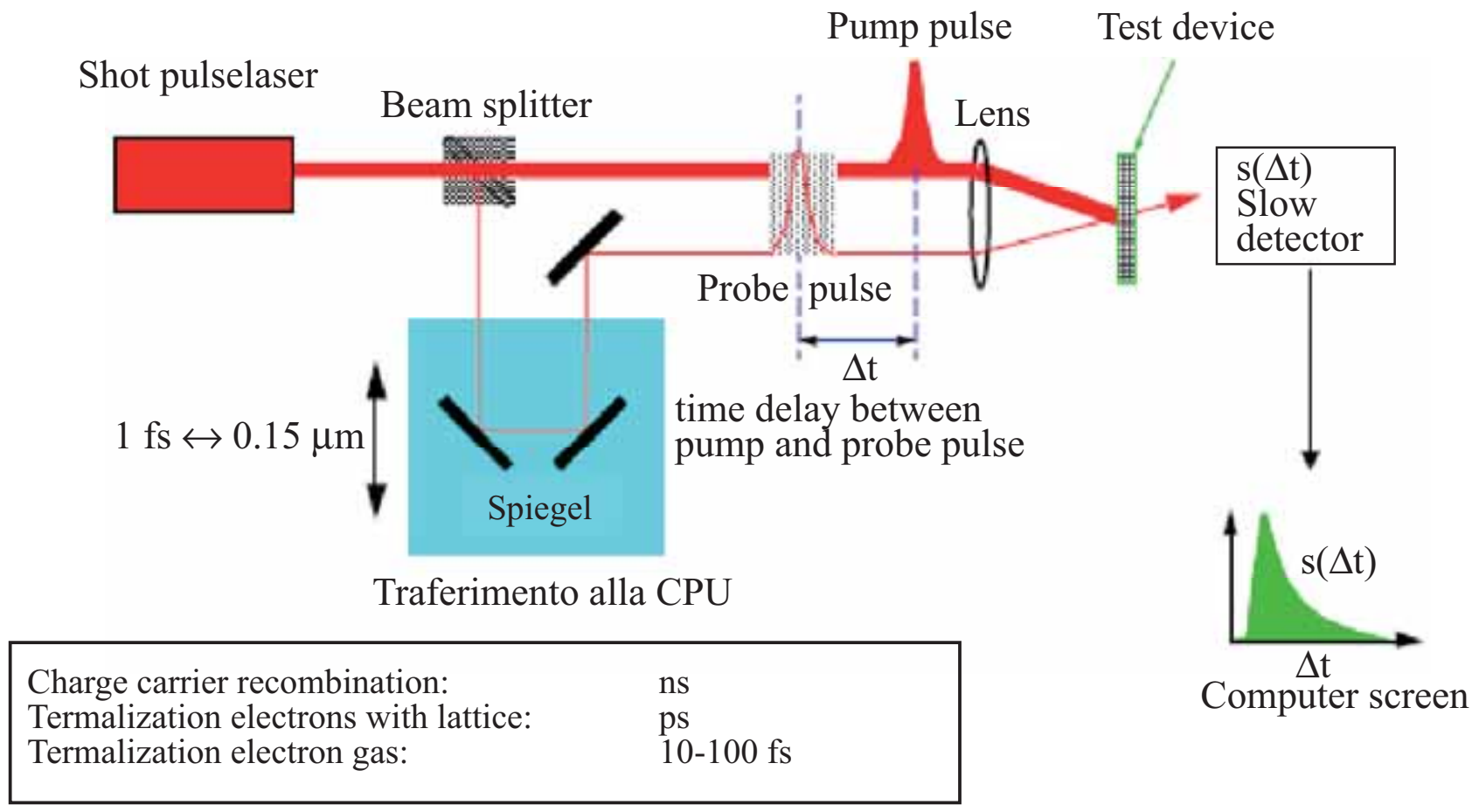
Shot pulselaser
Beam splitter
Pump pulse
Test device
Lens
s(Δt)
Slow
detector
Probe pulse
Δt
time delay between
pump and probe pulse
1 fs ↔ 0.15 μm
Spiegel
Traferimento alla CPU
s(Δt)
Δt
Computer screen
Charge carrier recombination: ns
Termalization electrons with lattice: ps
Termalization electron gas: 10-100 fs

Fig. 22

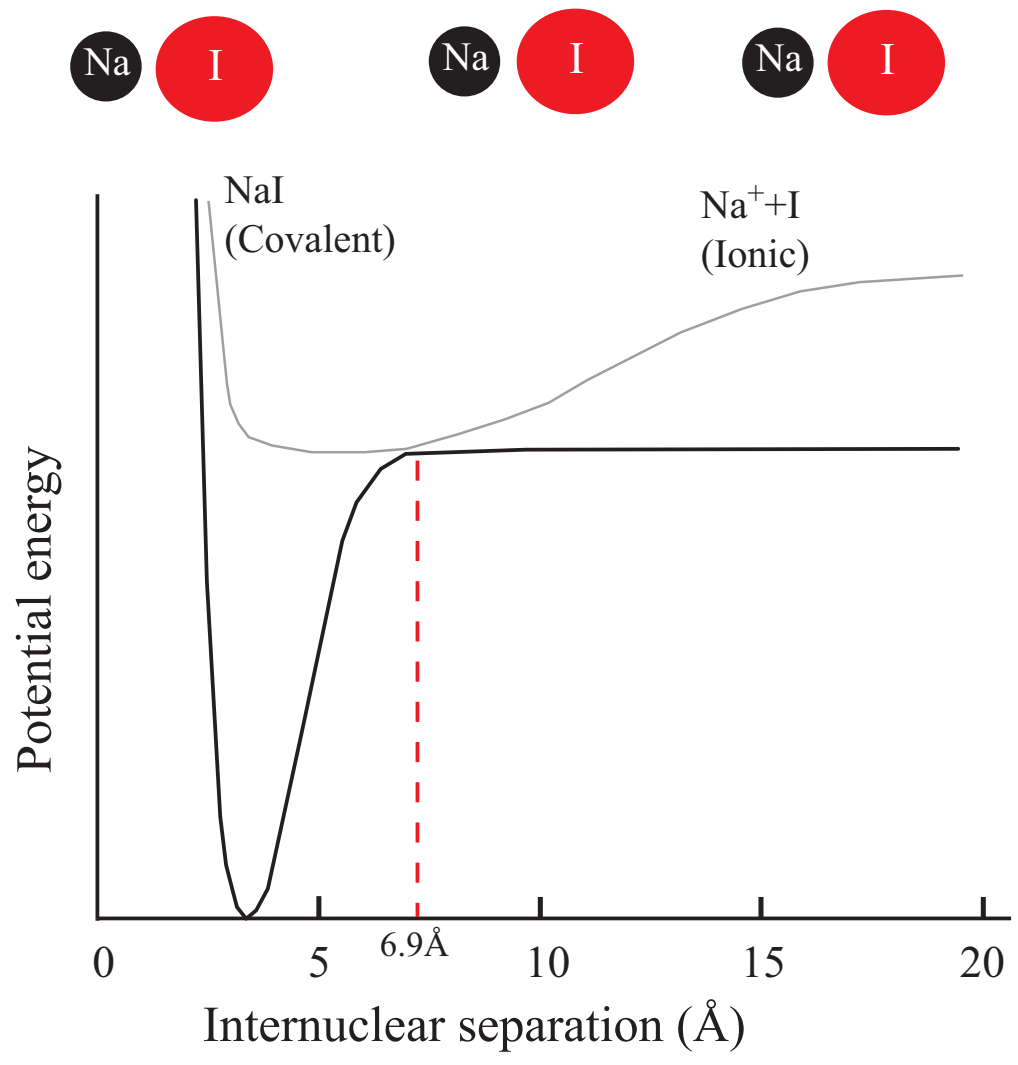

Fig. 23

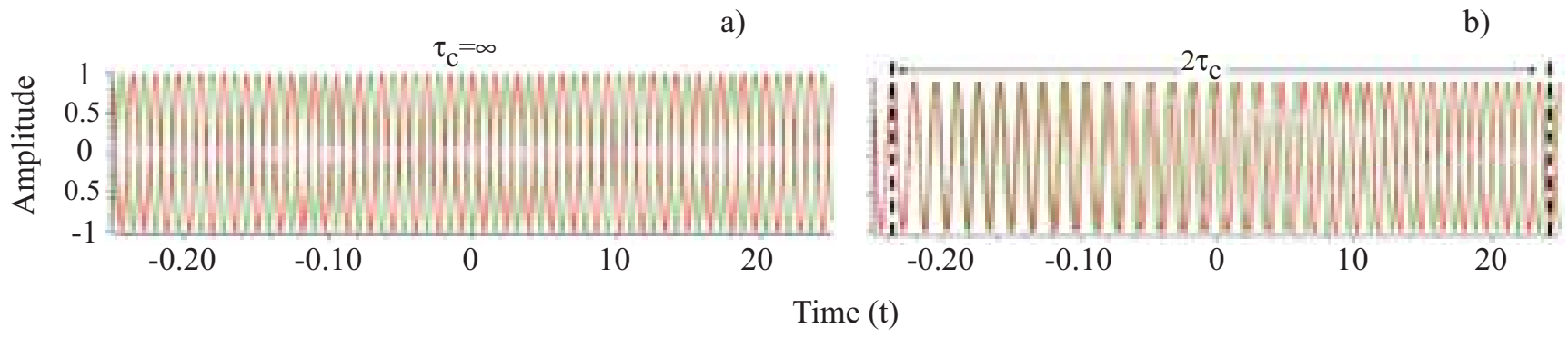

Fig. 24

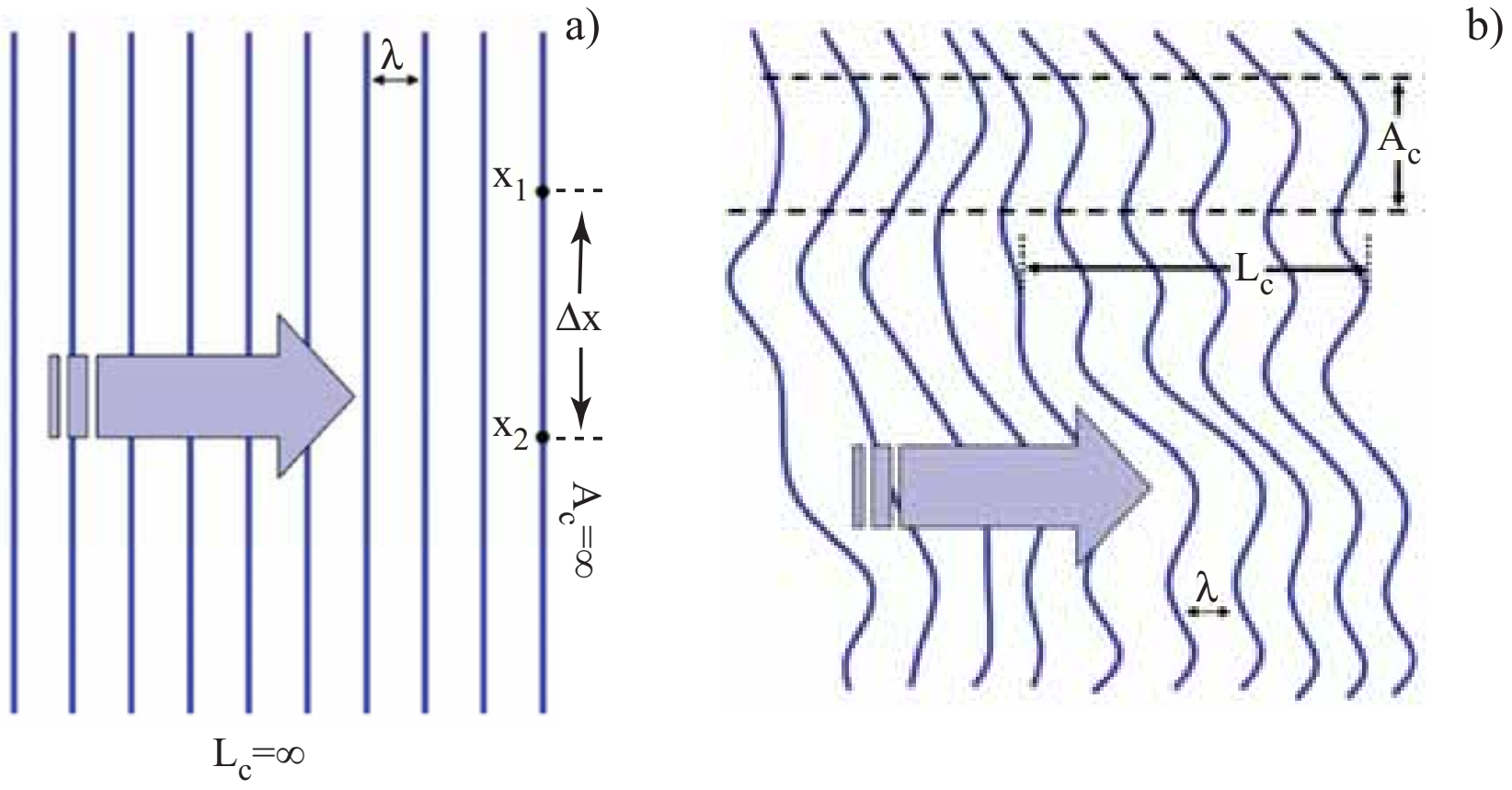

Fig. 25

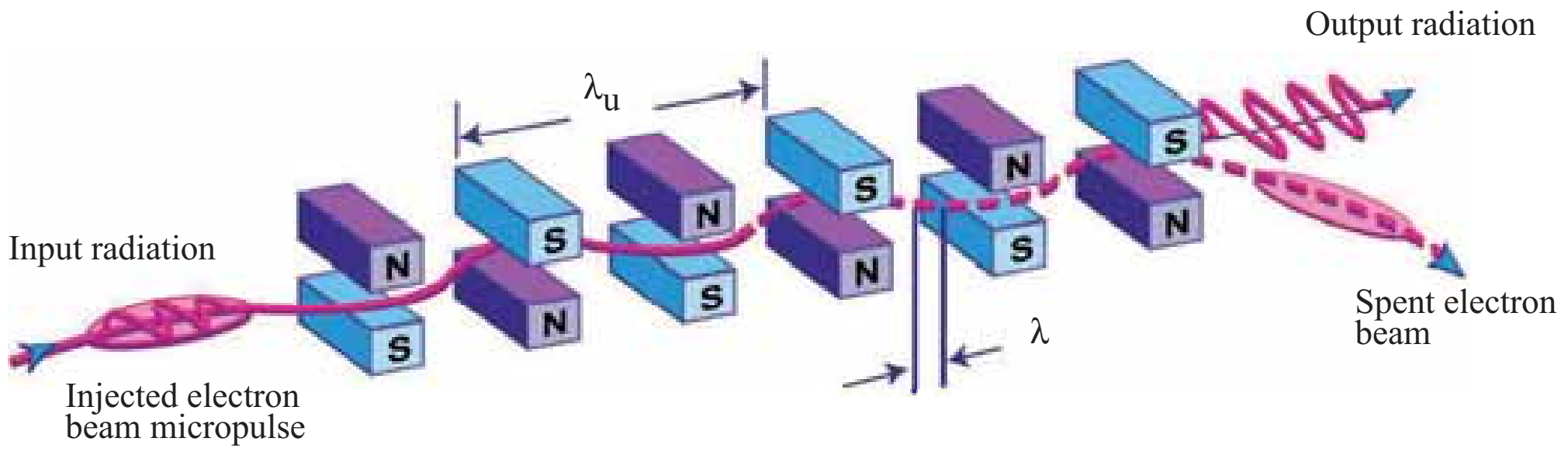

Fig. 26

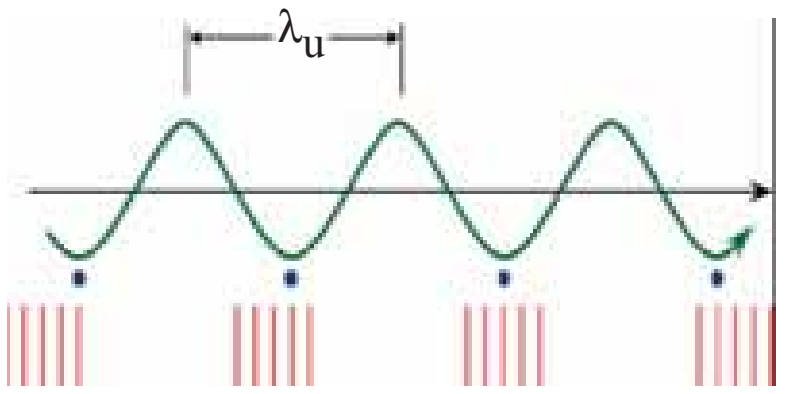

Fig. 27

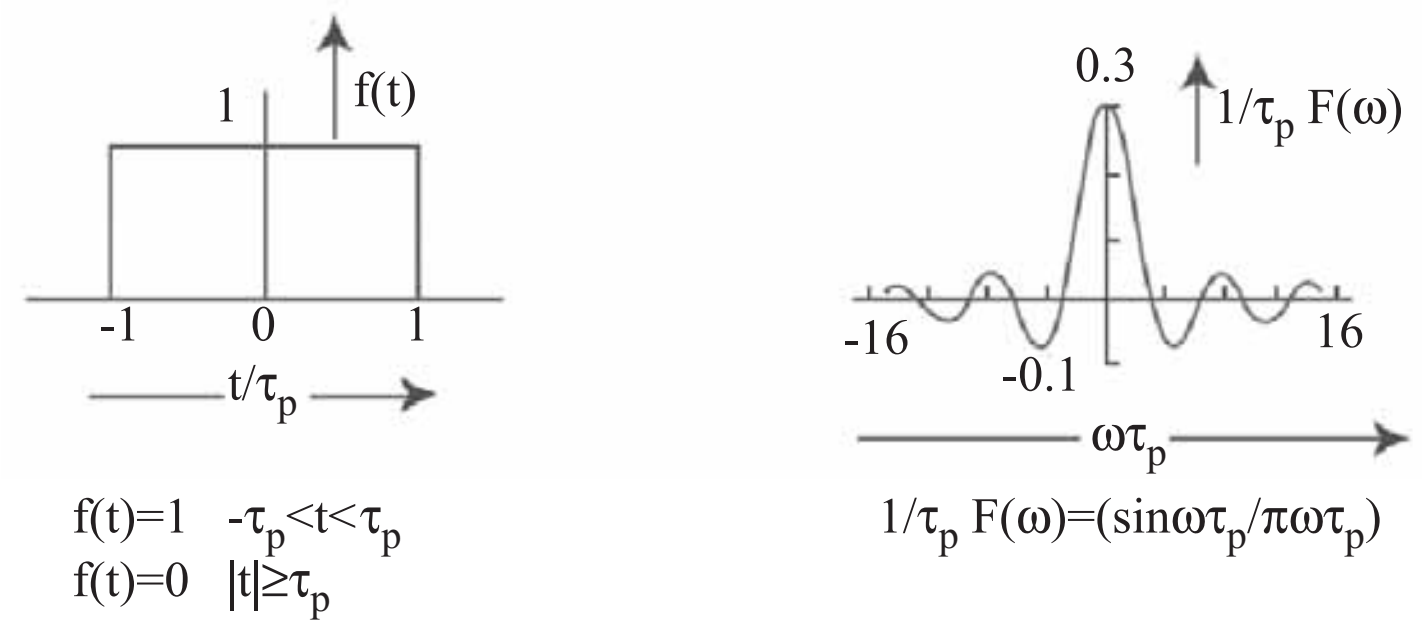

Fig. 28

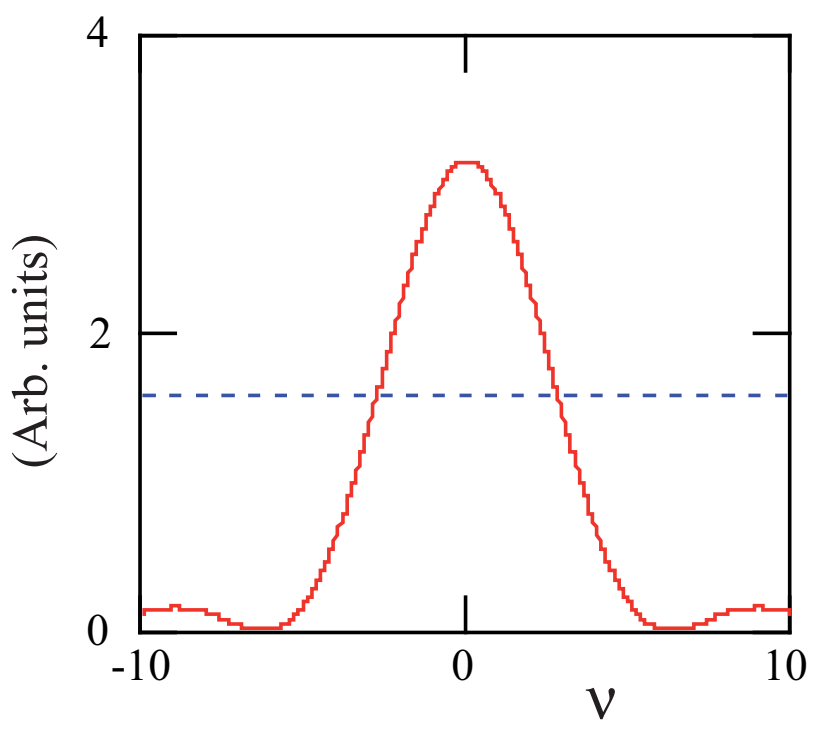

Fig. 29

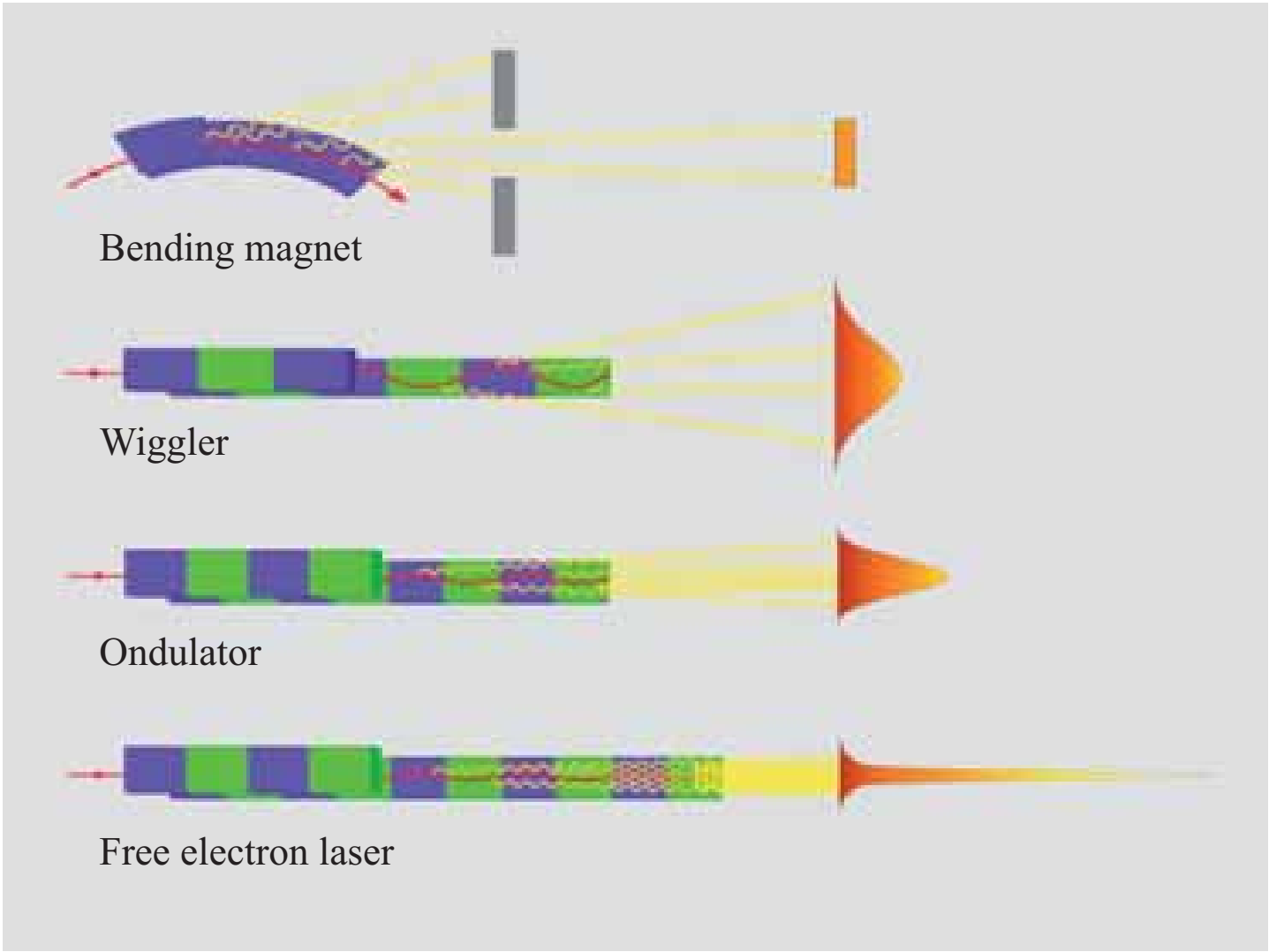

Fig. 30

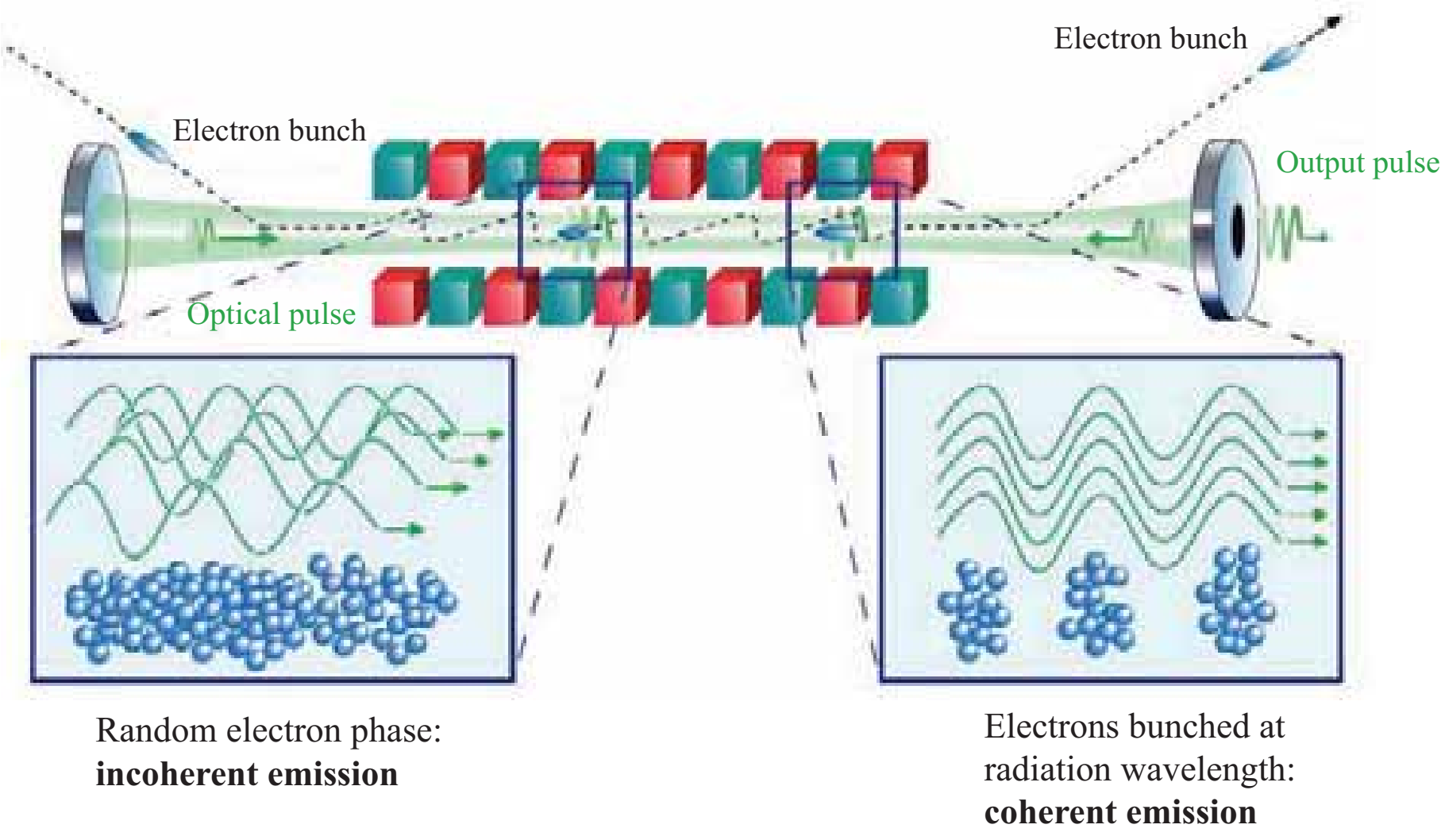

Fig. 31

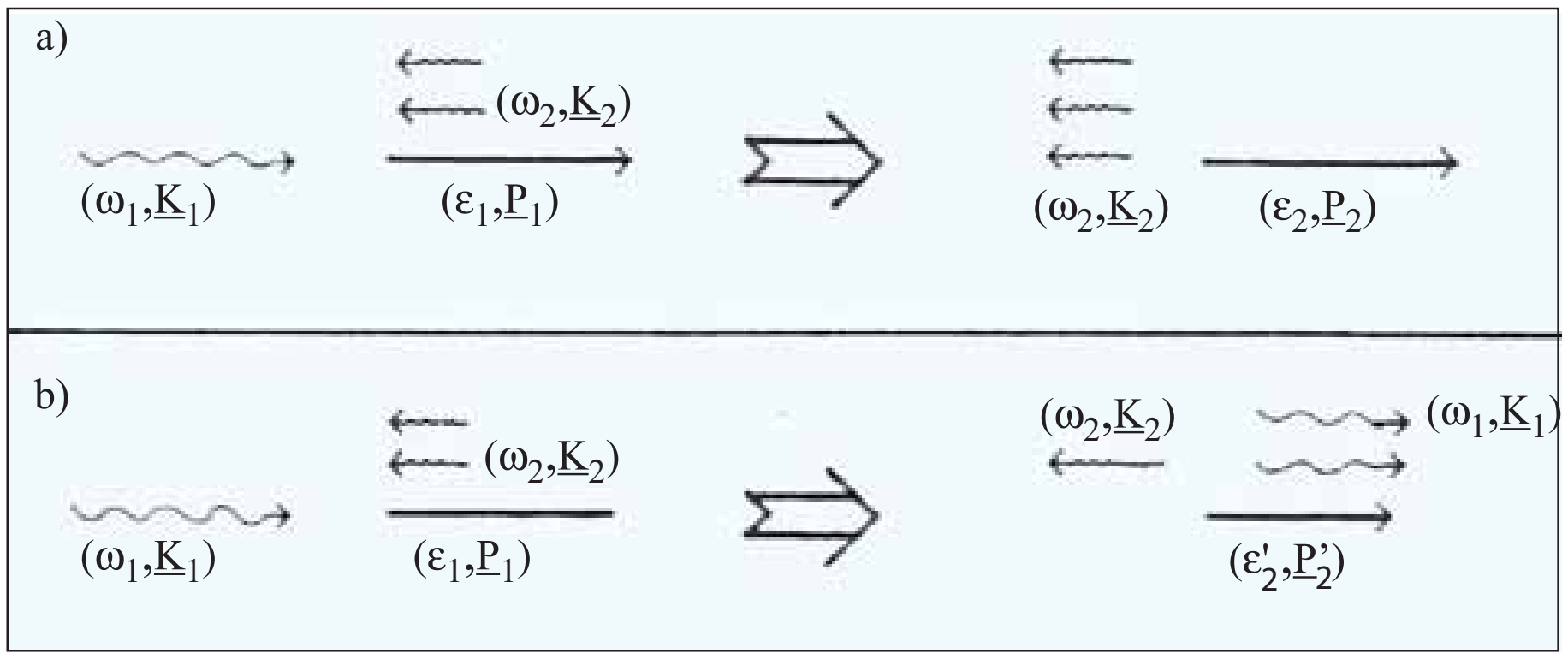

Fig. 32

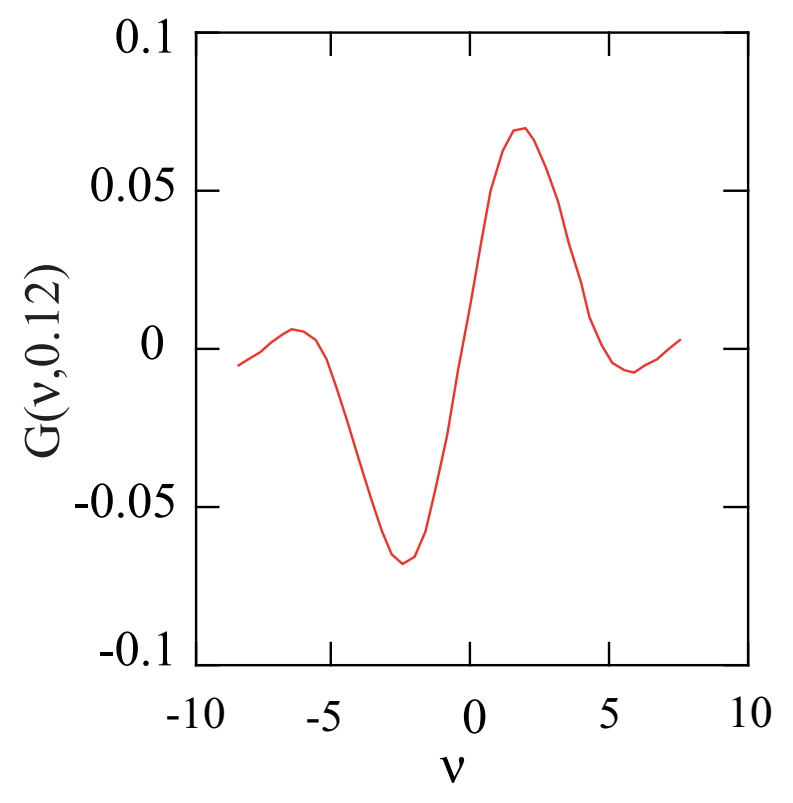

Fig. 33

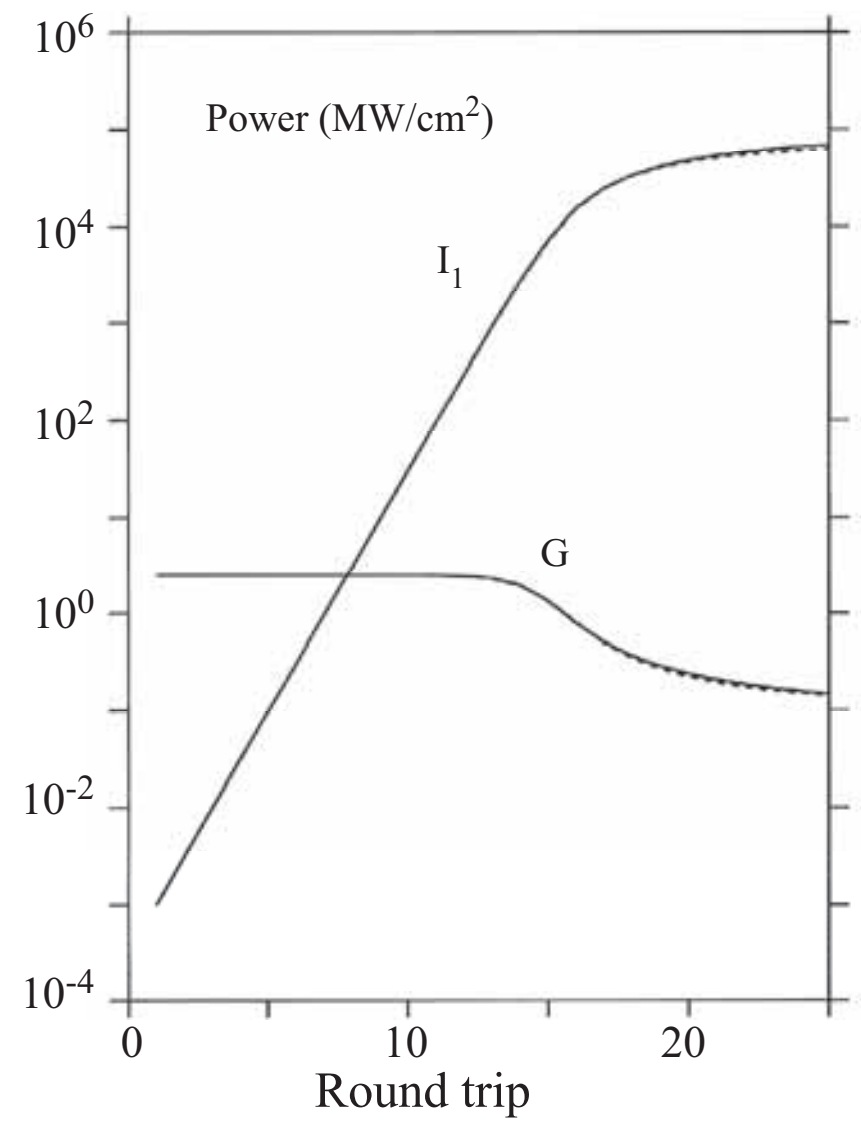

Fig. 34

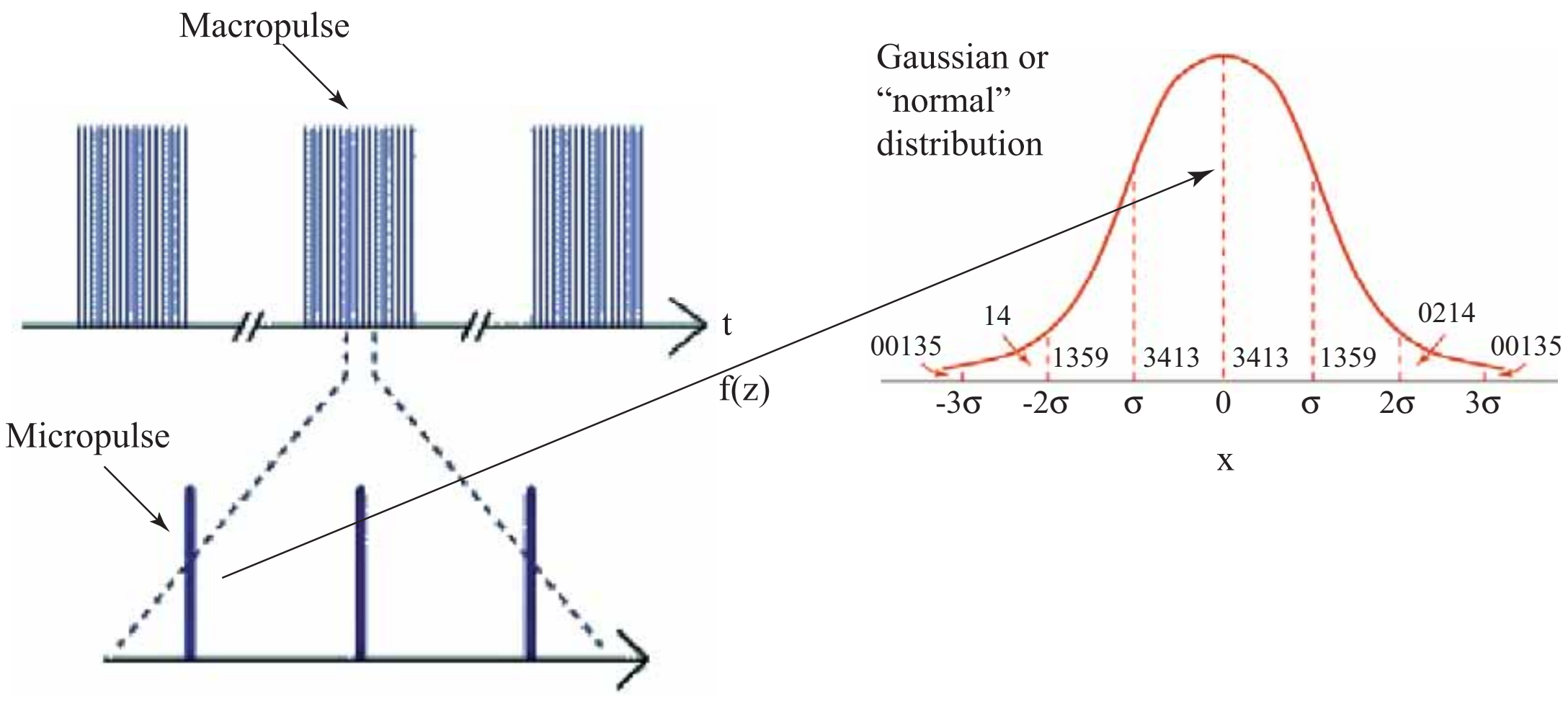

Fig. 35

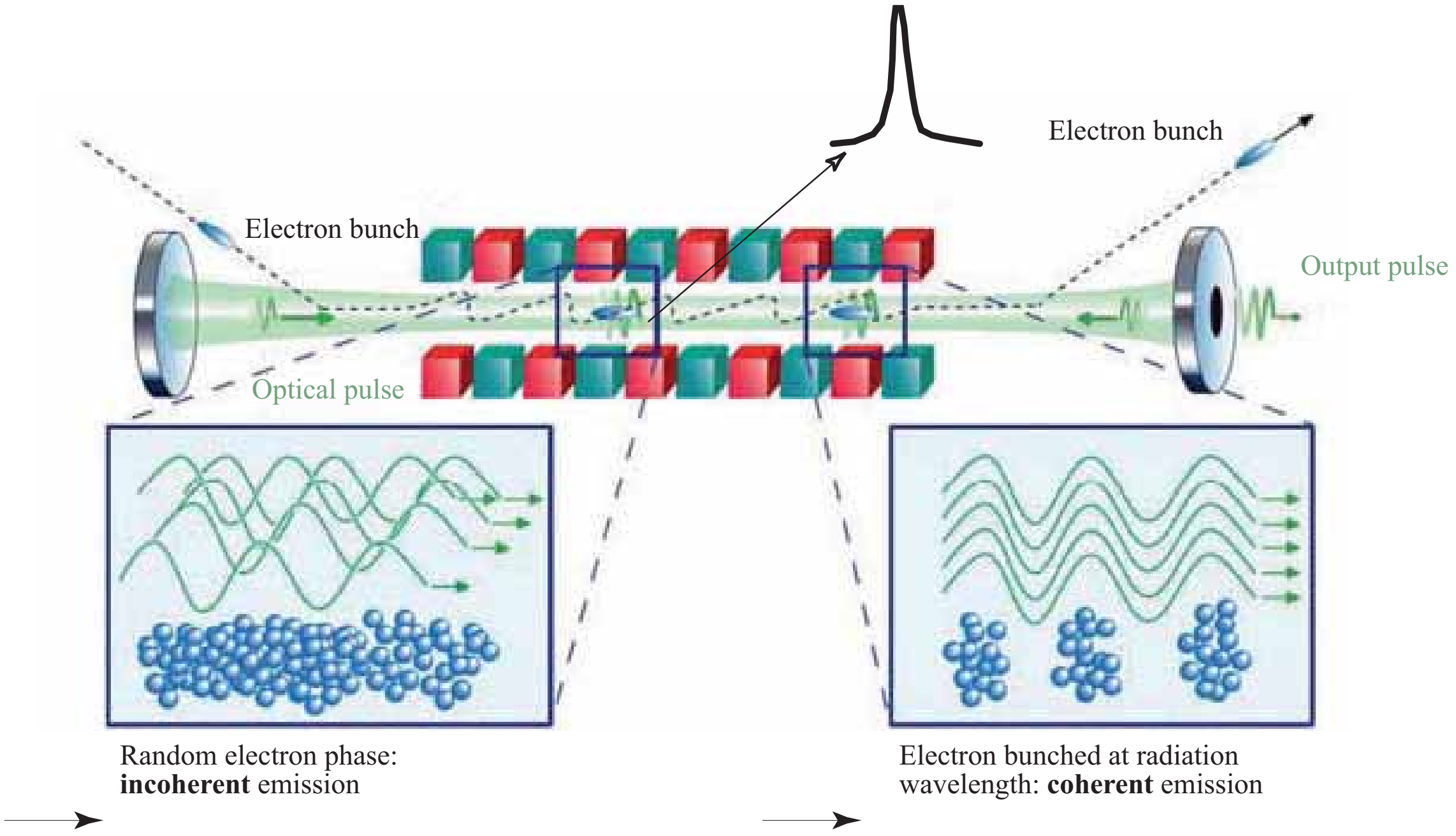

Fig. 36

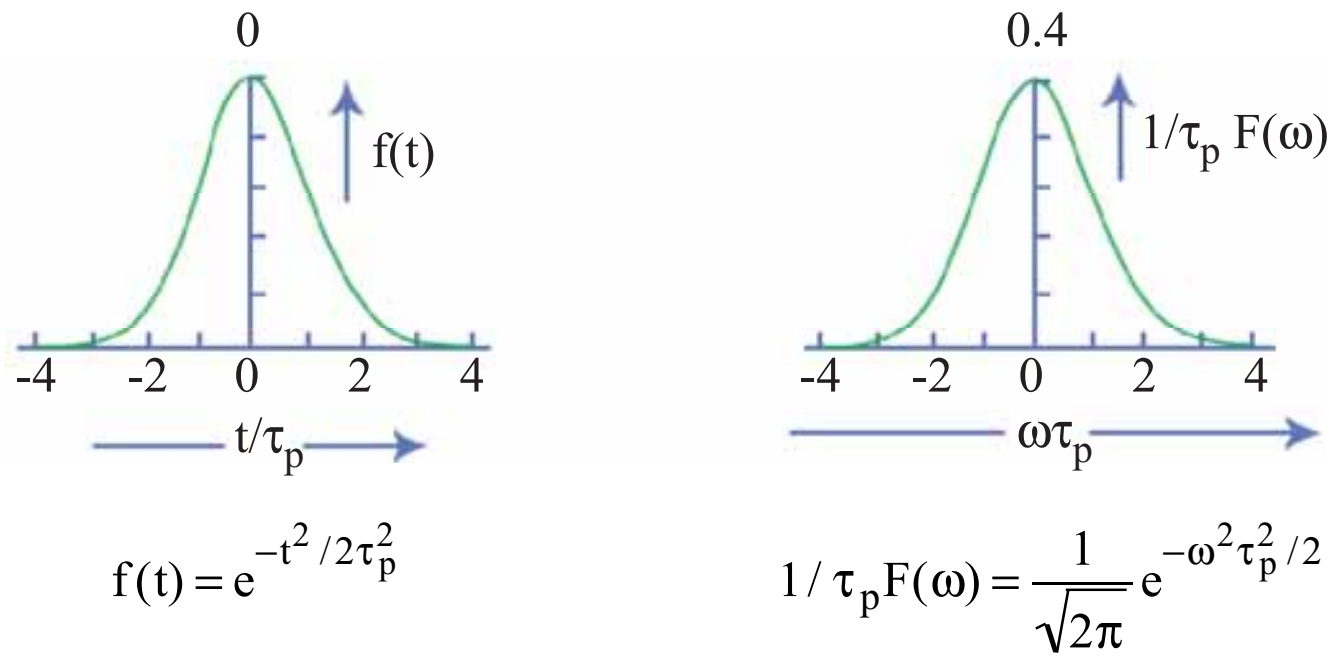

Fig. 37

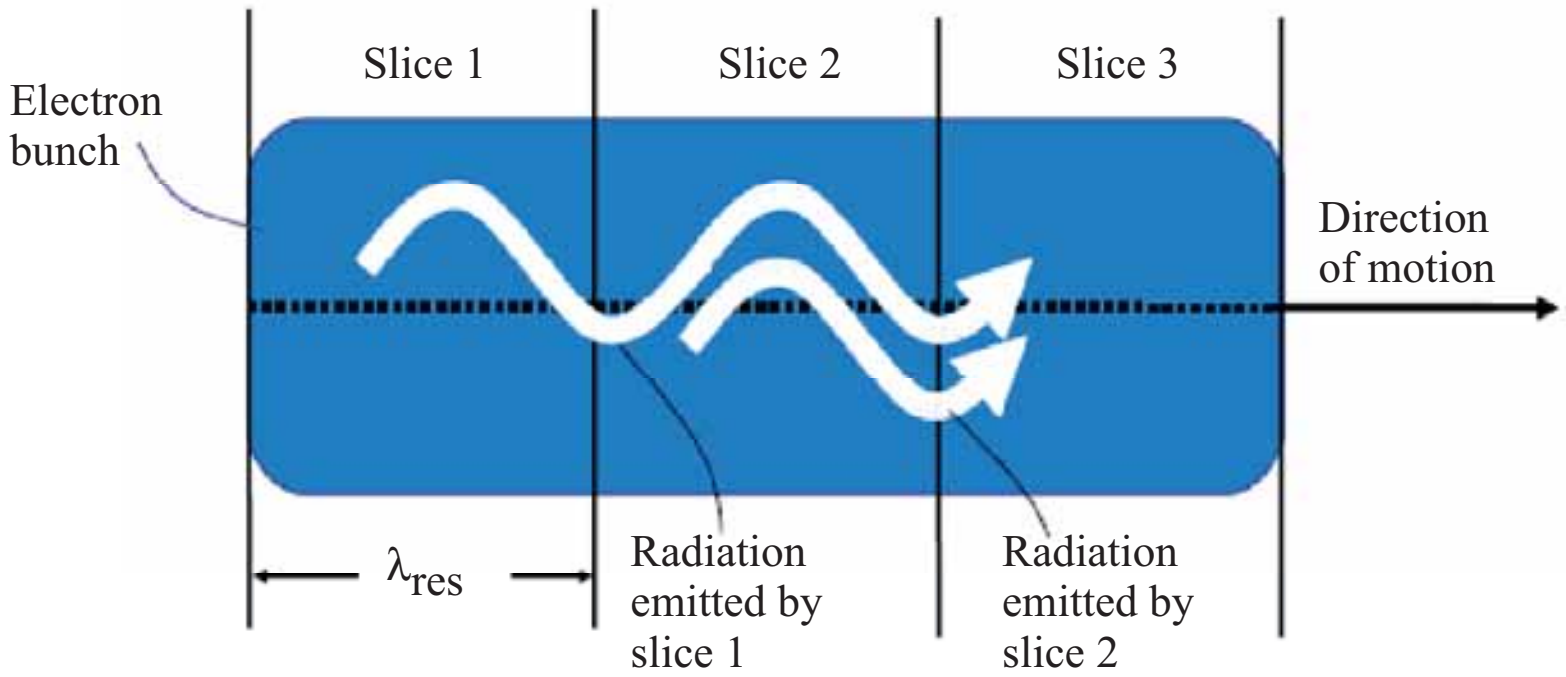

Fig. 38

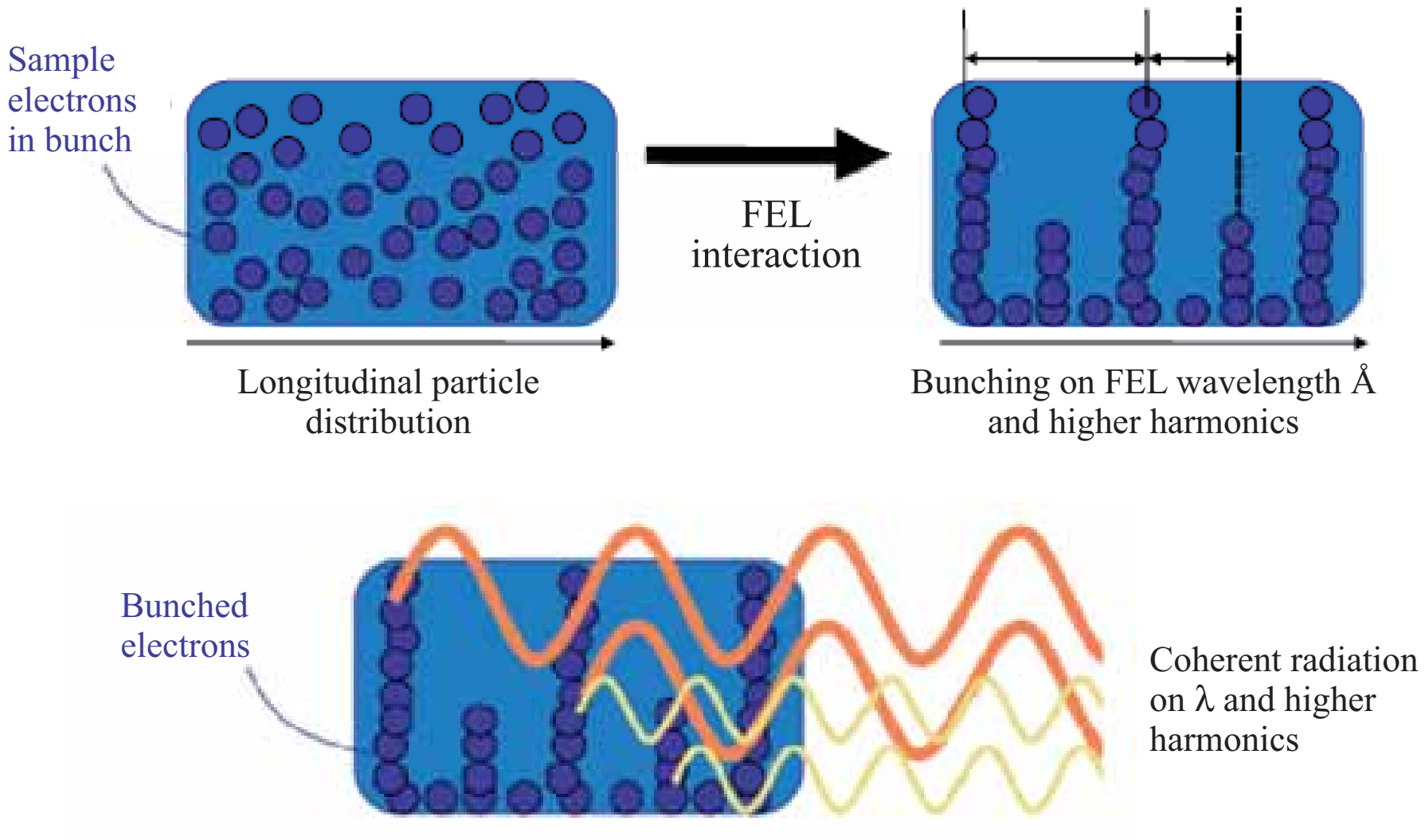

Fig. 39

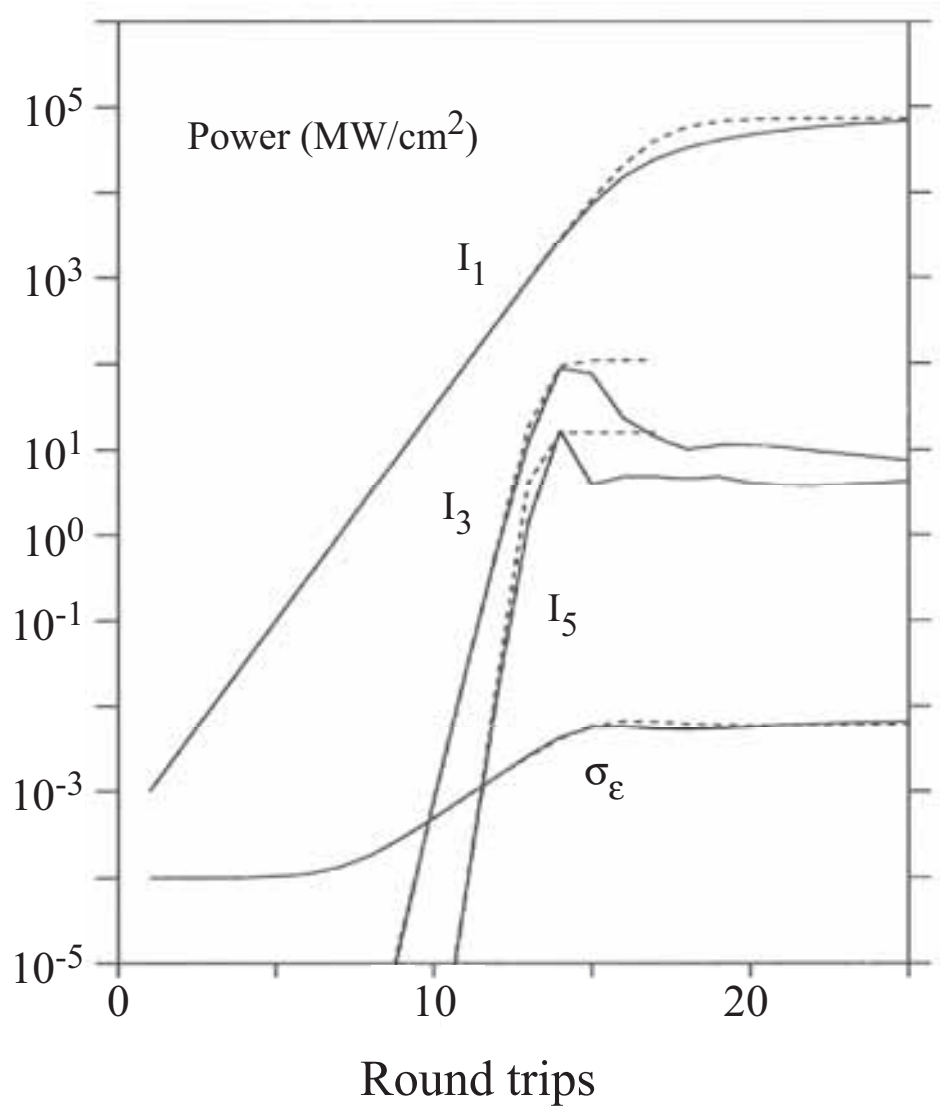

Fig. 40

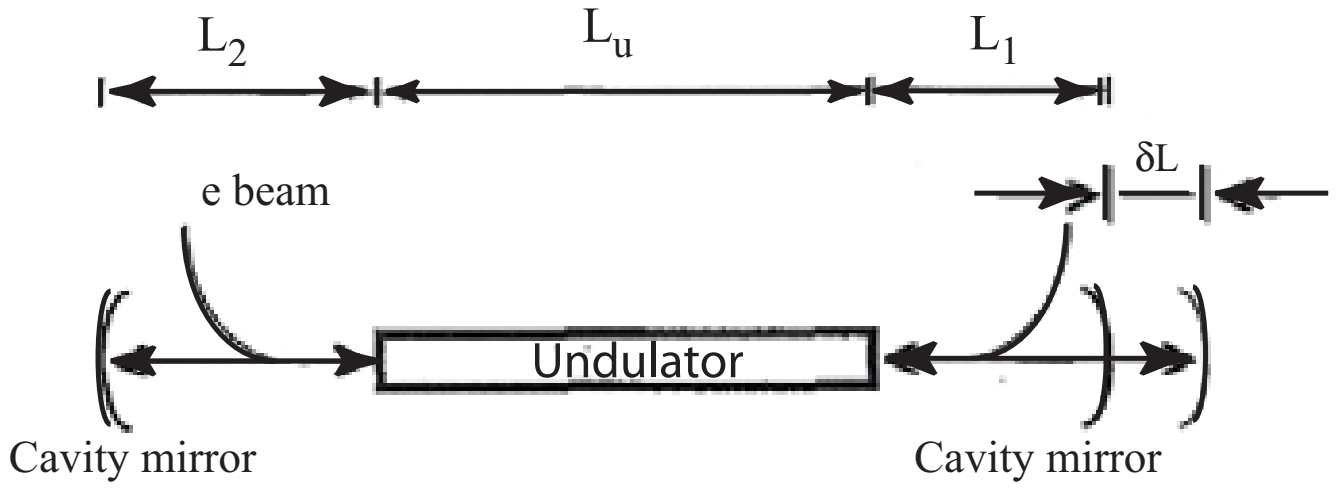

Fig. 41

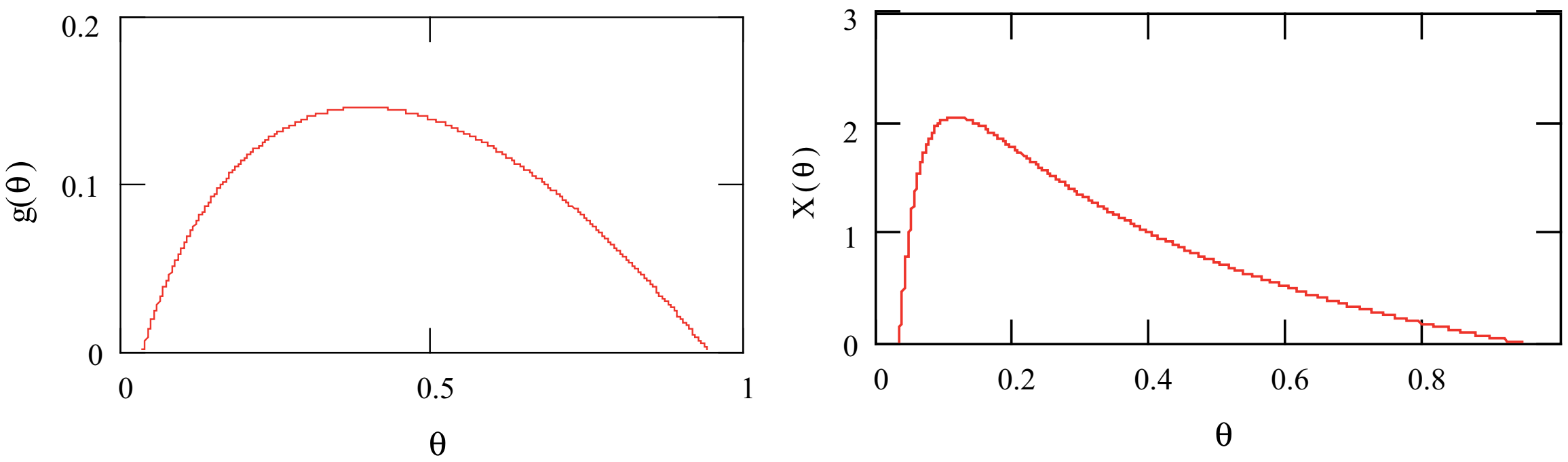

Fig. 42

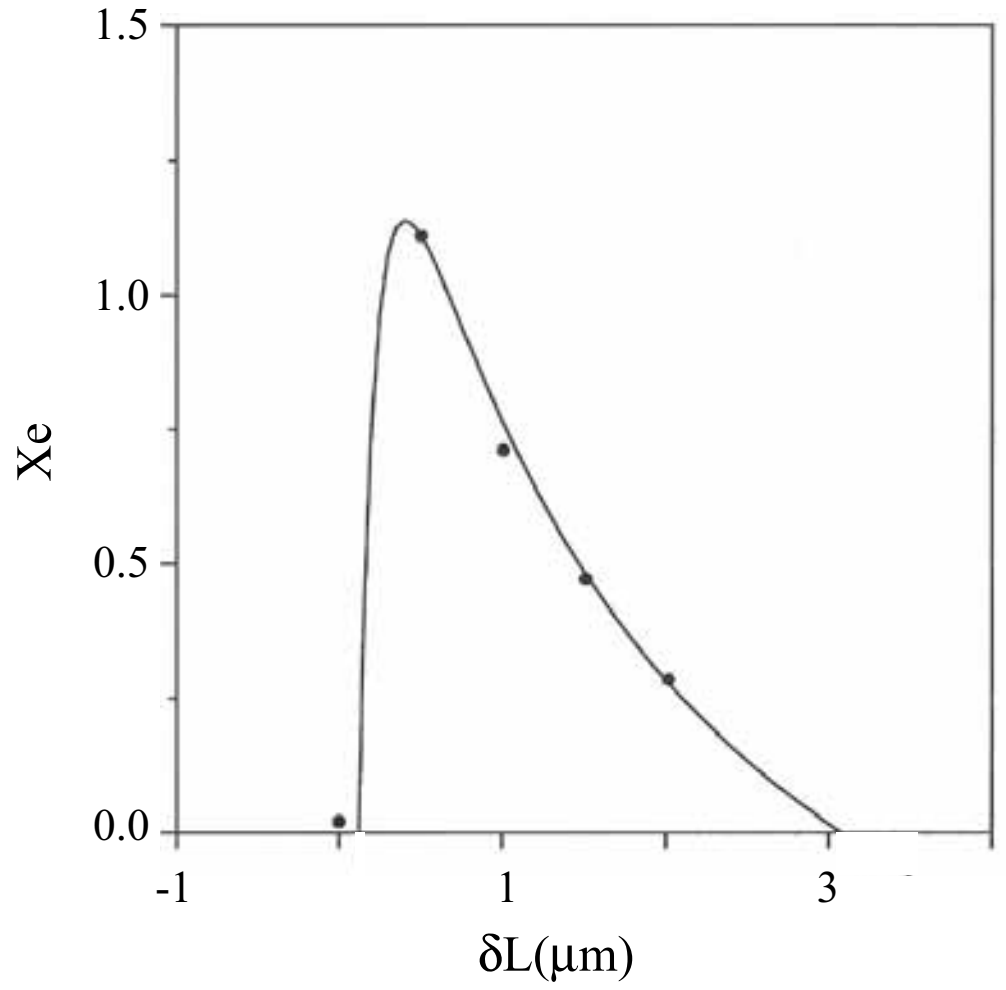

Fig. 43

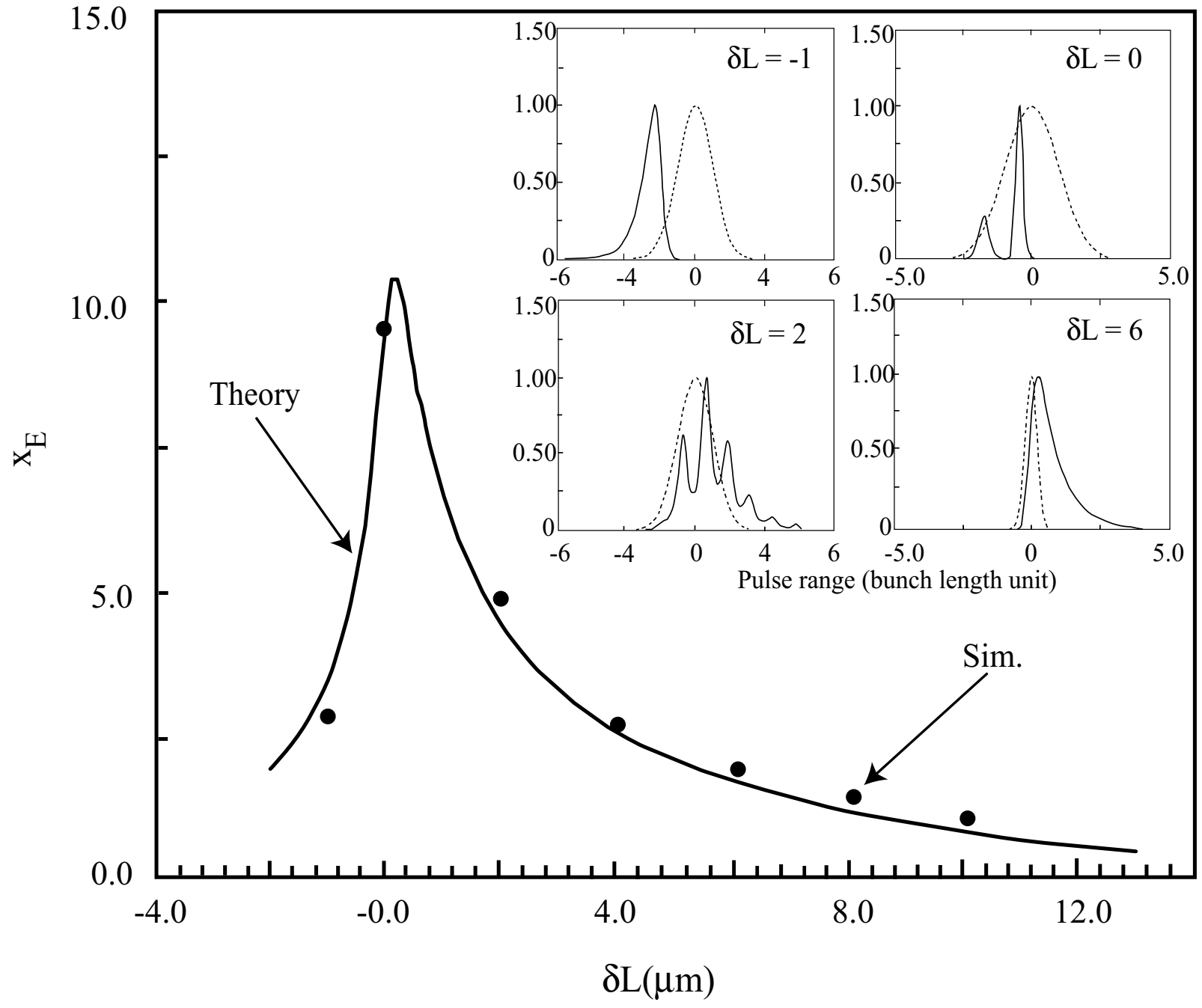

Fig. 44

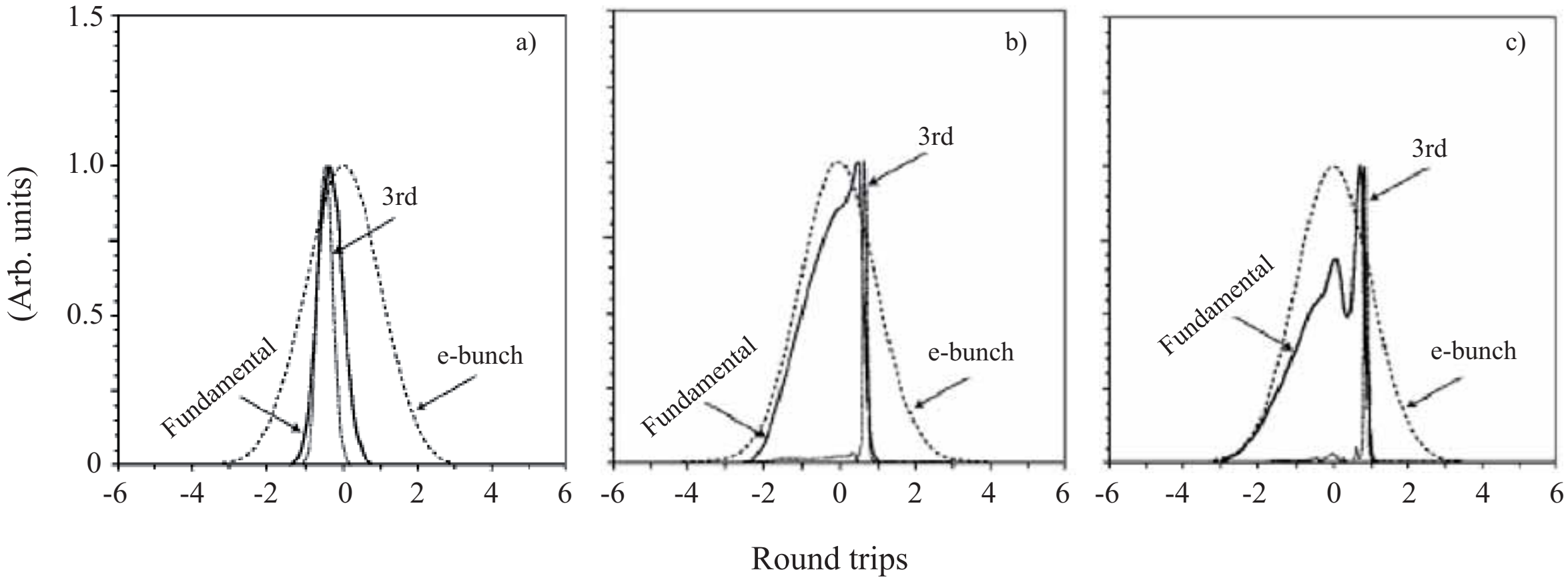

Fig. 45

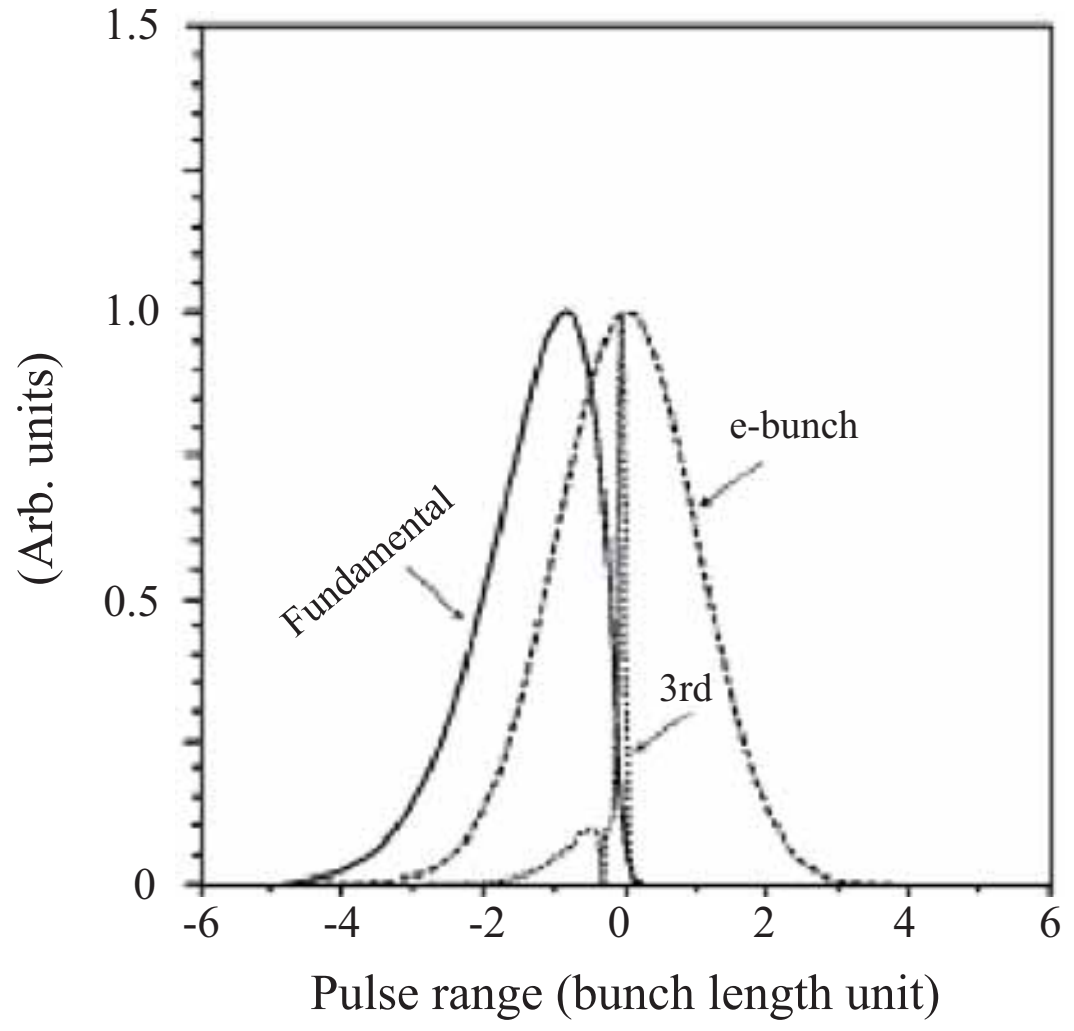

Fig. 46

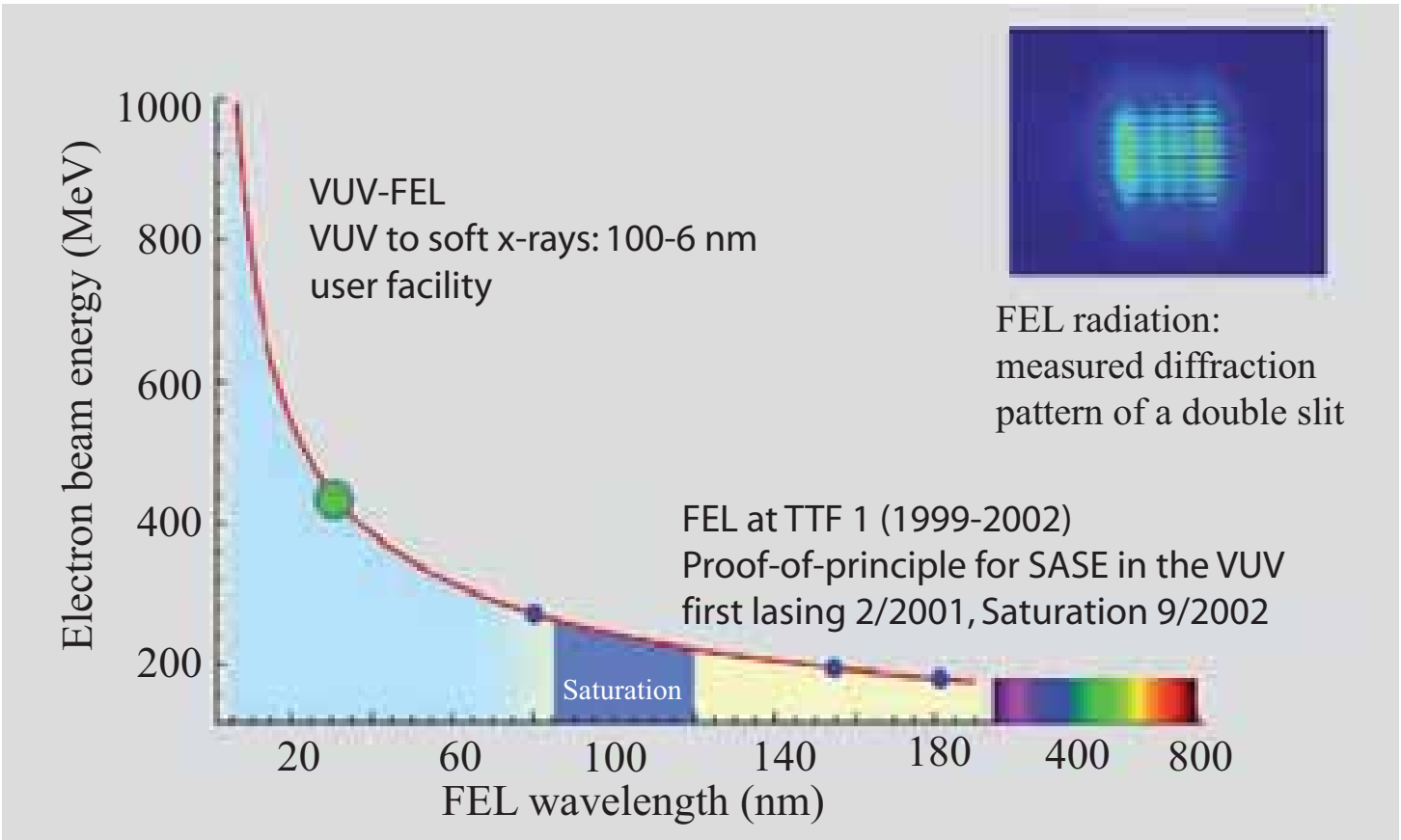

Fig. 47

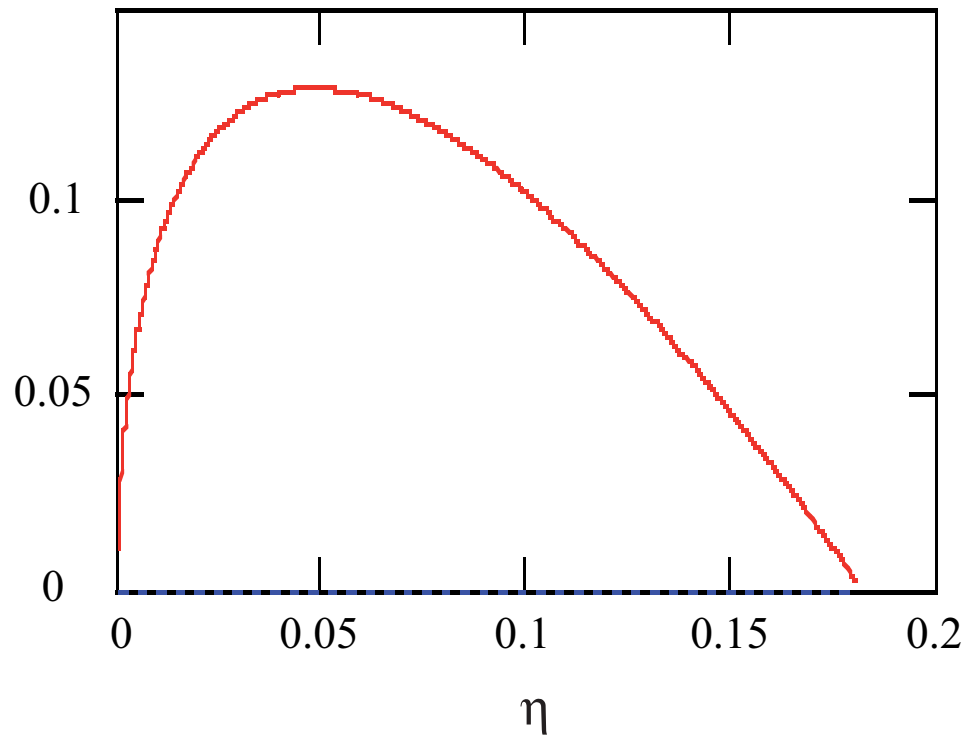

Fig. 48

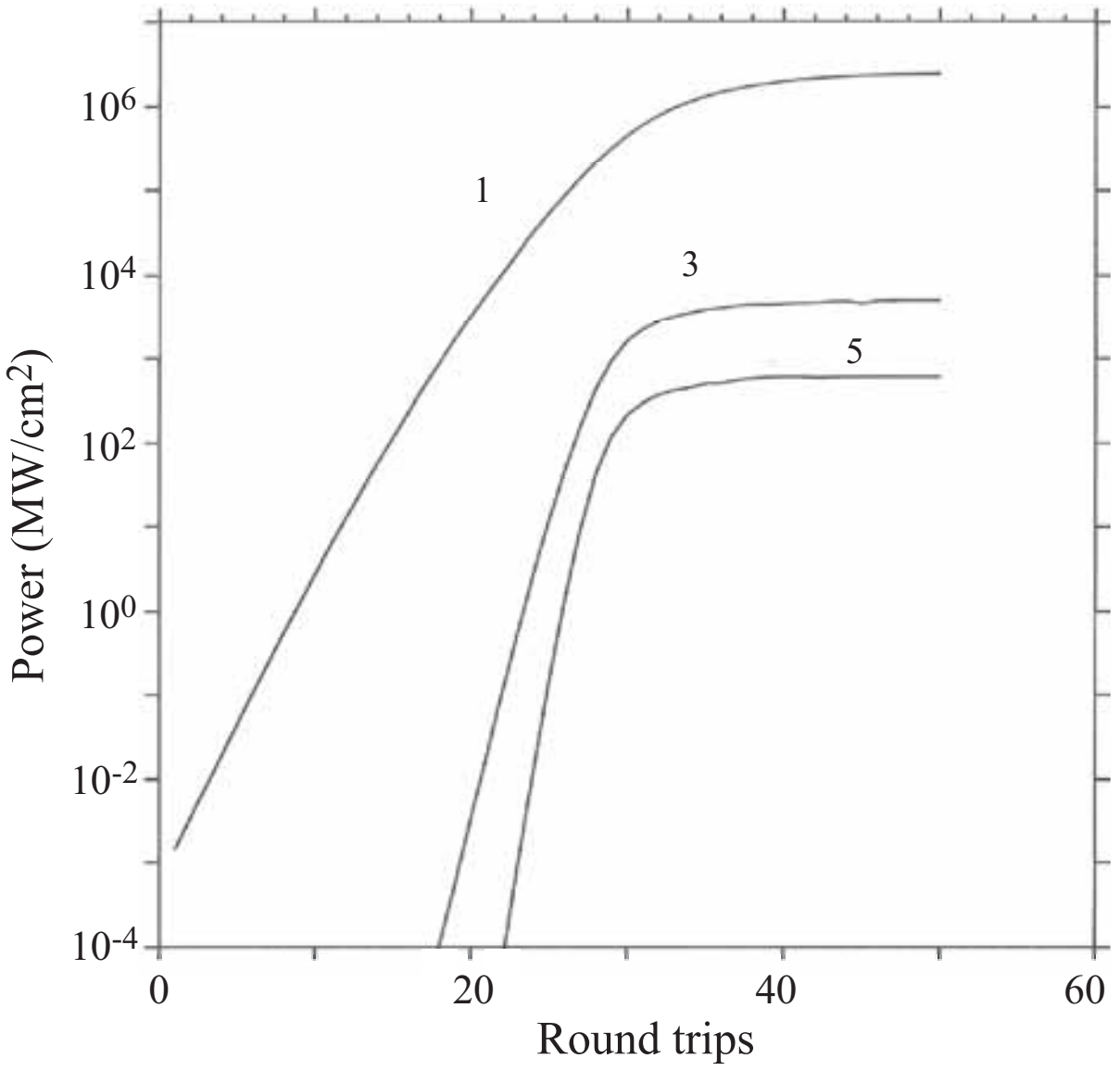

Fig. 49

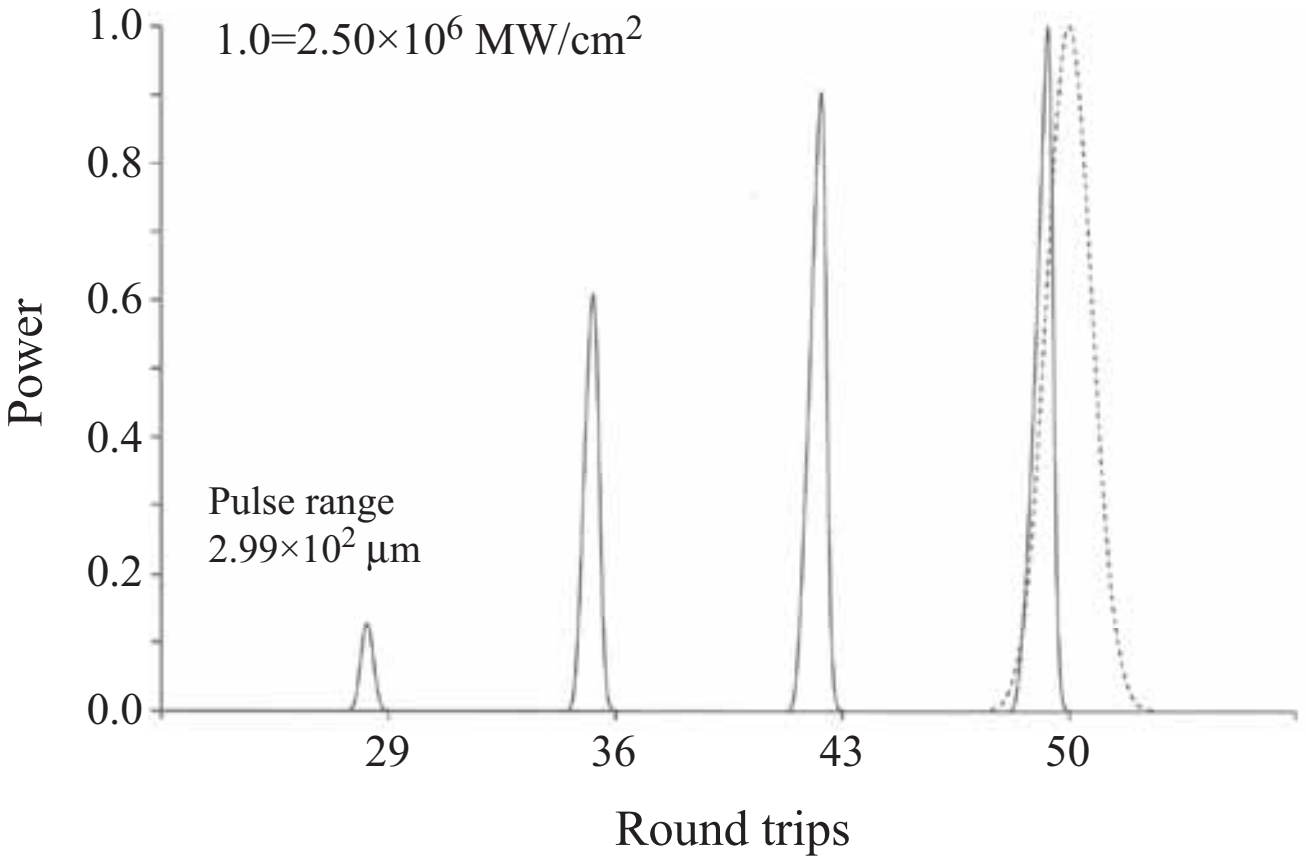

Fig. 50

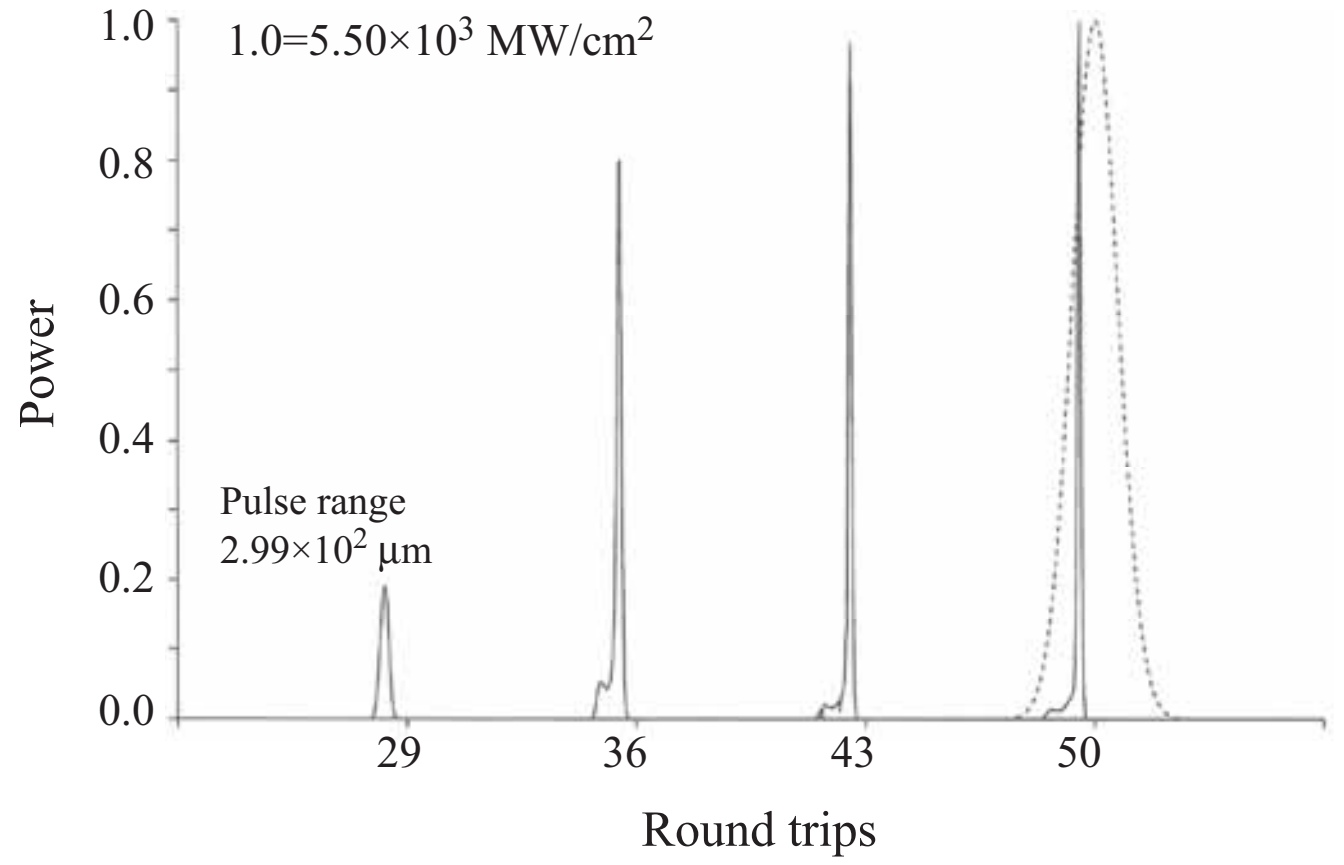

Fig. 51

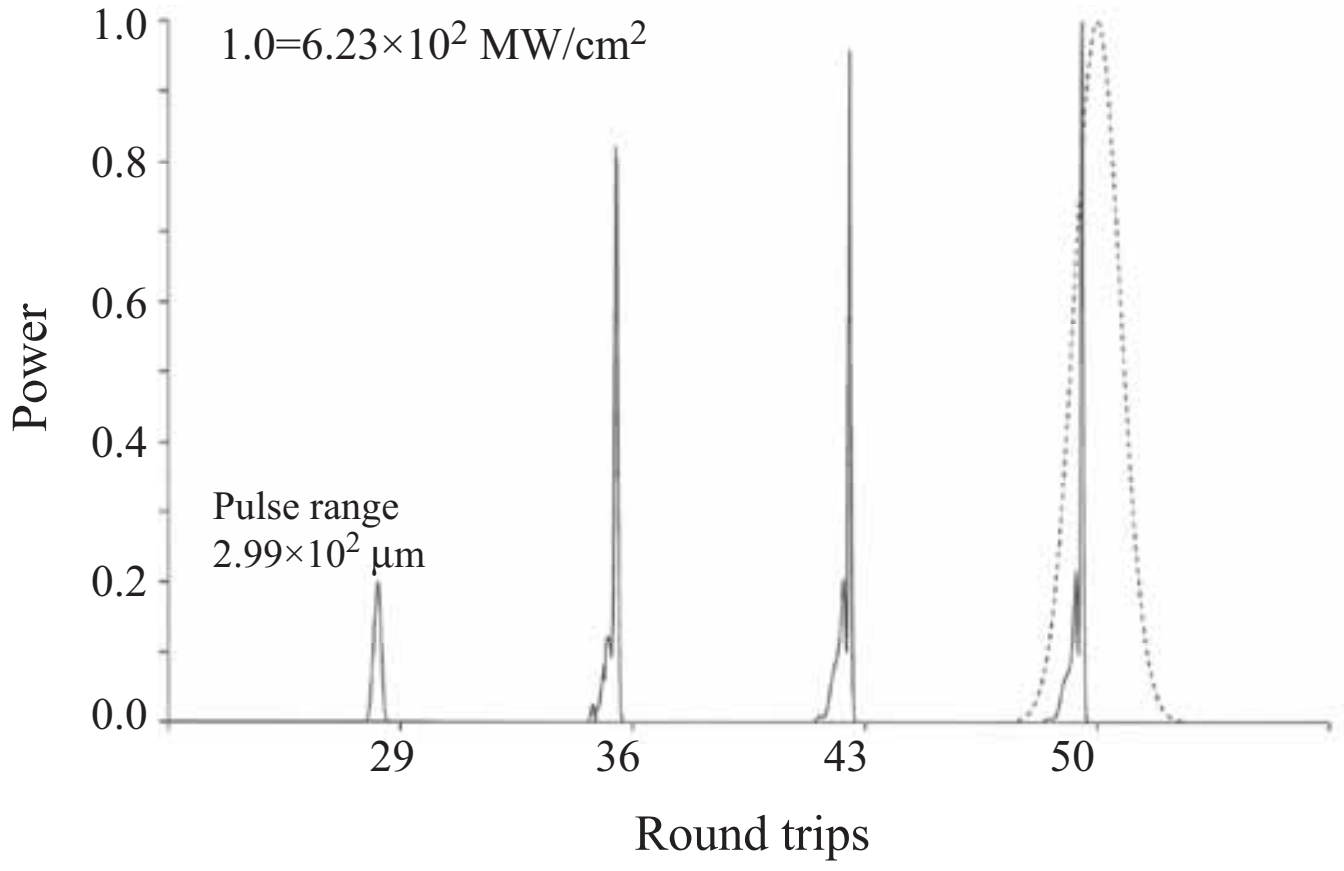

Fig. 52

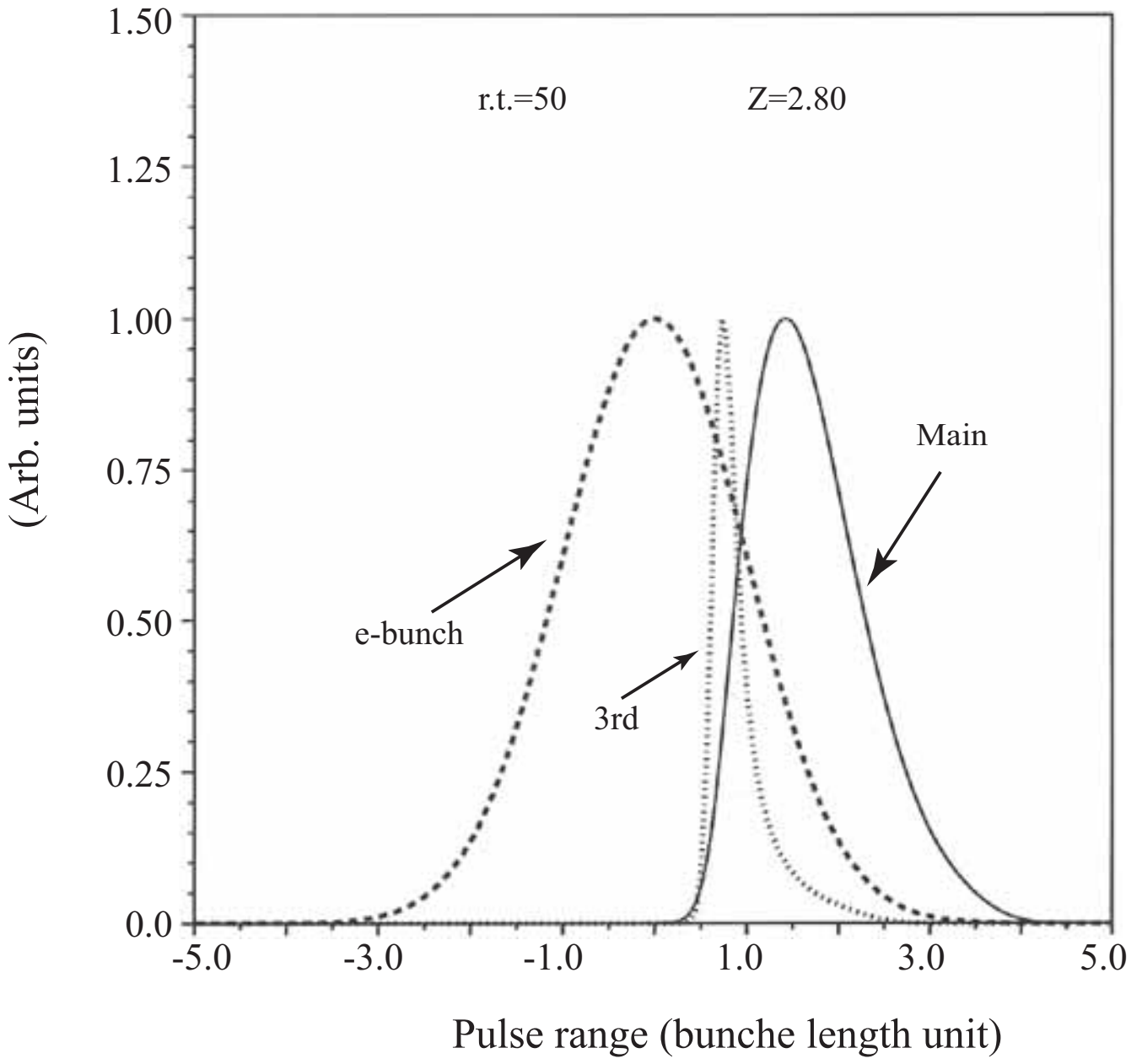

Fig. 53

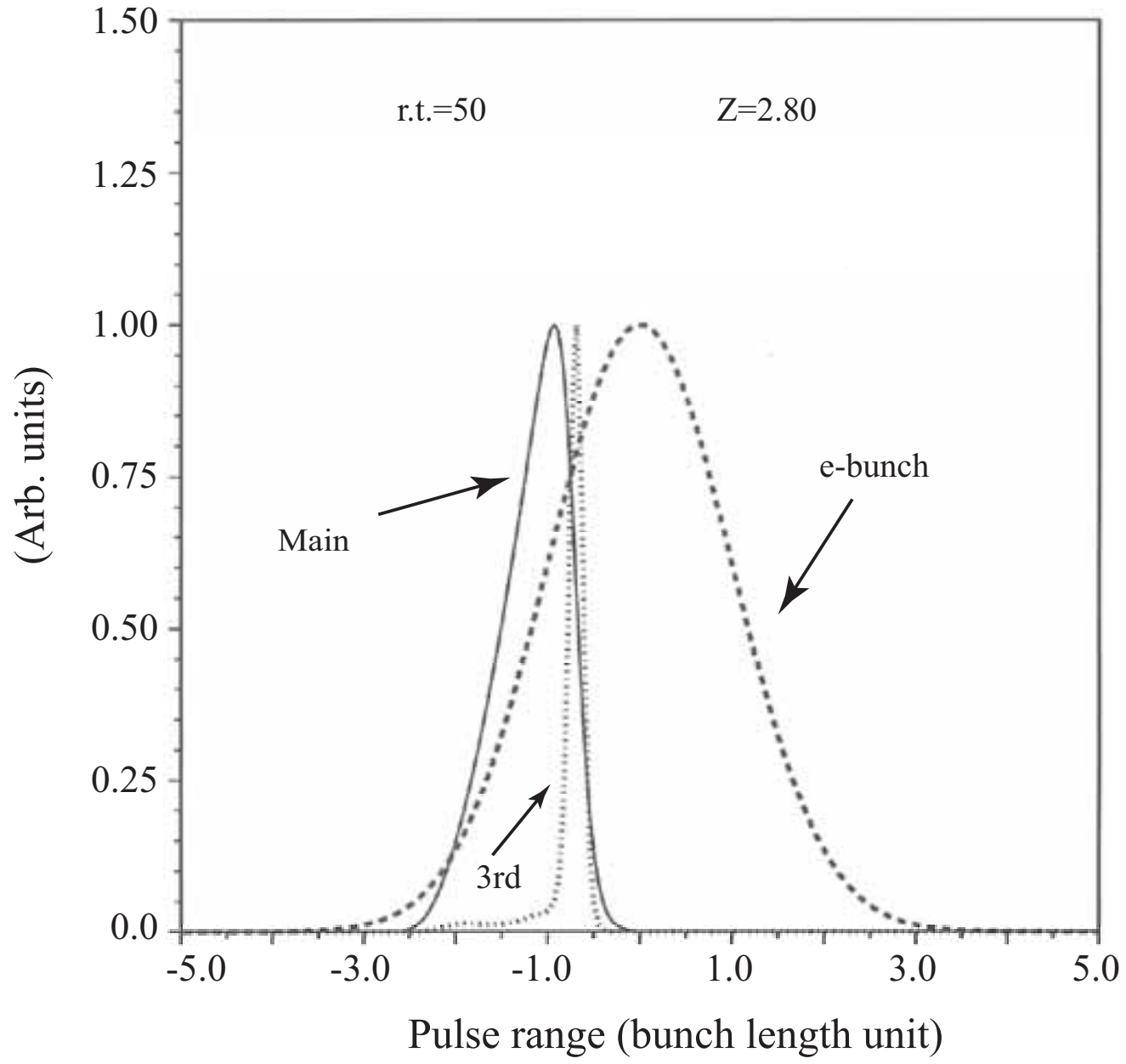

Fig. 54

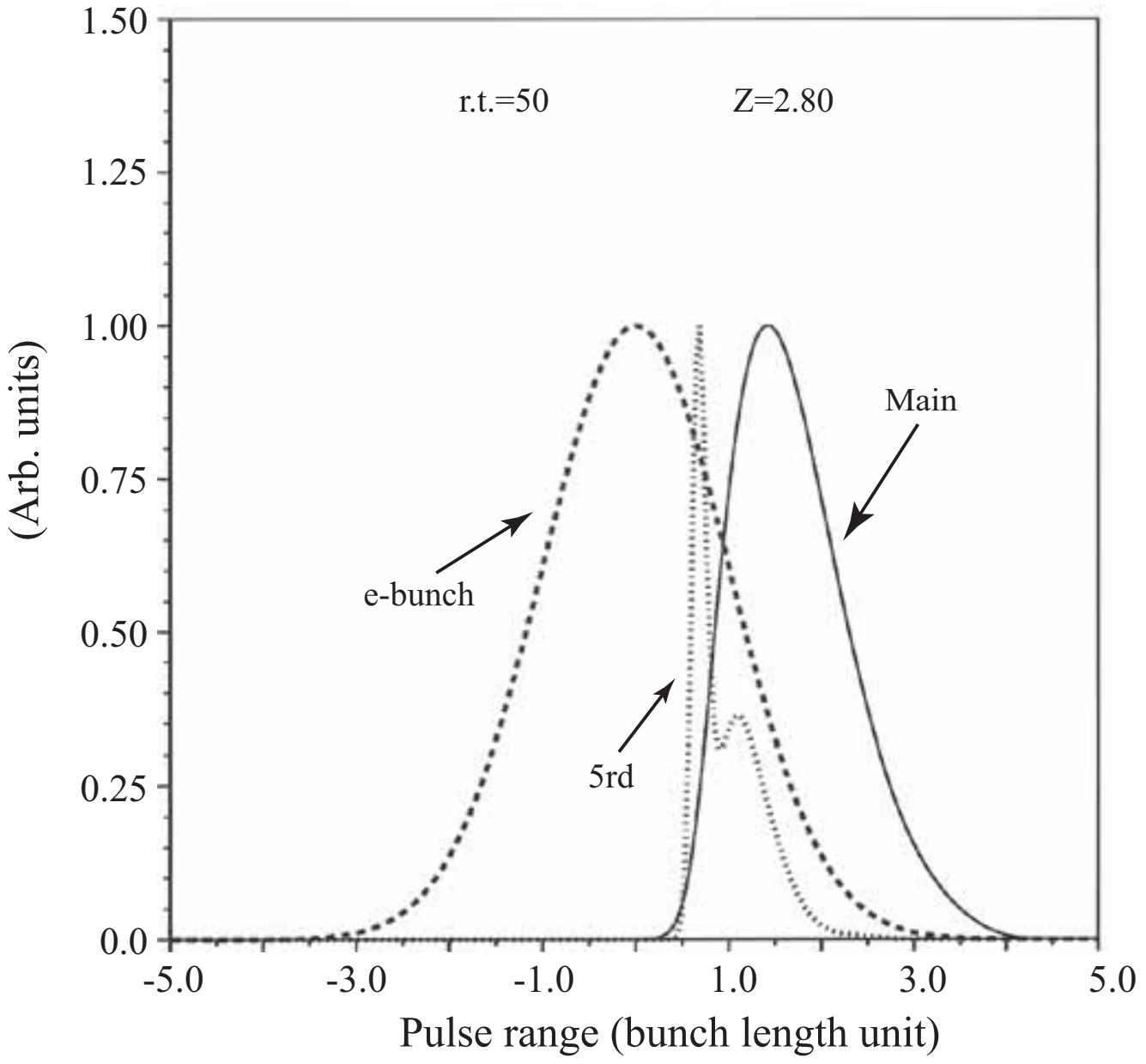

Fig. 55

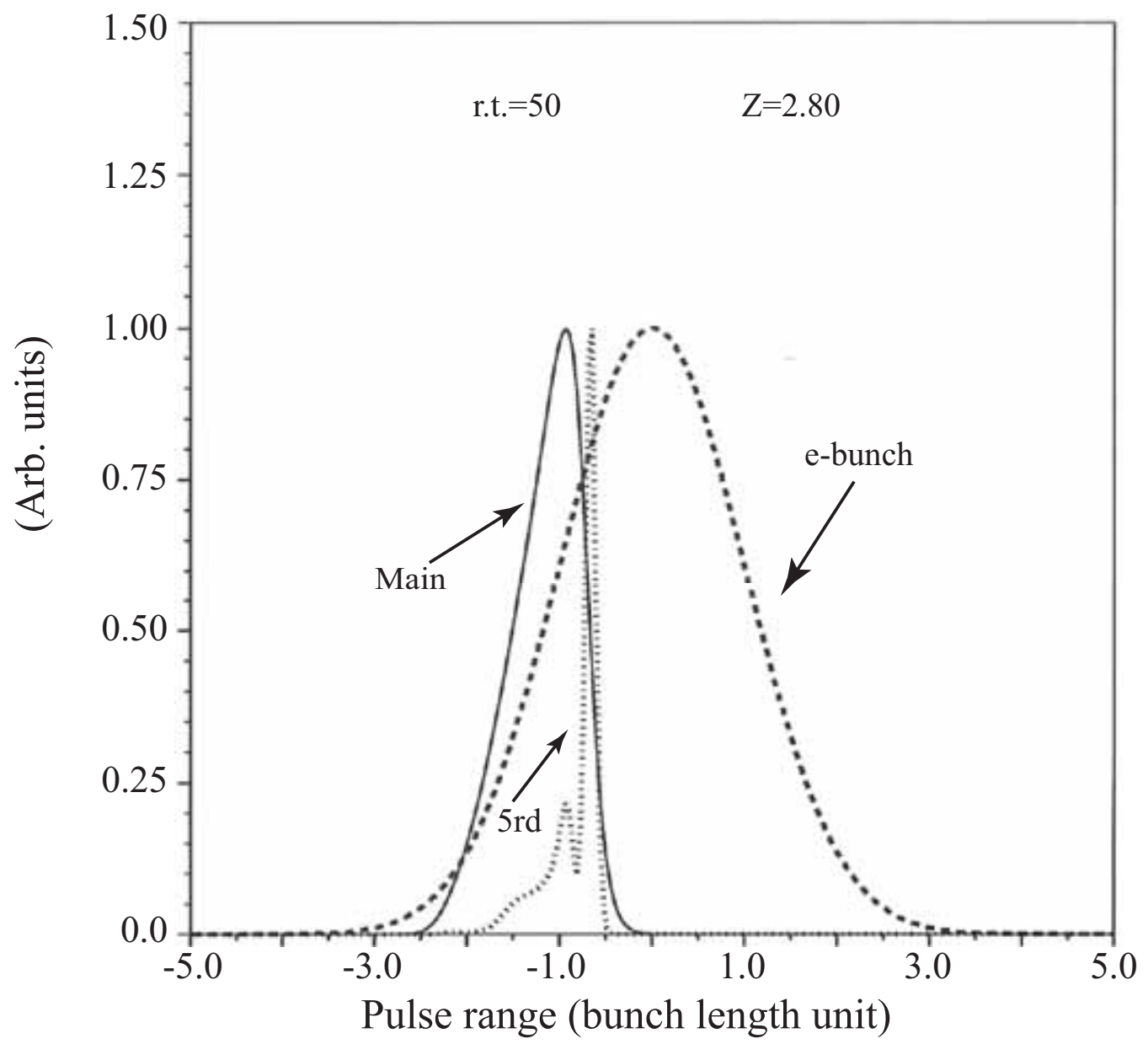

Fig. 56

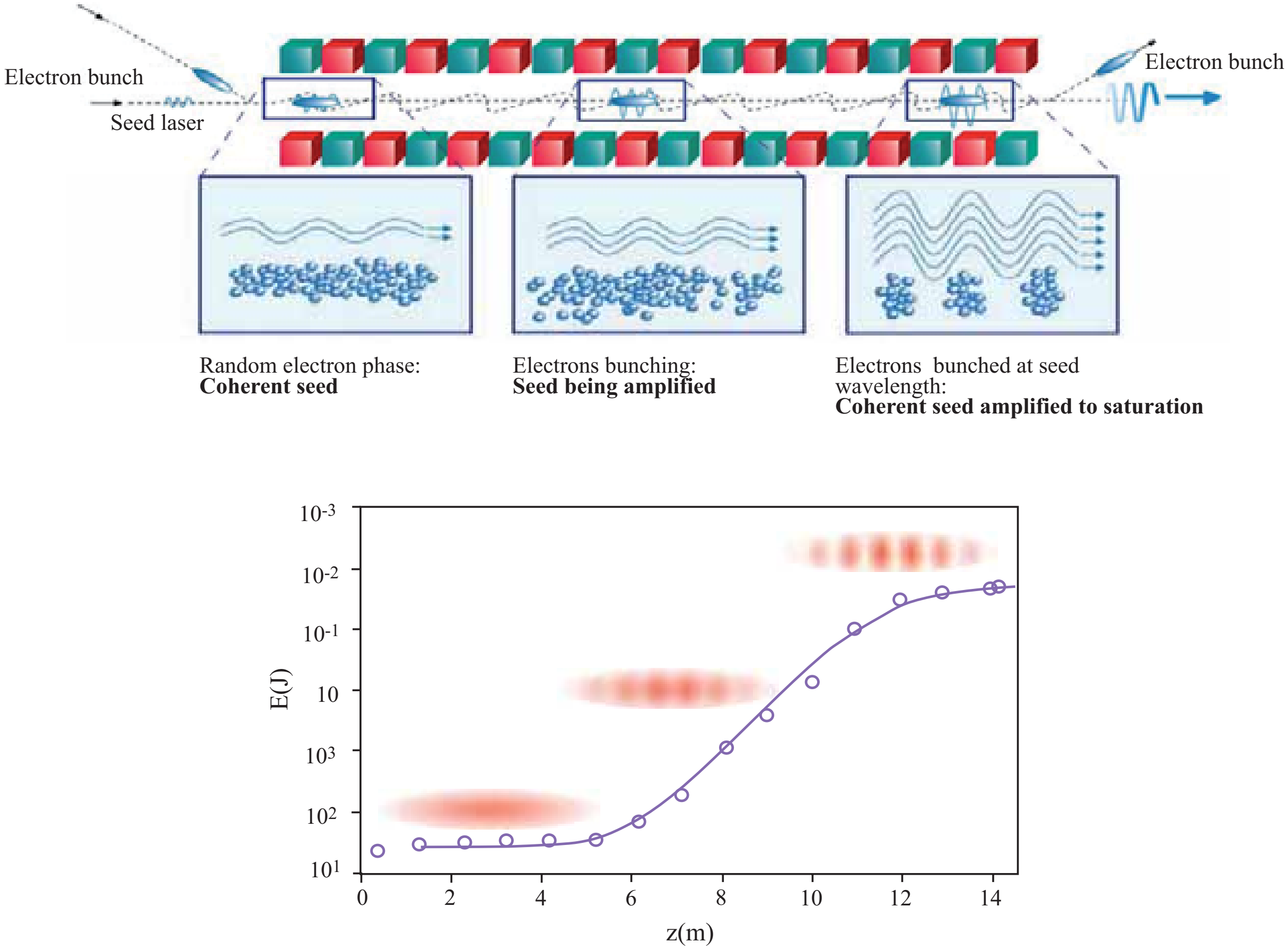

Fig. 57

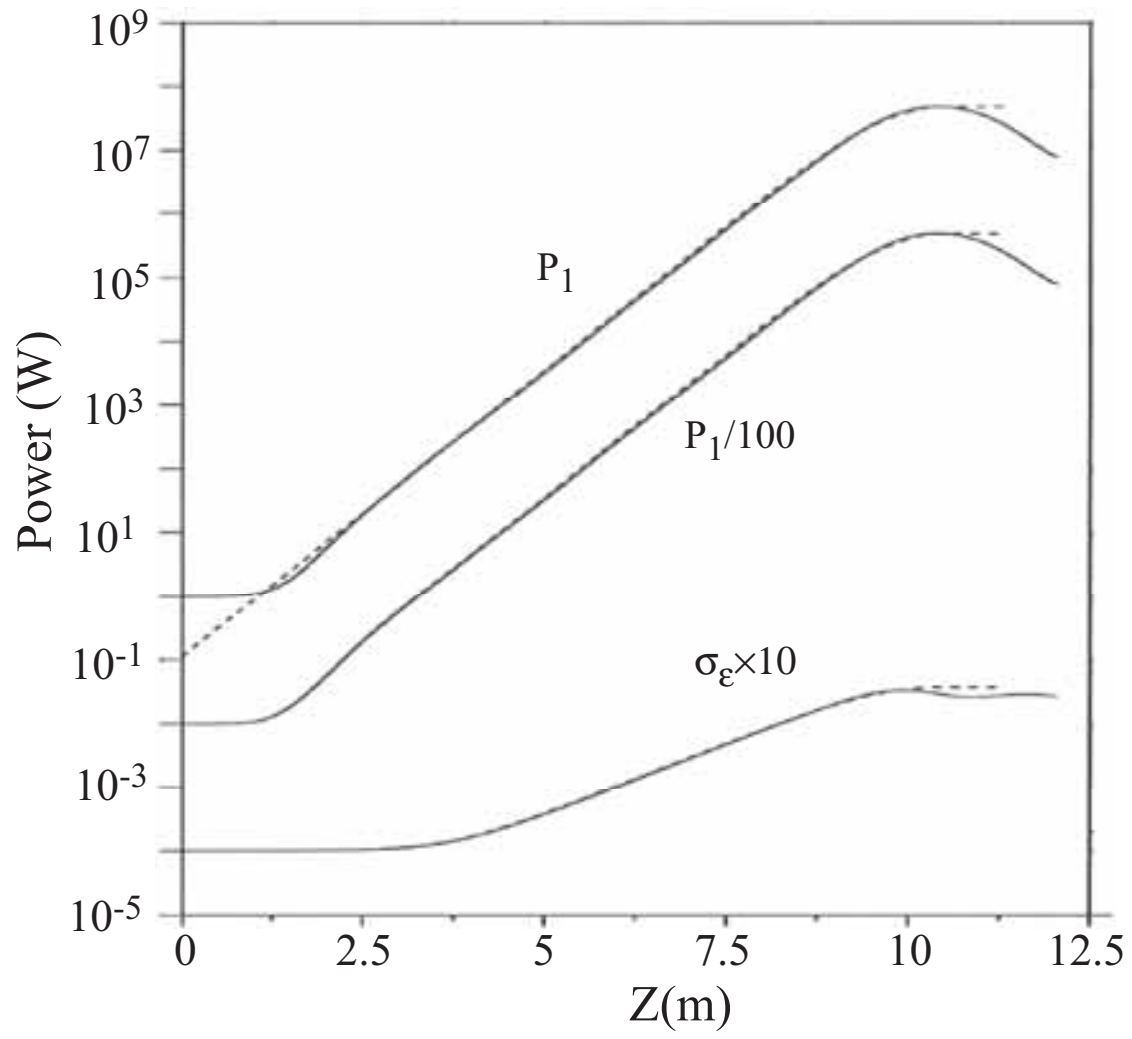

Fig. 58

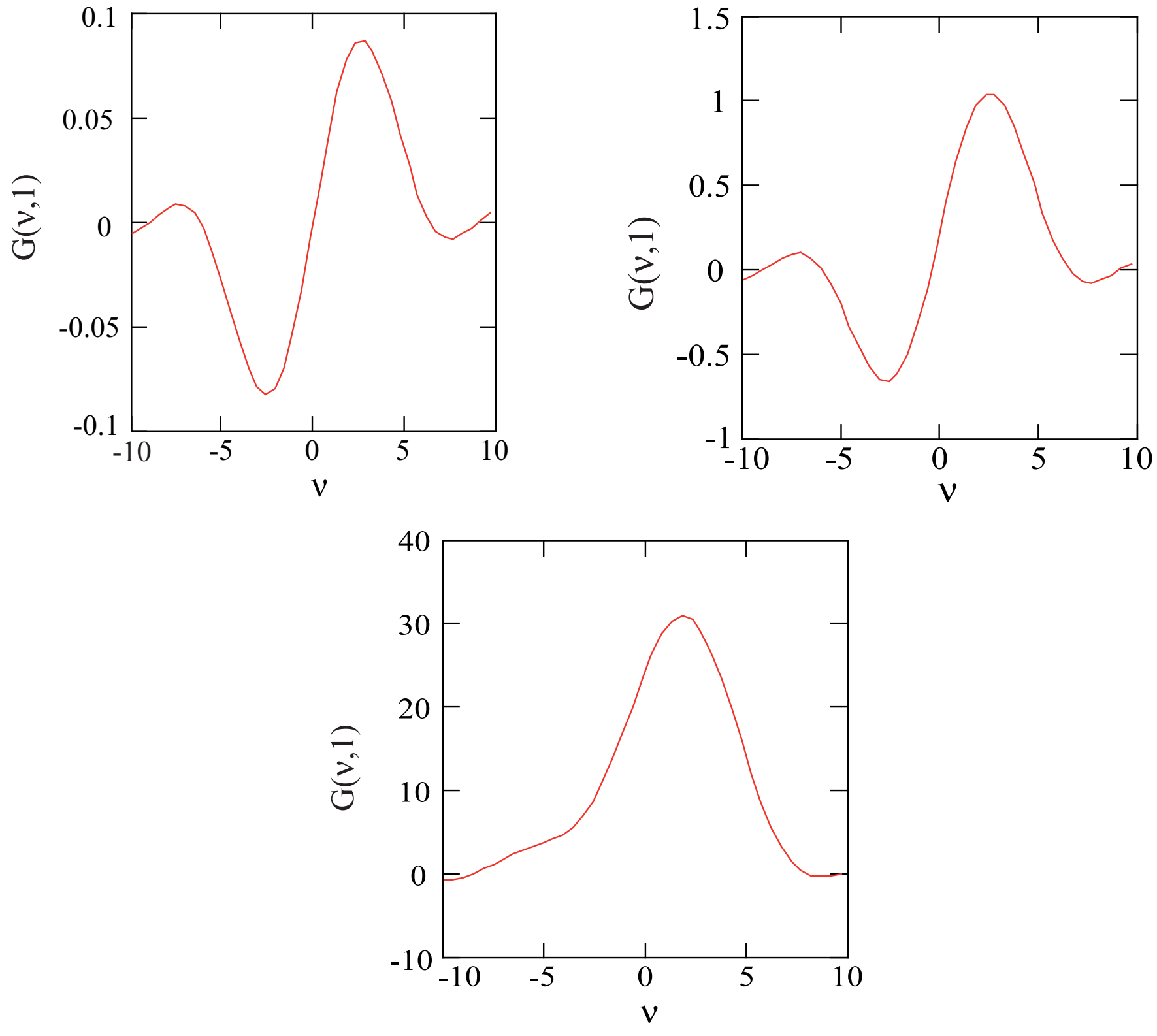

Fig. 59a,b,c

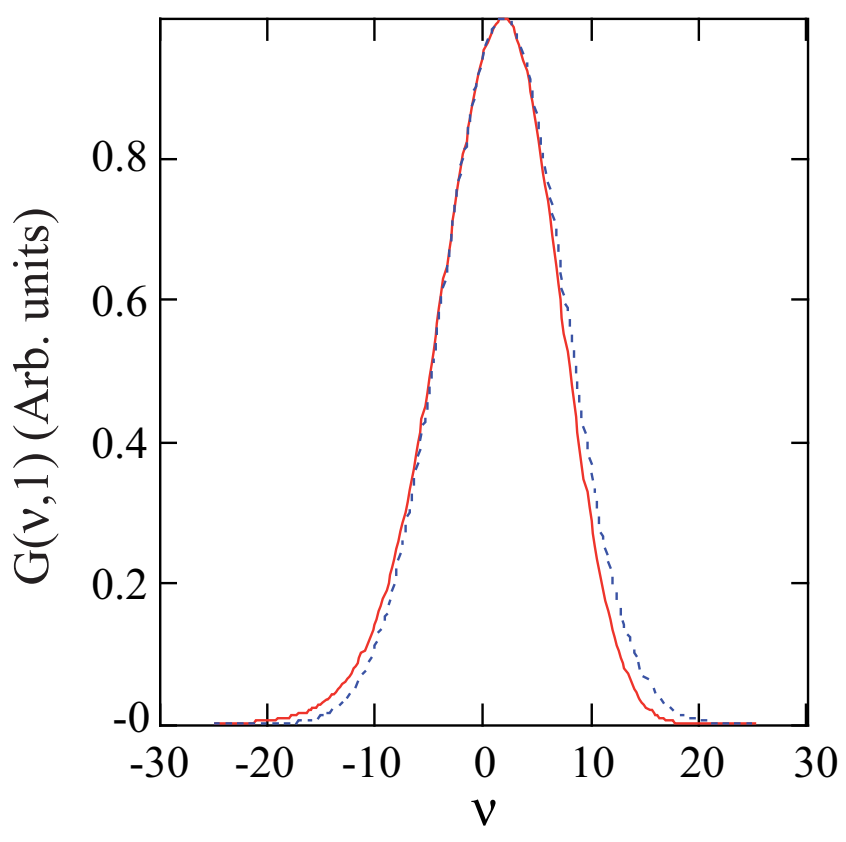

Fig. 60

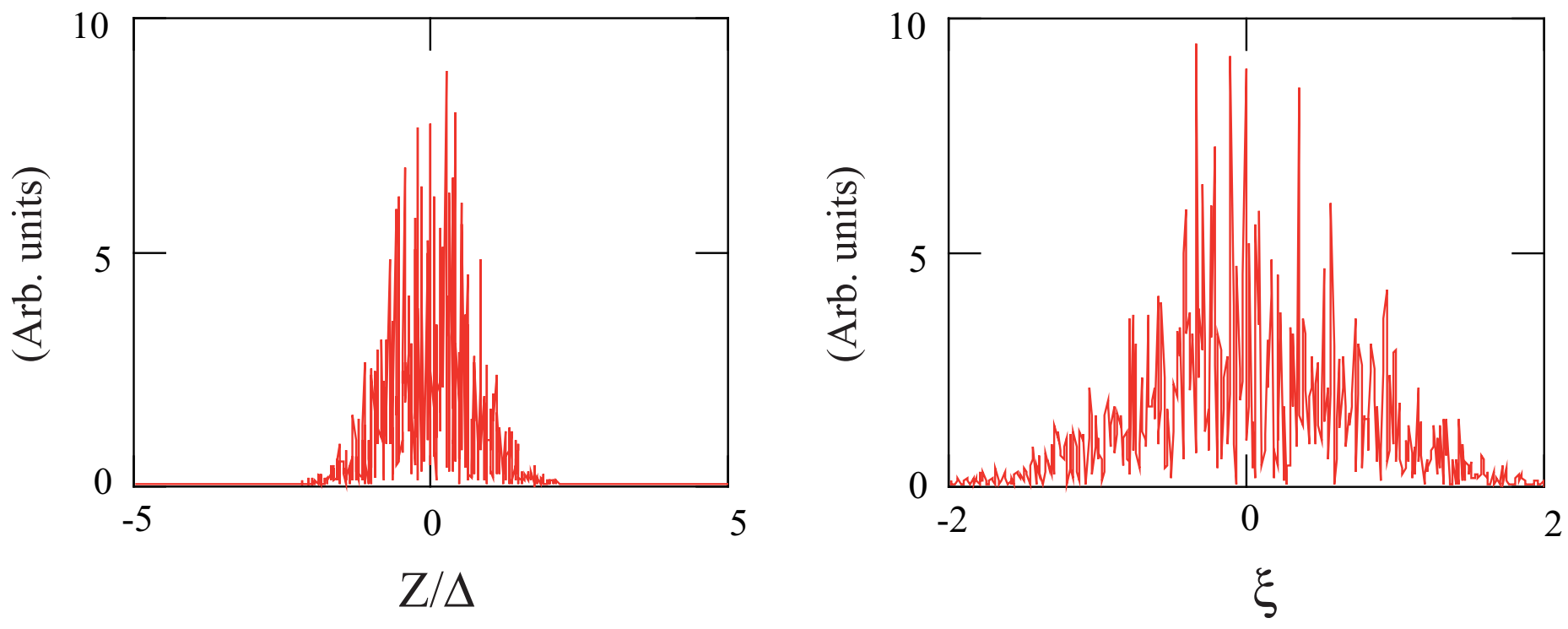

Fig. 61

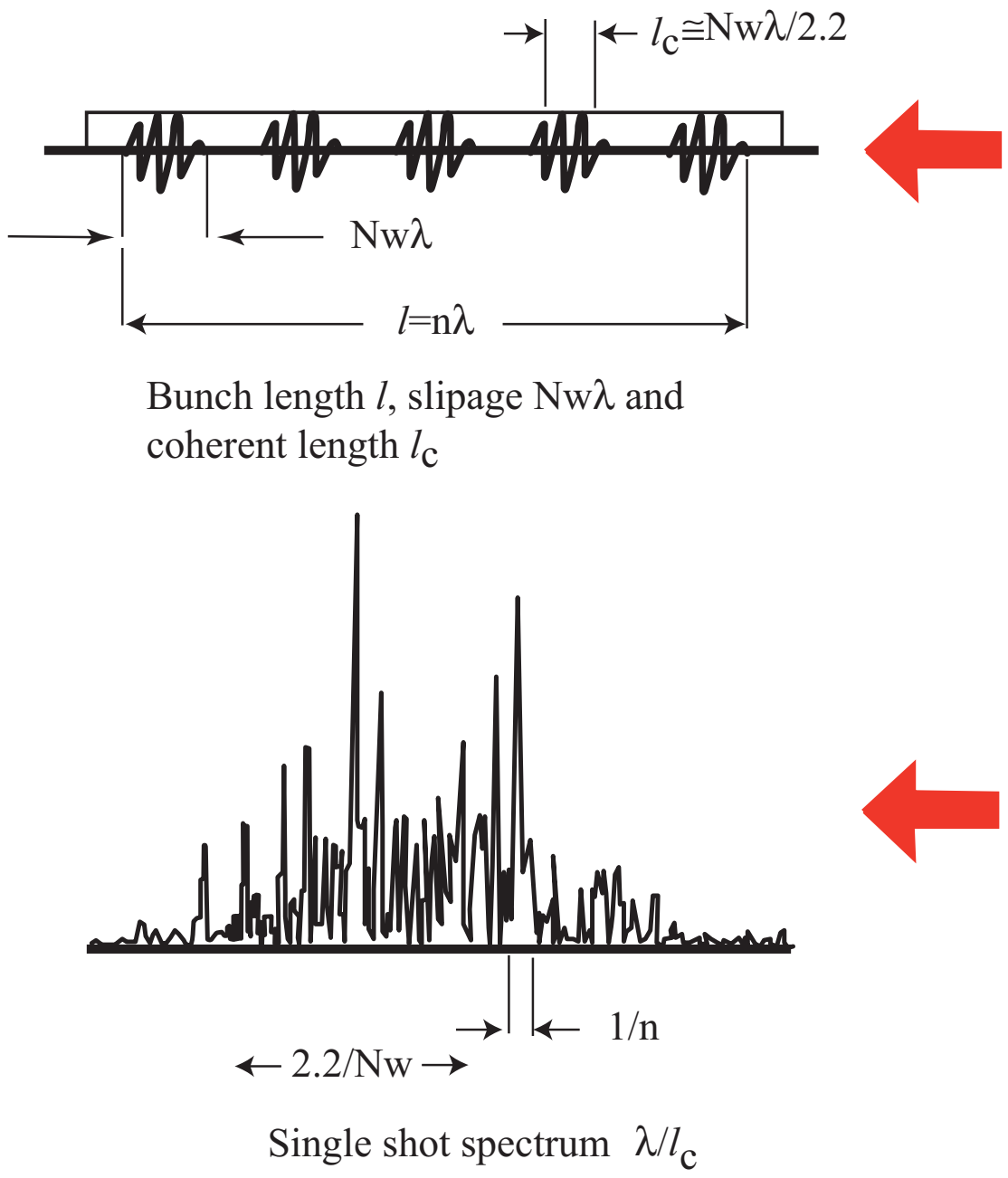

Fig. 62

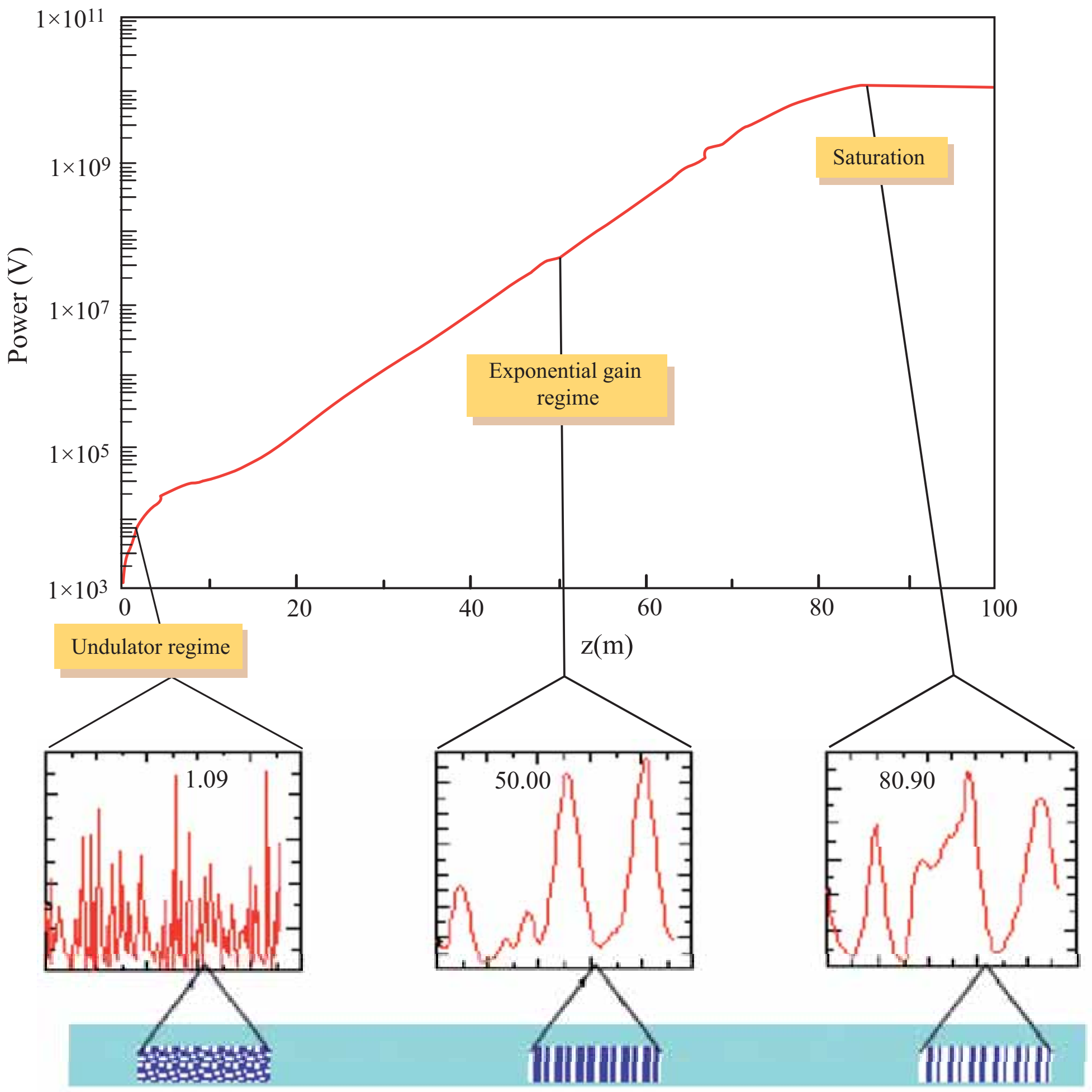

Fig. 63

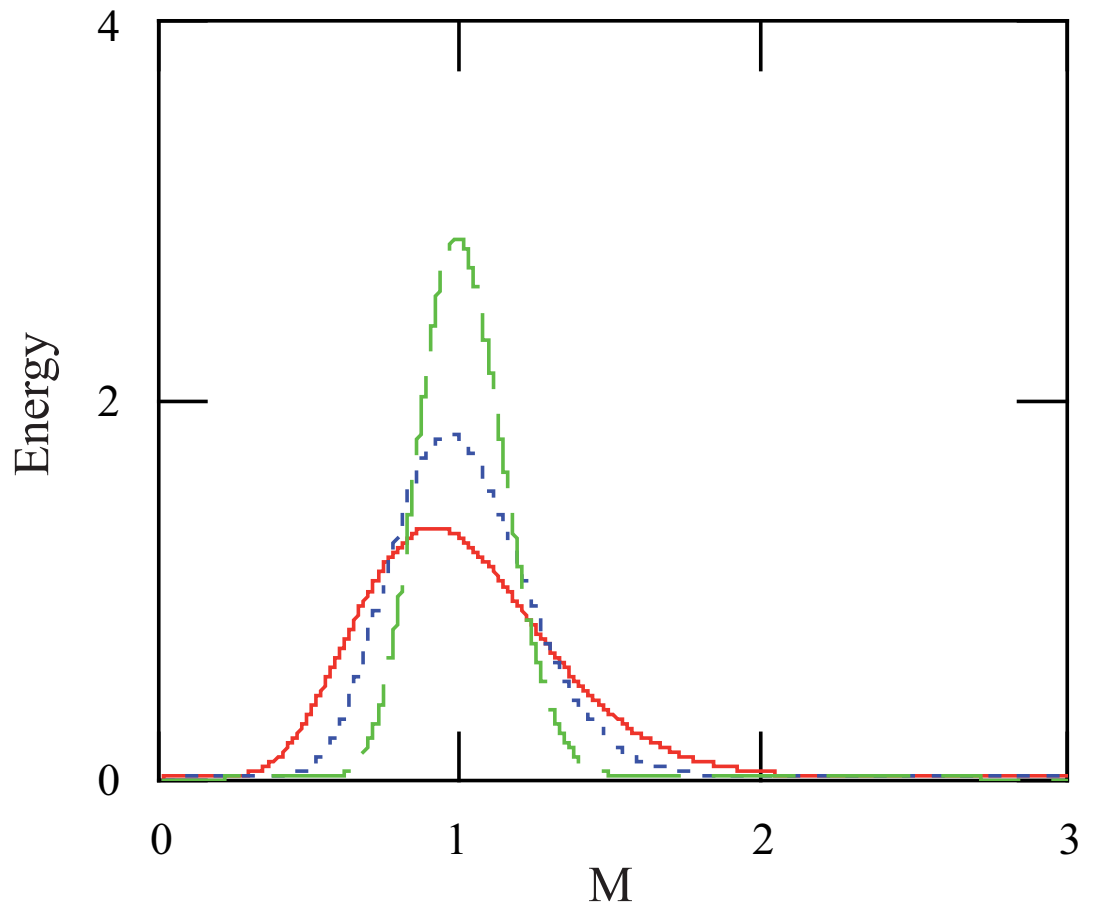

Fig. 64

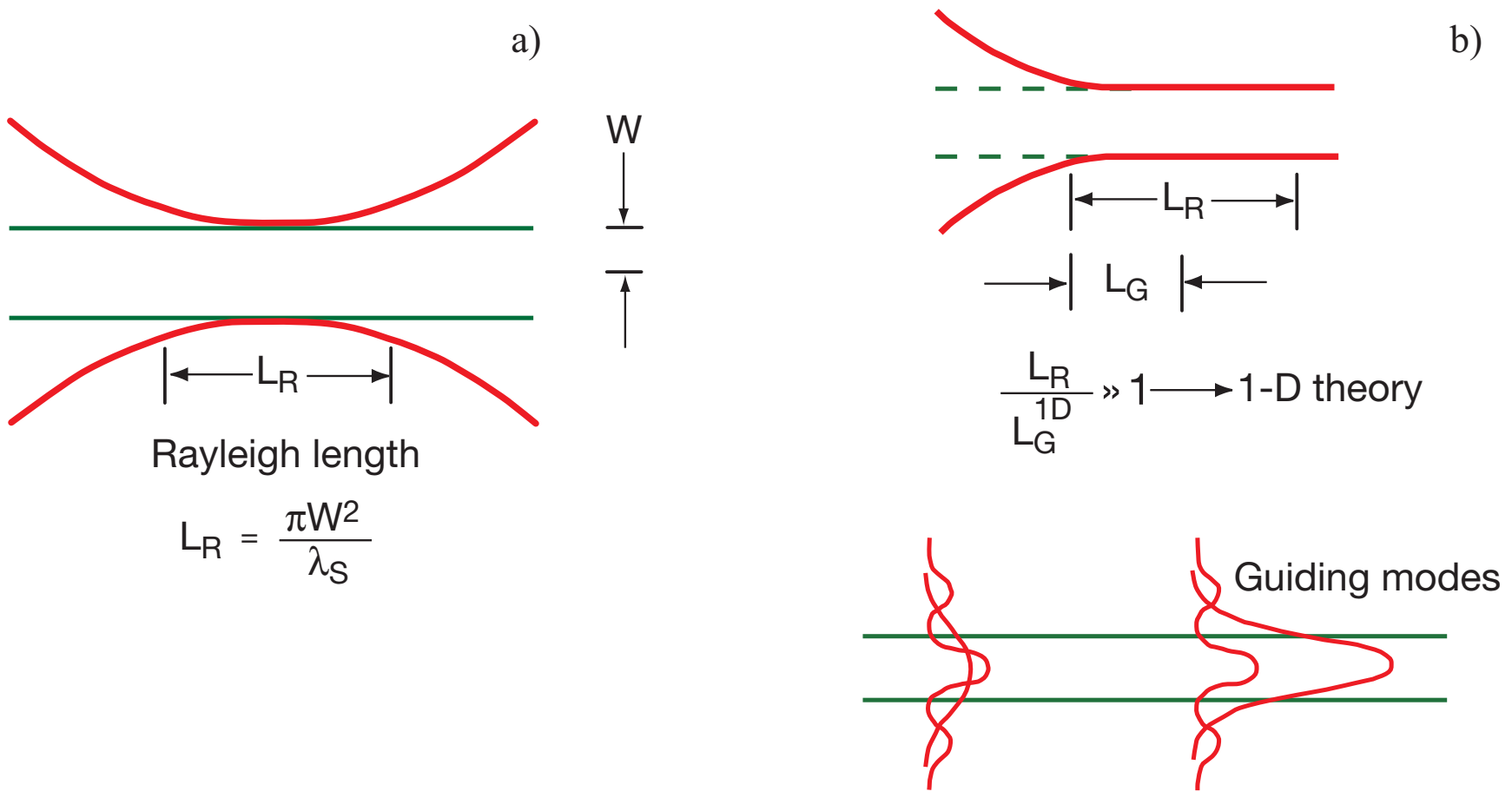

Fig. 65

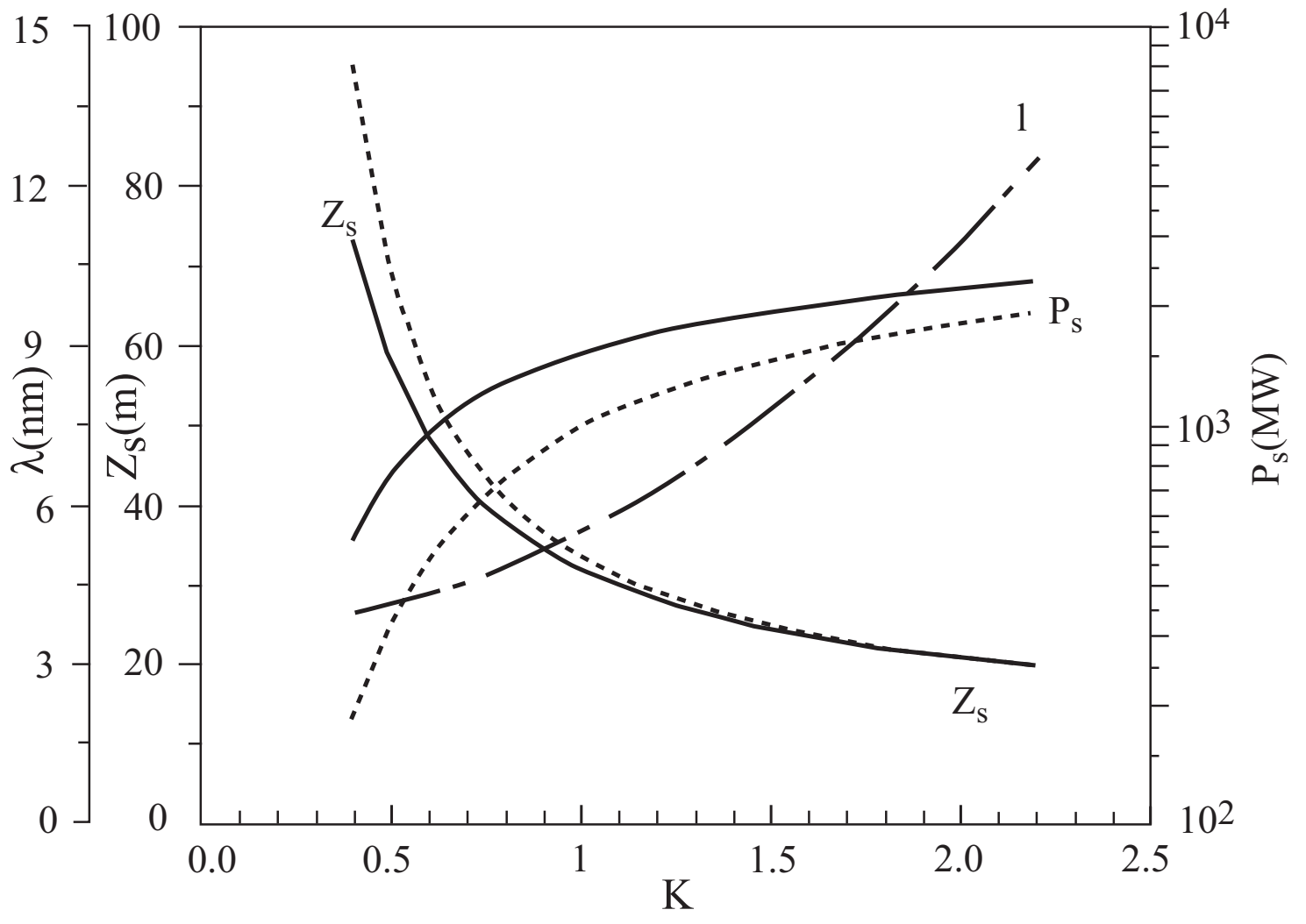

Fig. 66

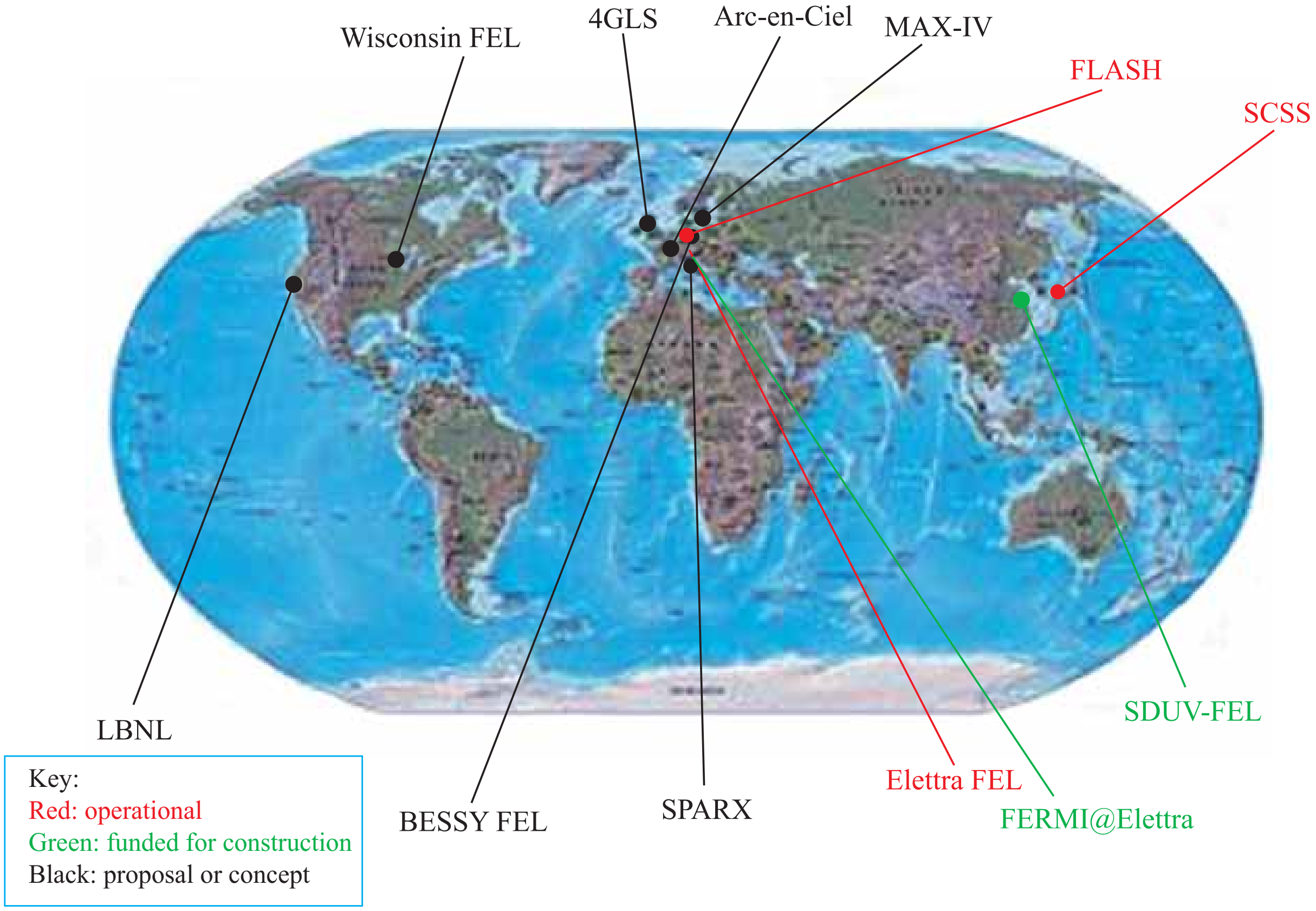

Fig. 67

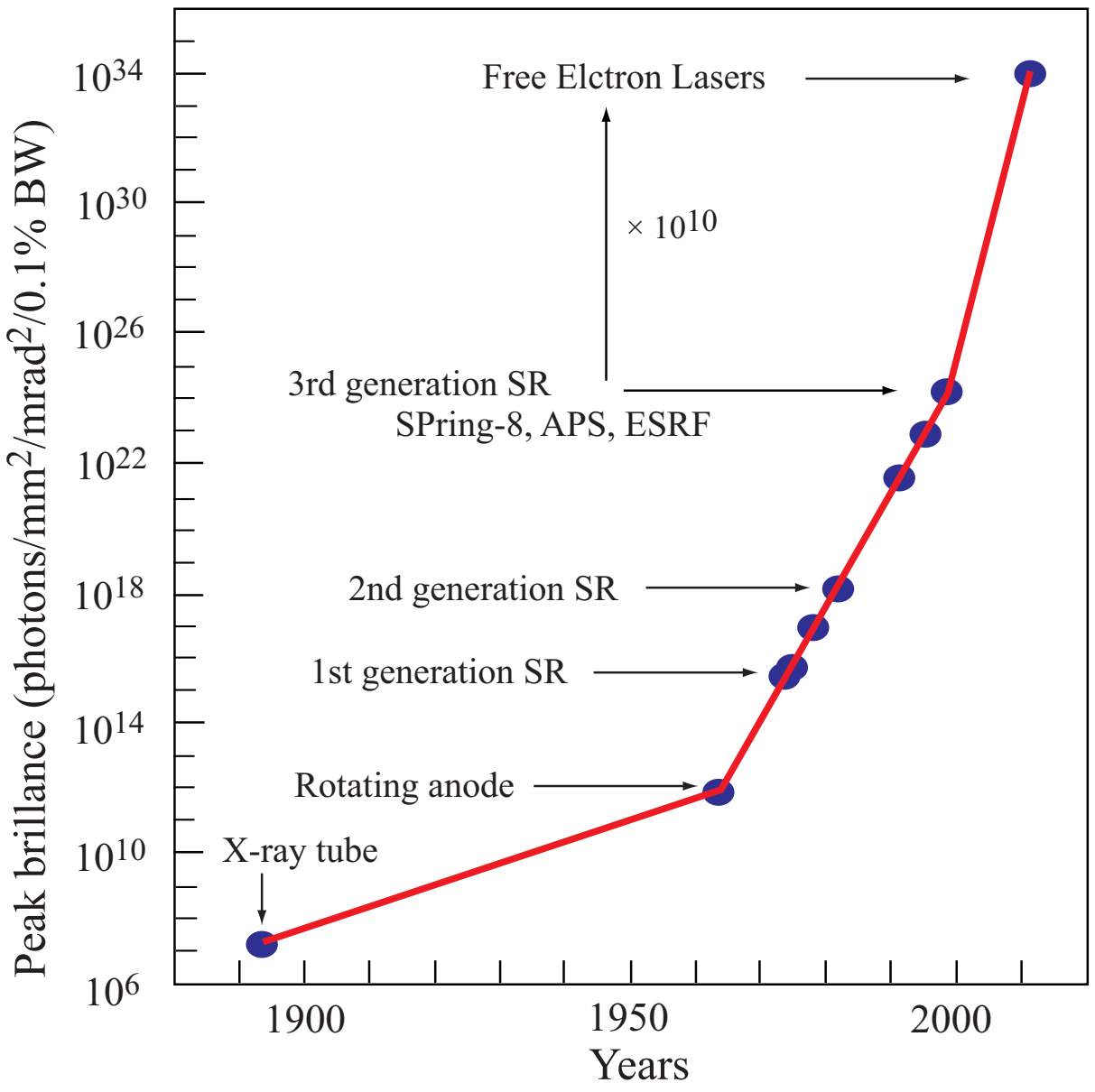

Fig. 68

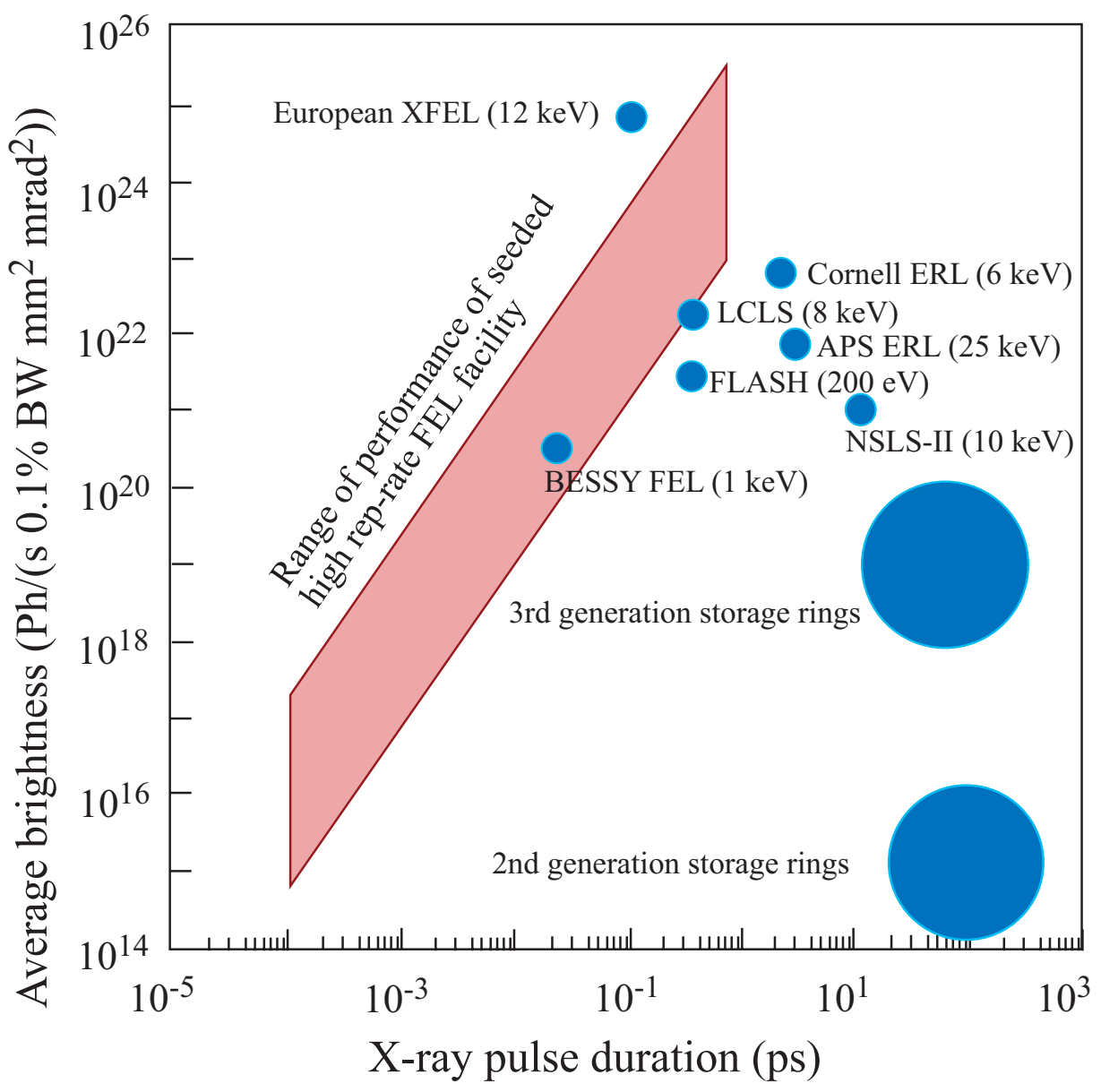

Average brightness (Ph/(s 0.1% BW mm² mrad²))
X-ray pulse duration (ps)
European XFEL (12 keV)
Range of performance of seeded high rep-rate FEL facility
Cornell ERL (6 keV)
LCLS (8 keV)
APS ERL (25 keV)
FLASH (200 eV)
NSLS-II (10 keV)
BESSY FEL (1 keV)
3rd generation storage rings
2nd generation storage rings

Fig. 69

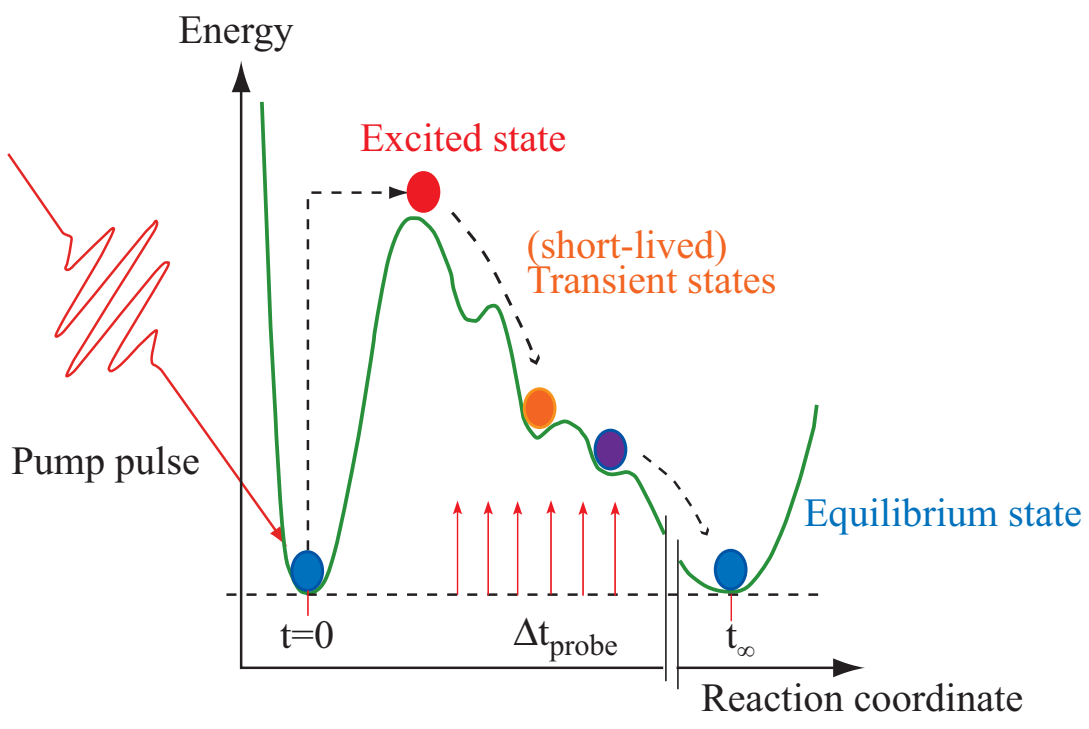

Fig. 70

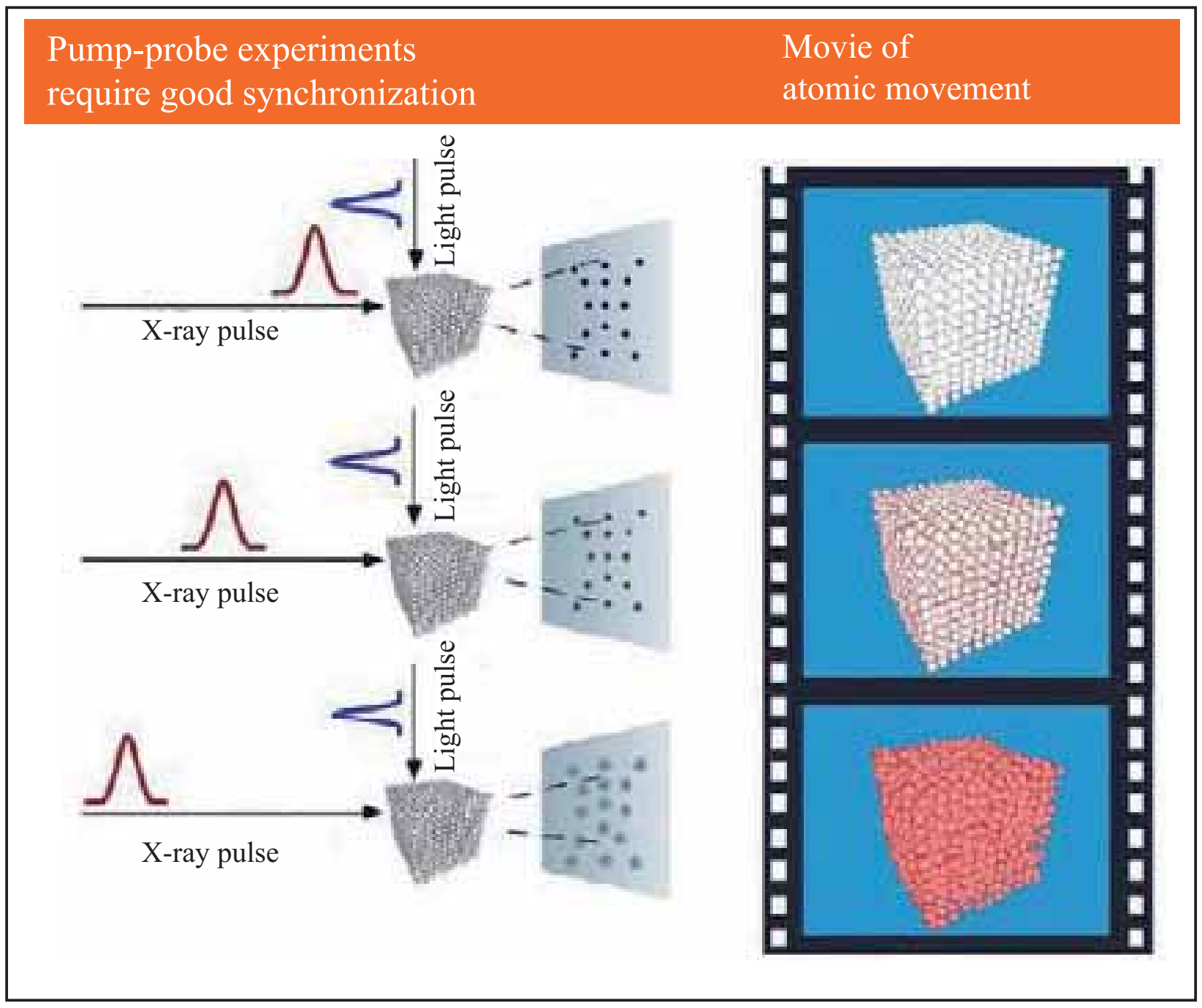

Fig. 71

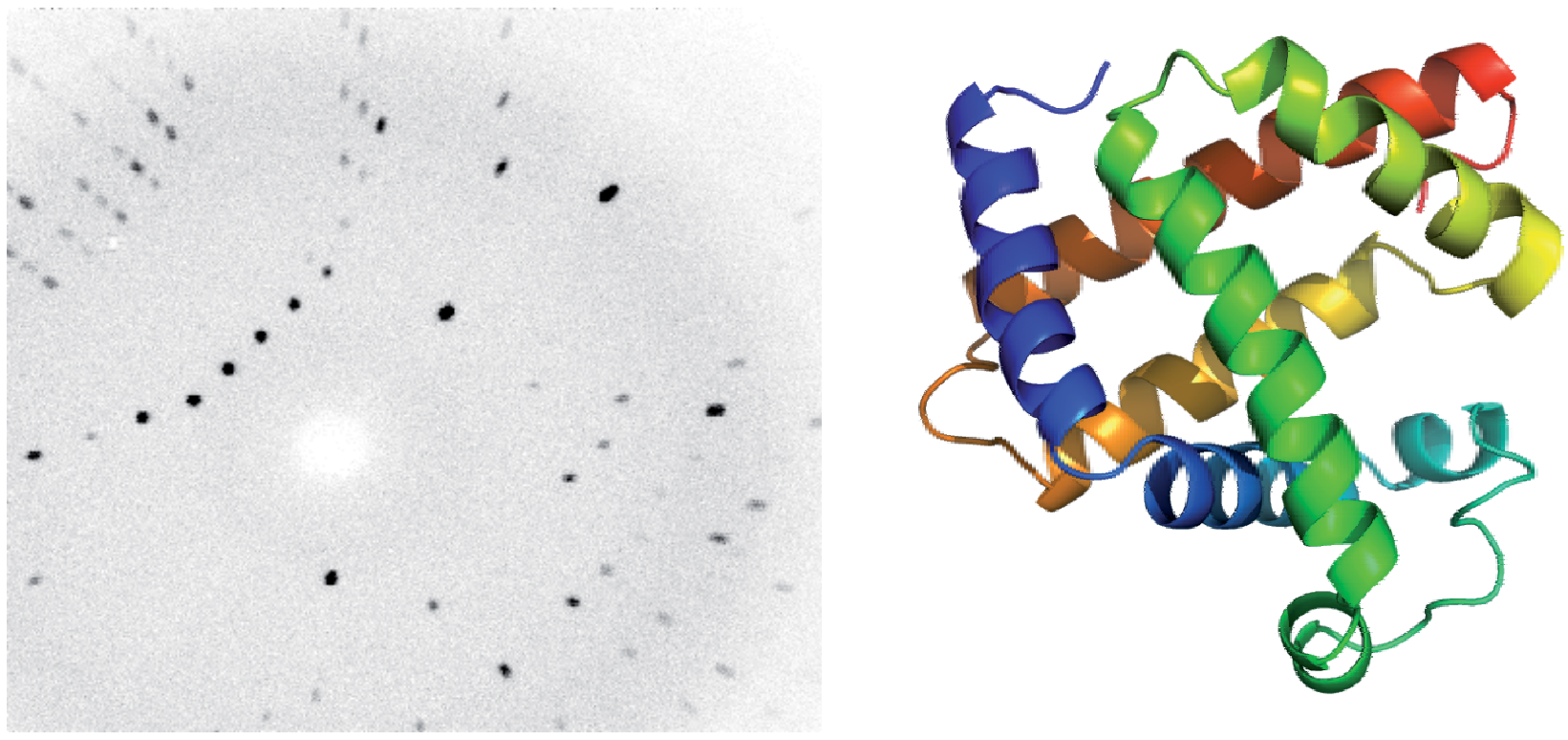

Fig. 72

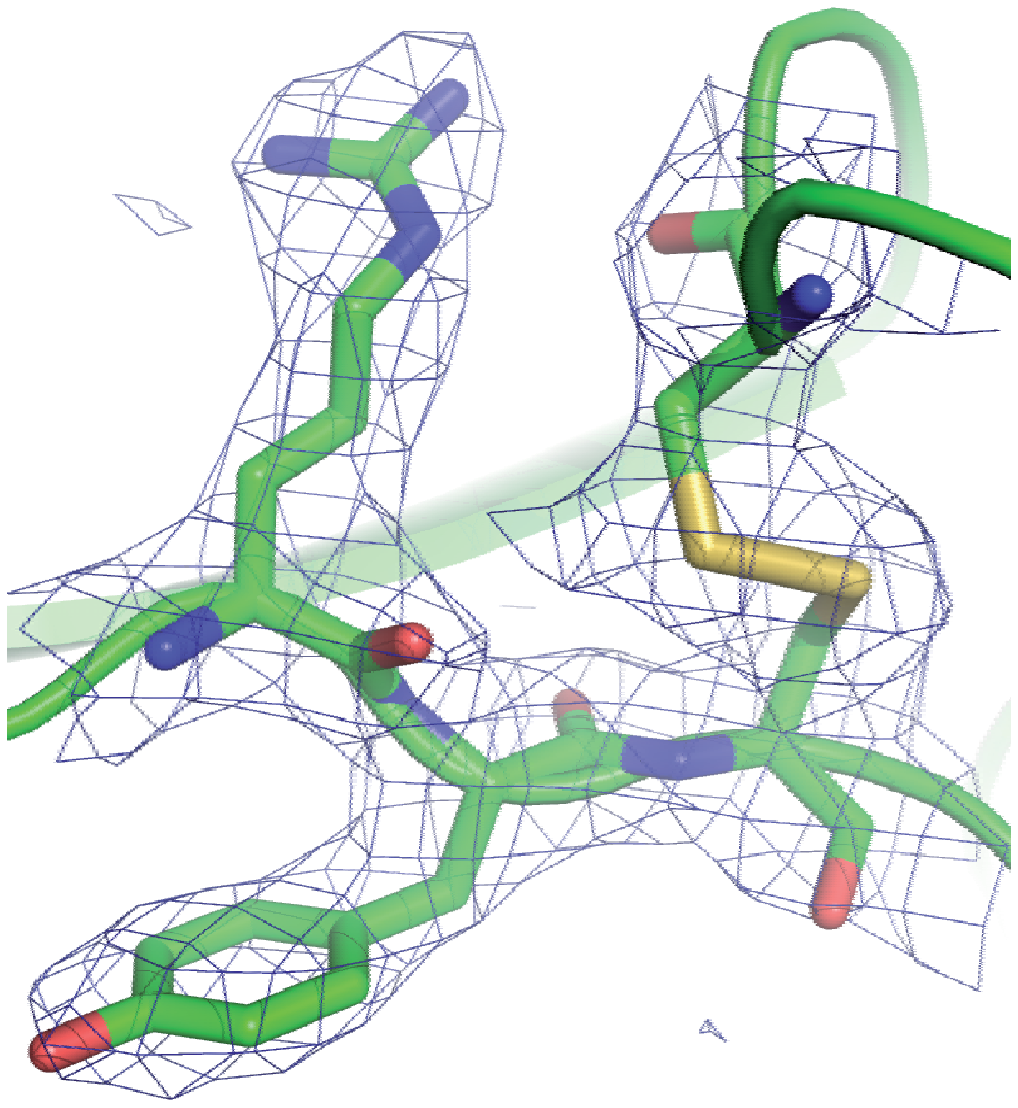

Fig. 73

Structural Biology

Ultra-high resolution

- Precise location and identification
- Hydrogen atoms
- Atomic motions

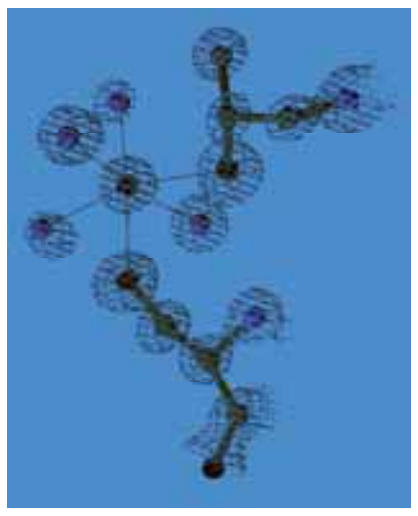

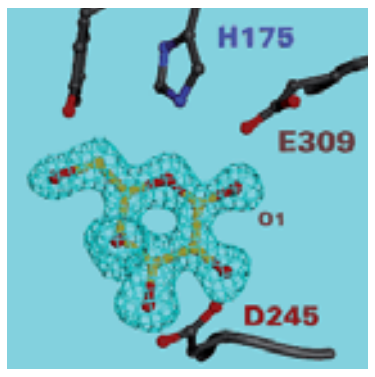

Time-resolved

- Watch proteins work

Single Molecule/particle

- Structure/function without crystals

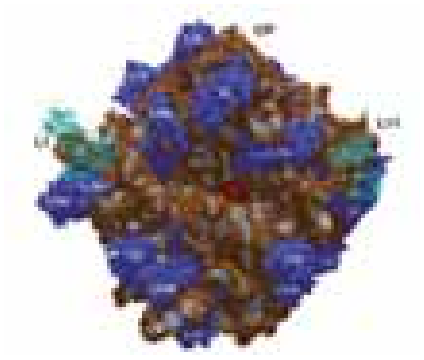

Fig. 74

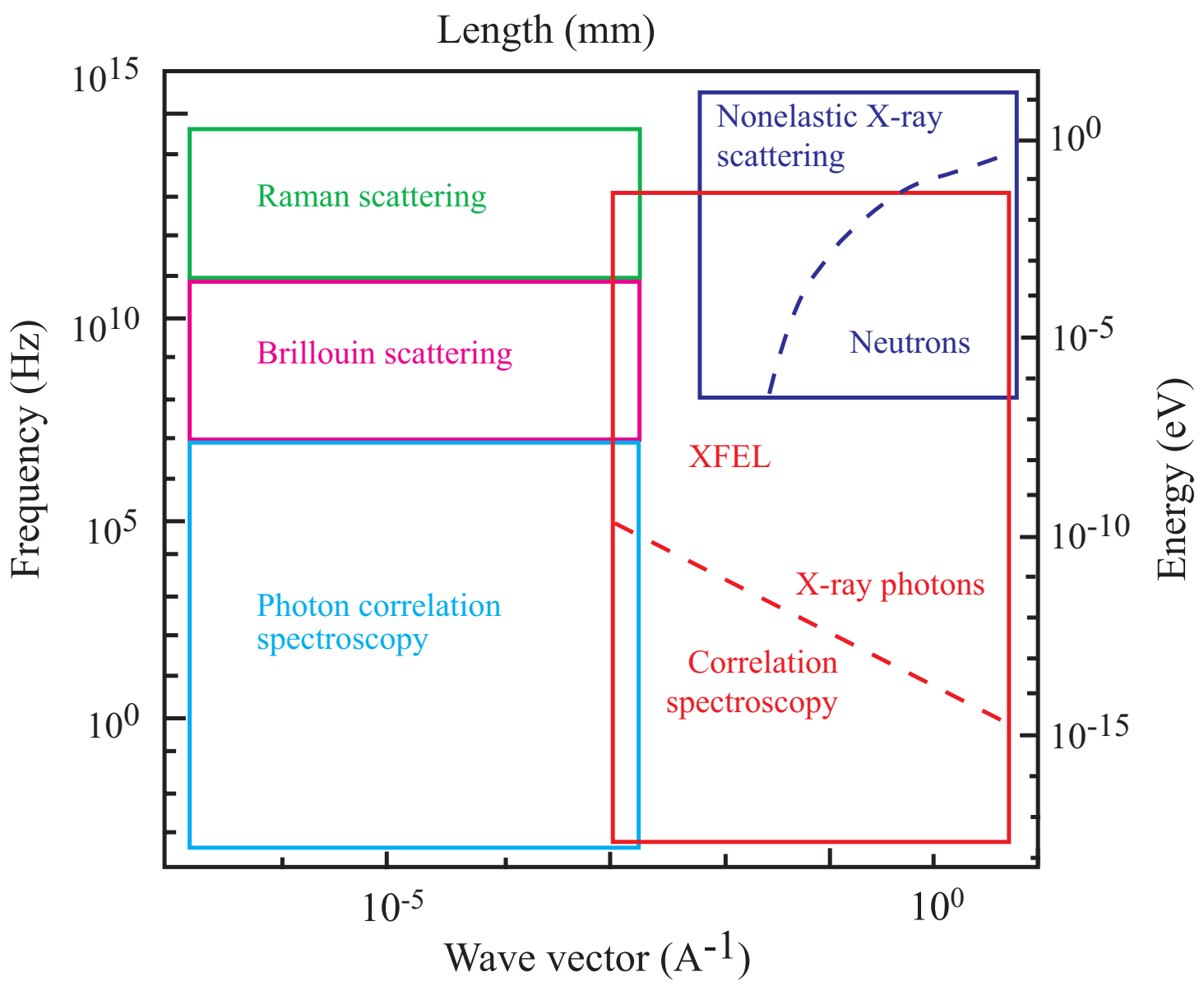

fig. 75

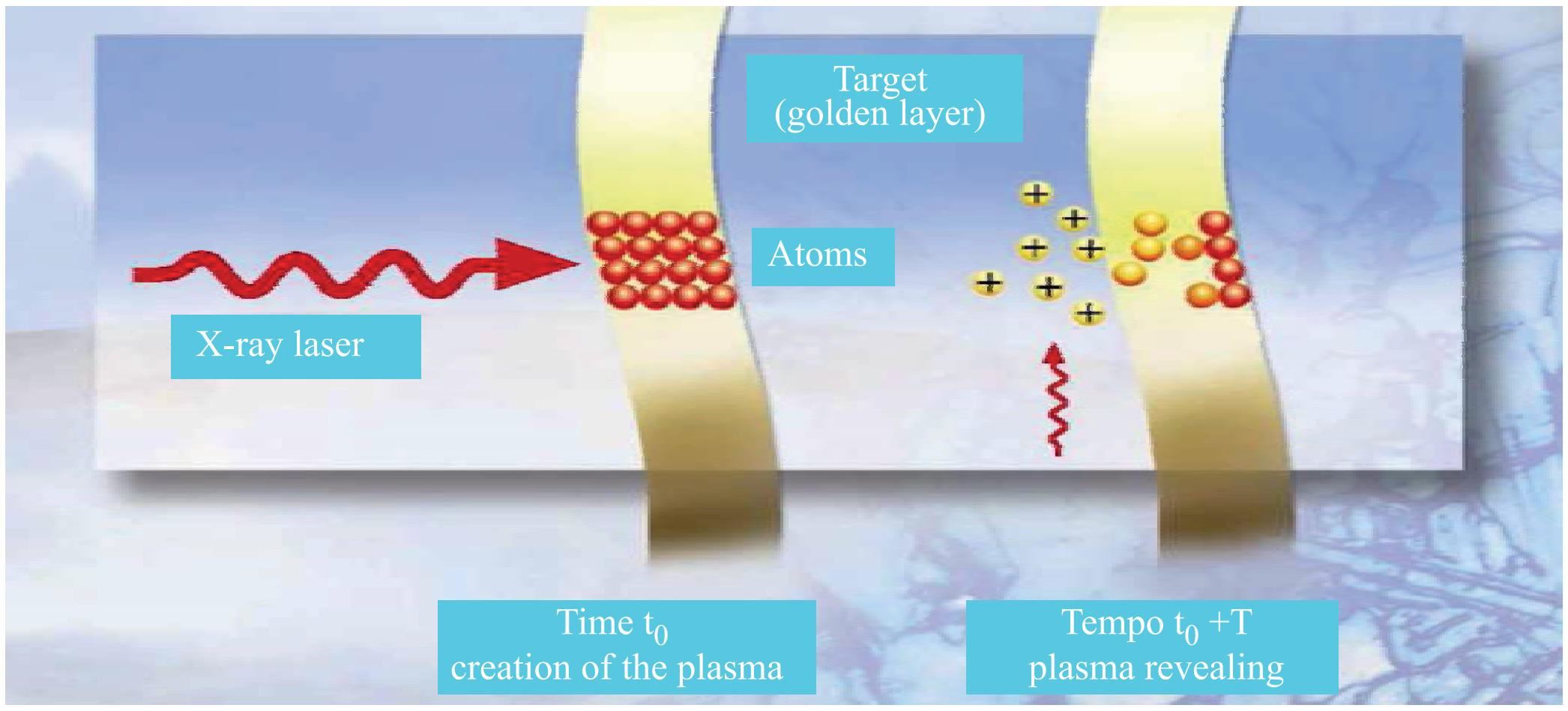

Target (golden layer)
Atoms
X-ray laser
Time t0 creation of the plasma
Tempo t0 +T plasma revealing

Fig. 76

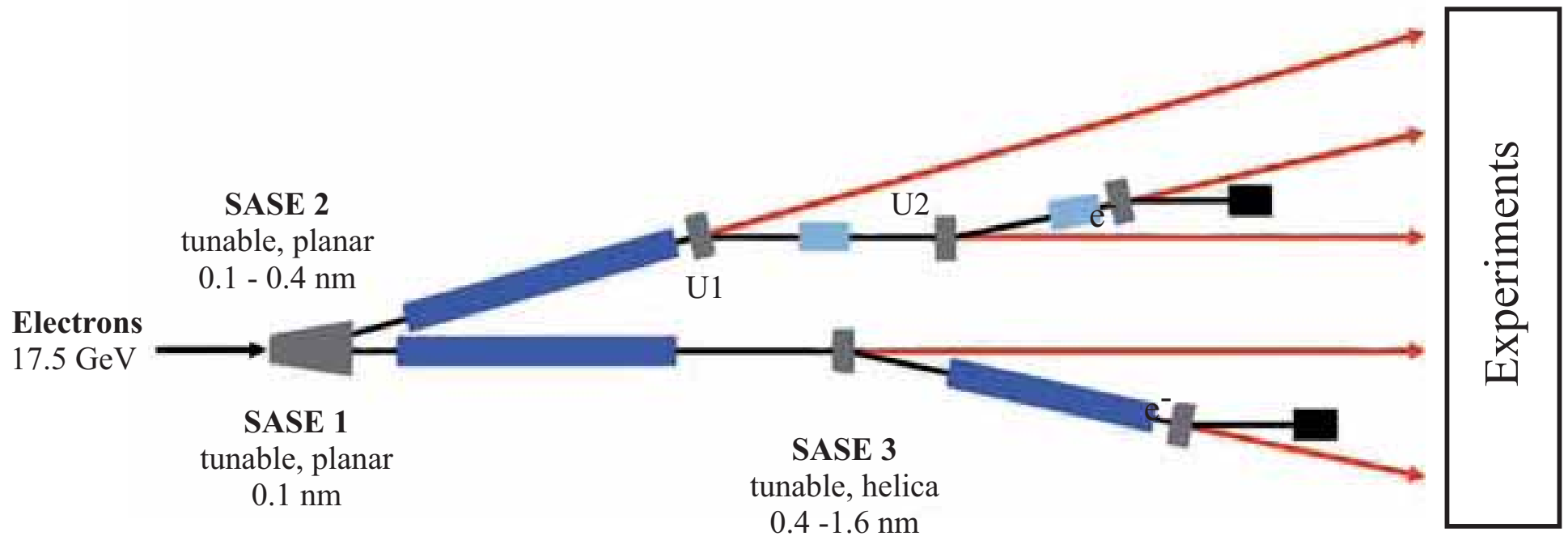

Fig. 77

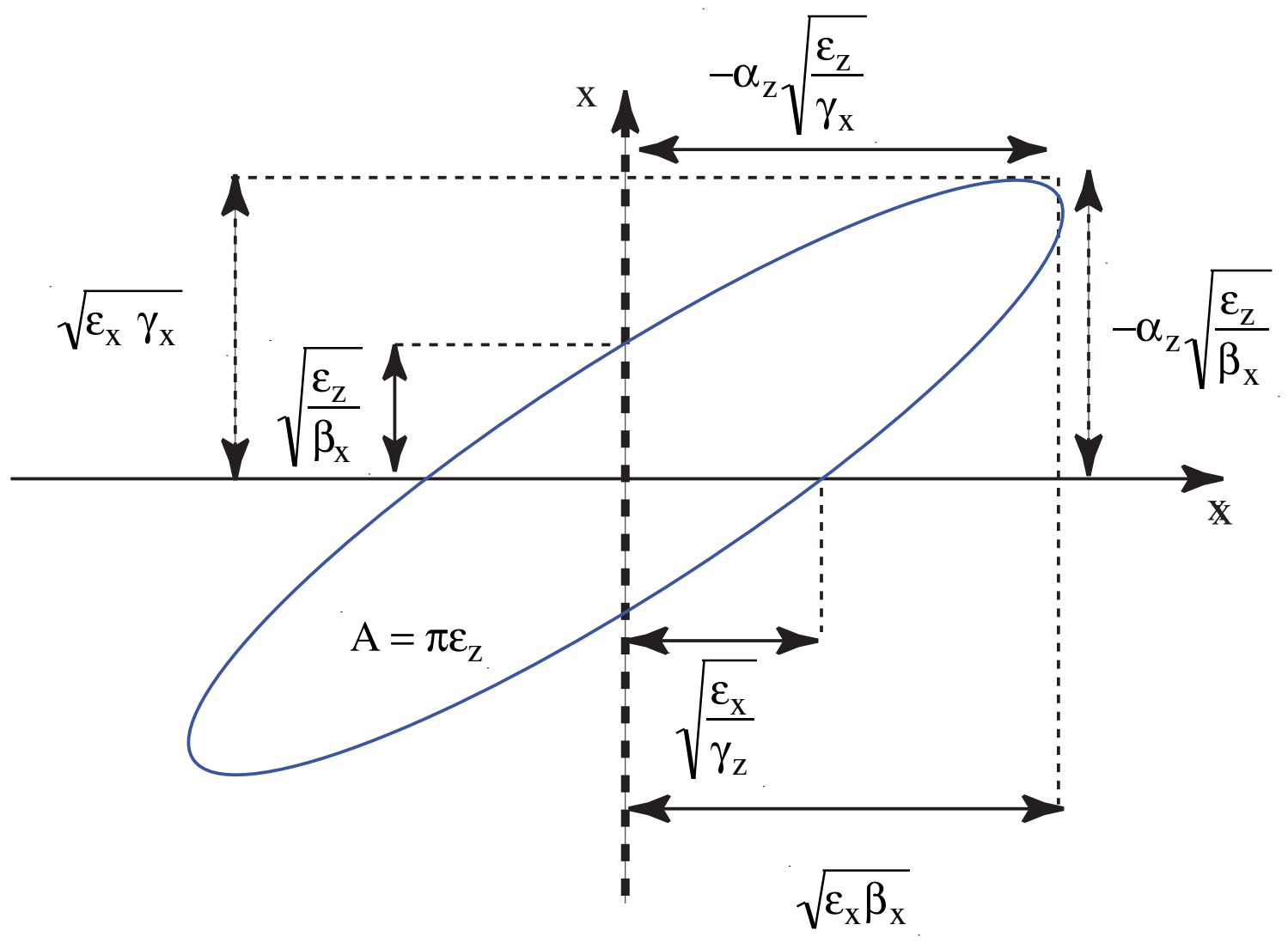

Fig. 78

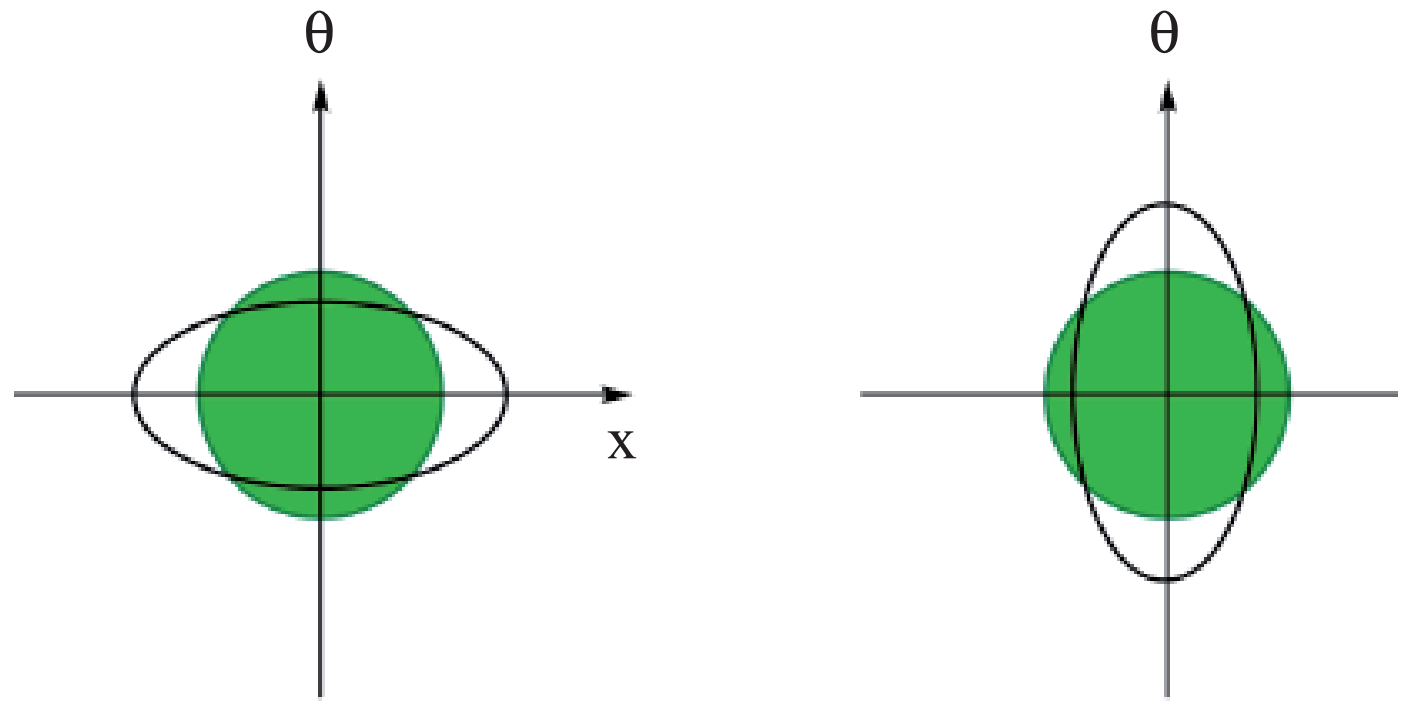

Fig. 79

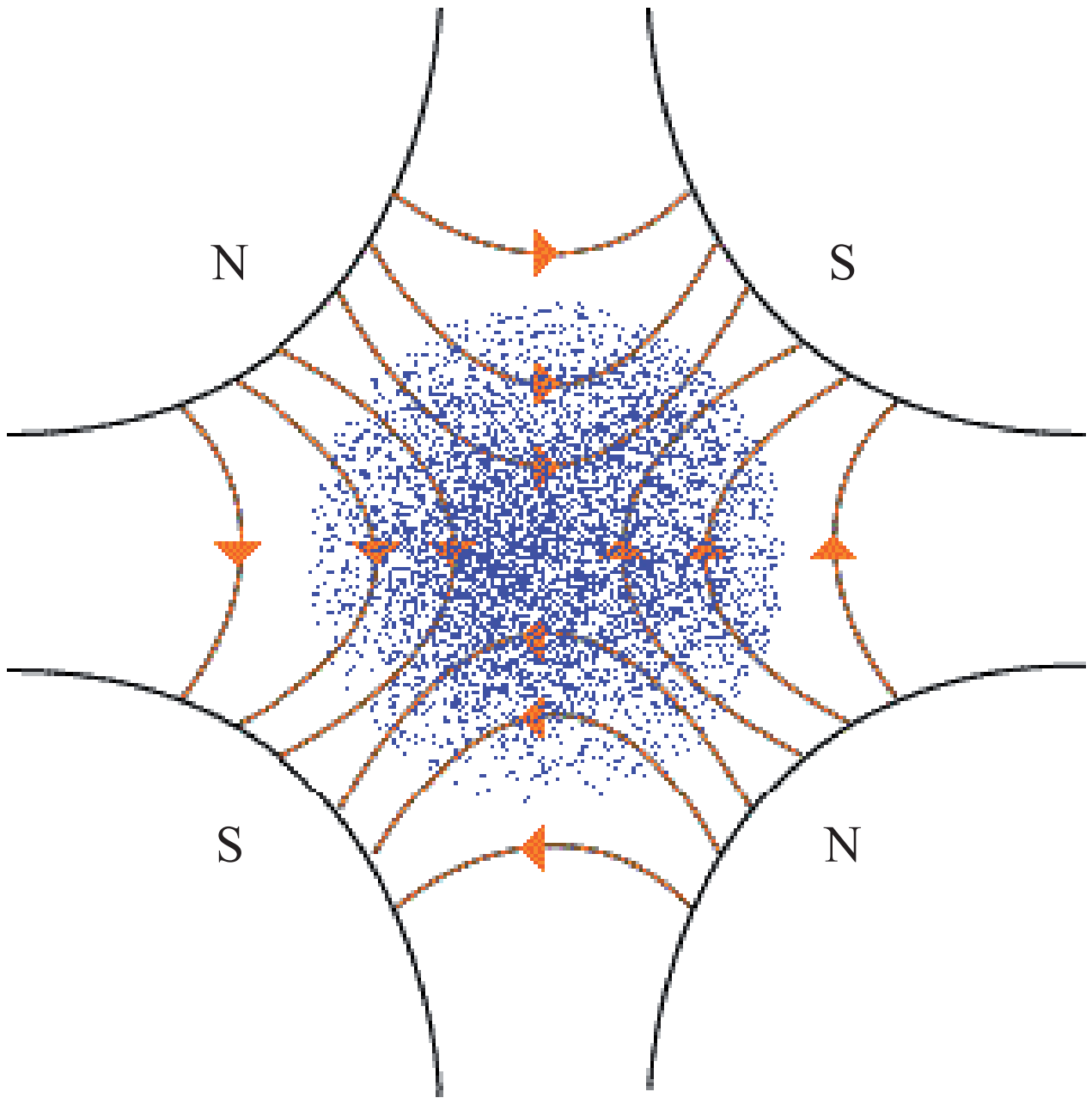

fig. 80

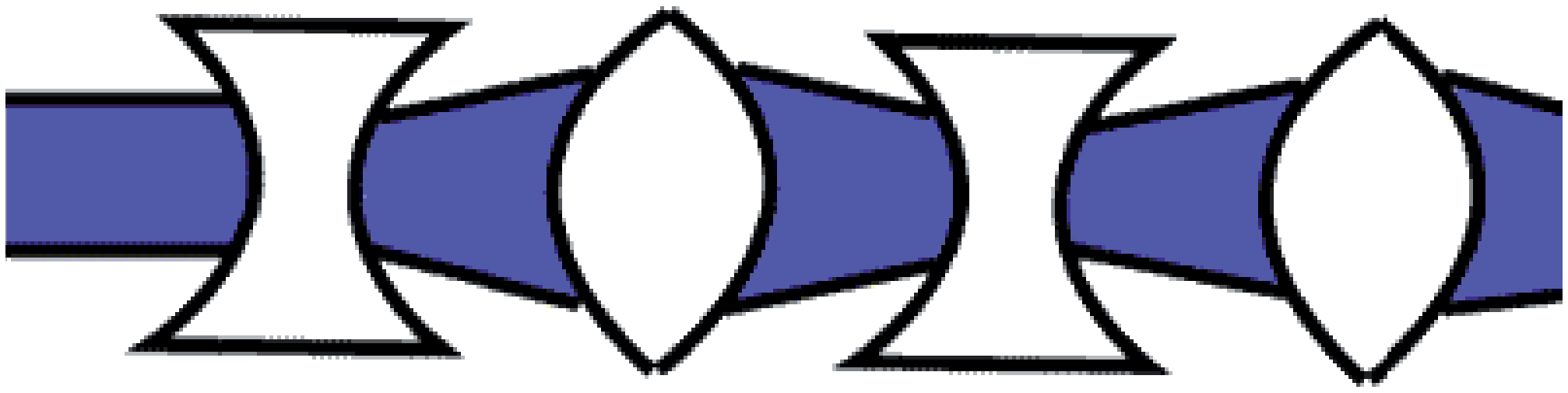

fig. 81

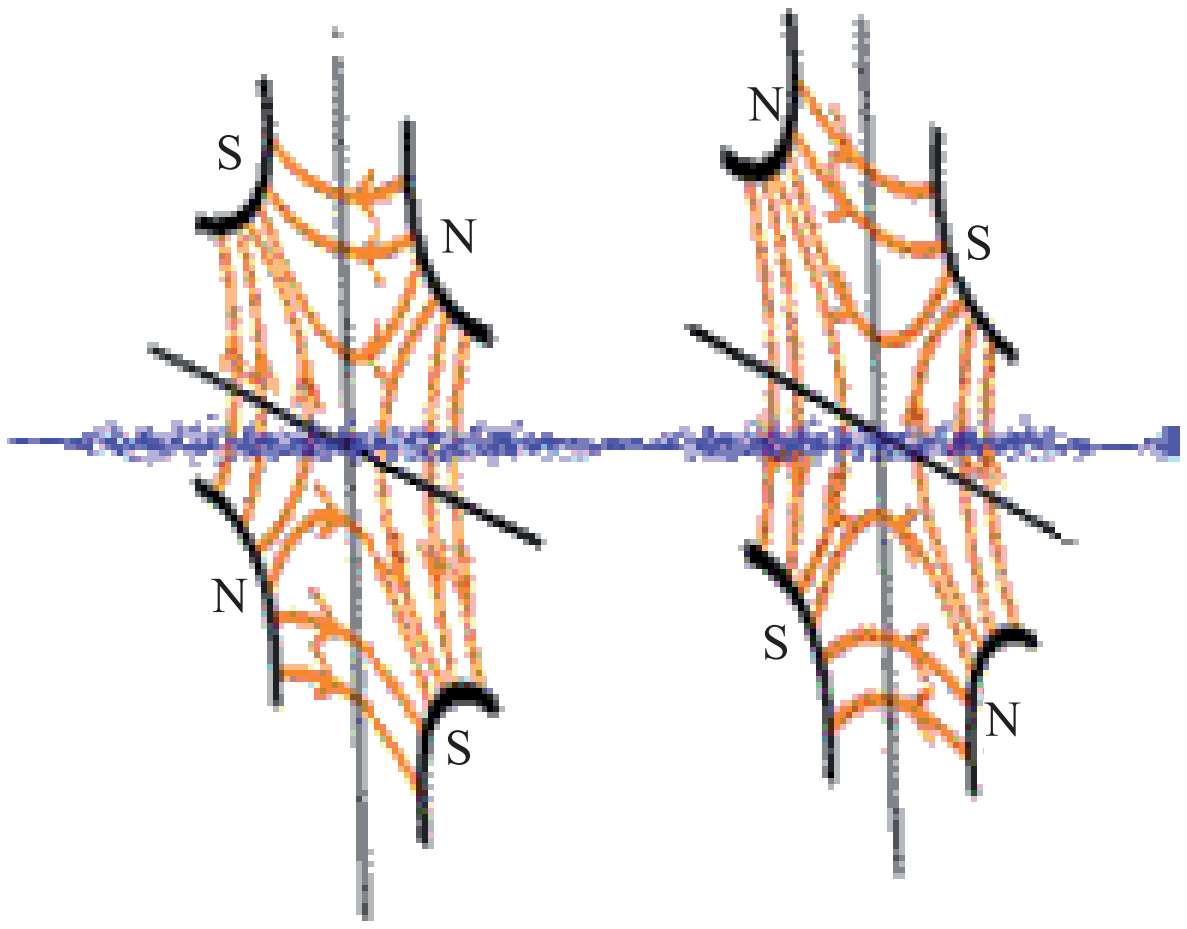

Fig. 82

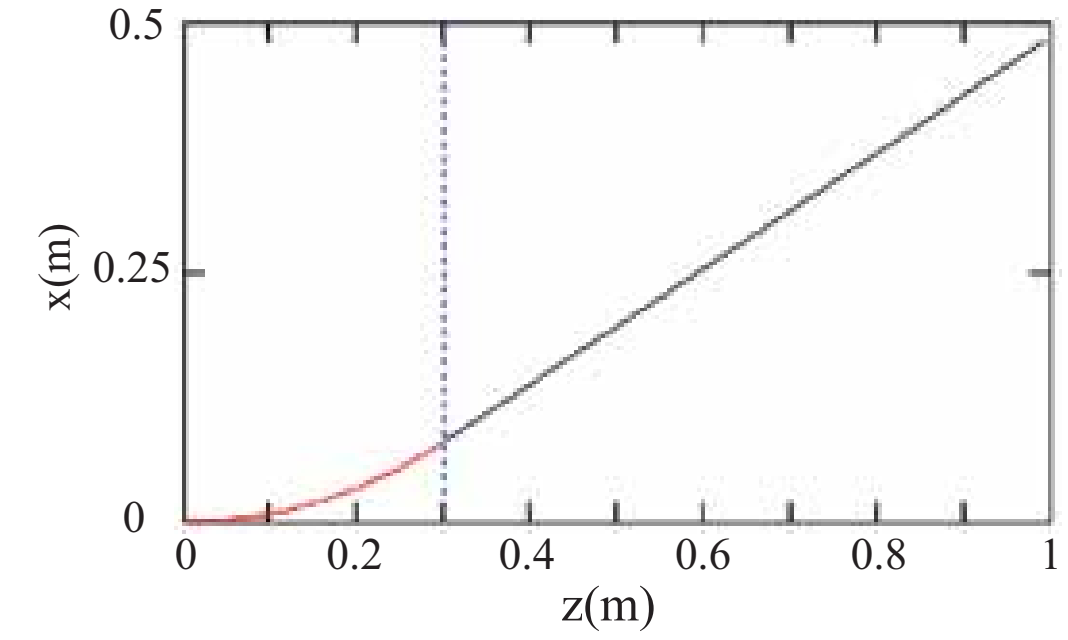

fig. 83

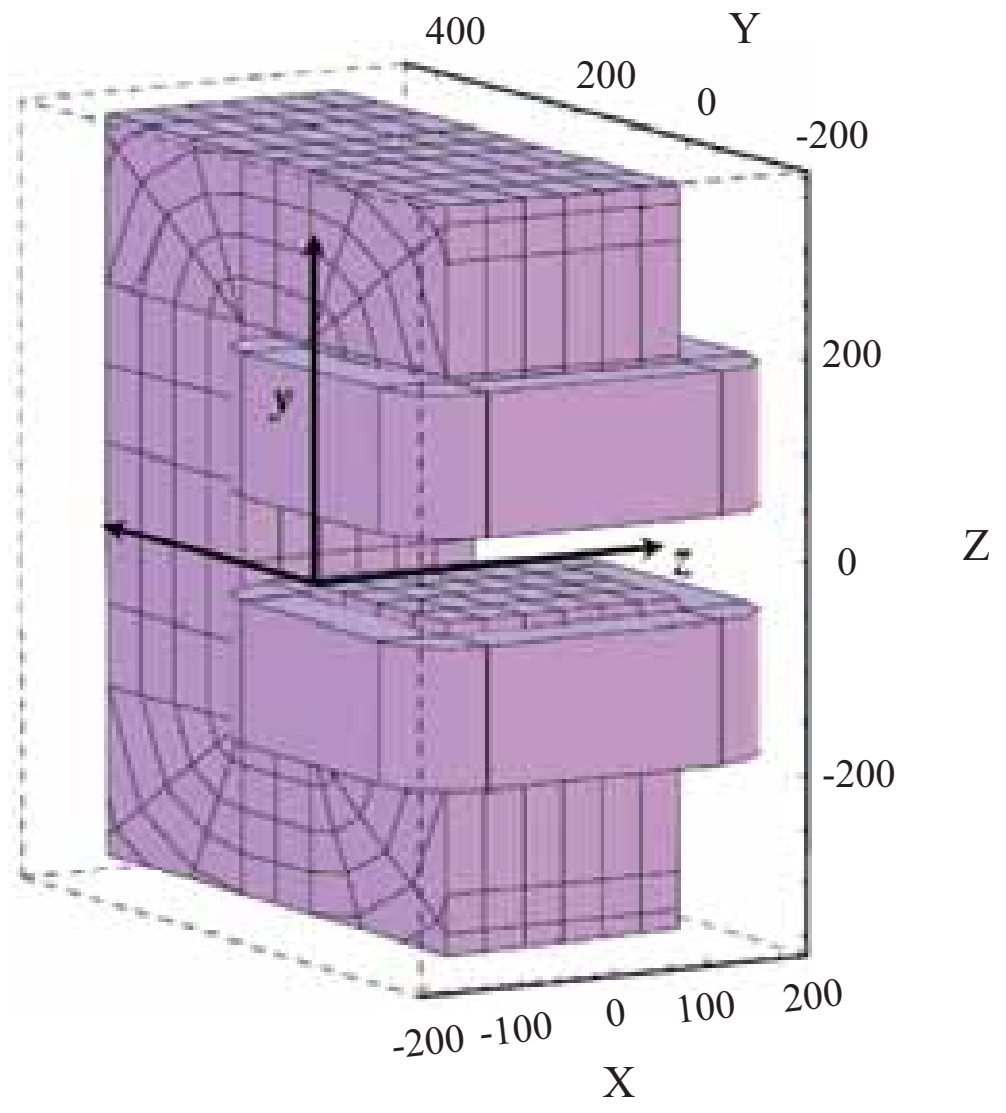

Fig. 84

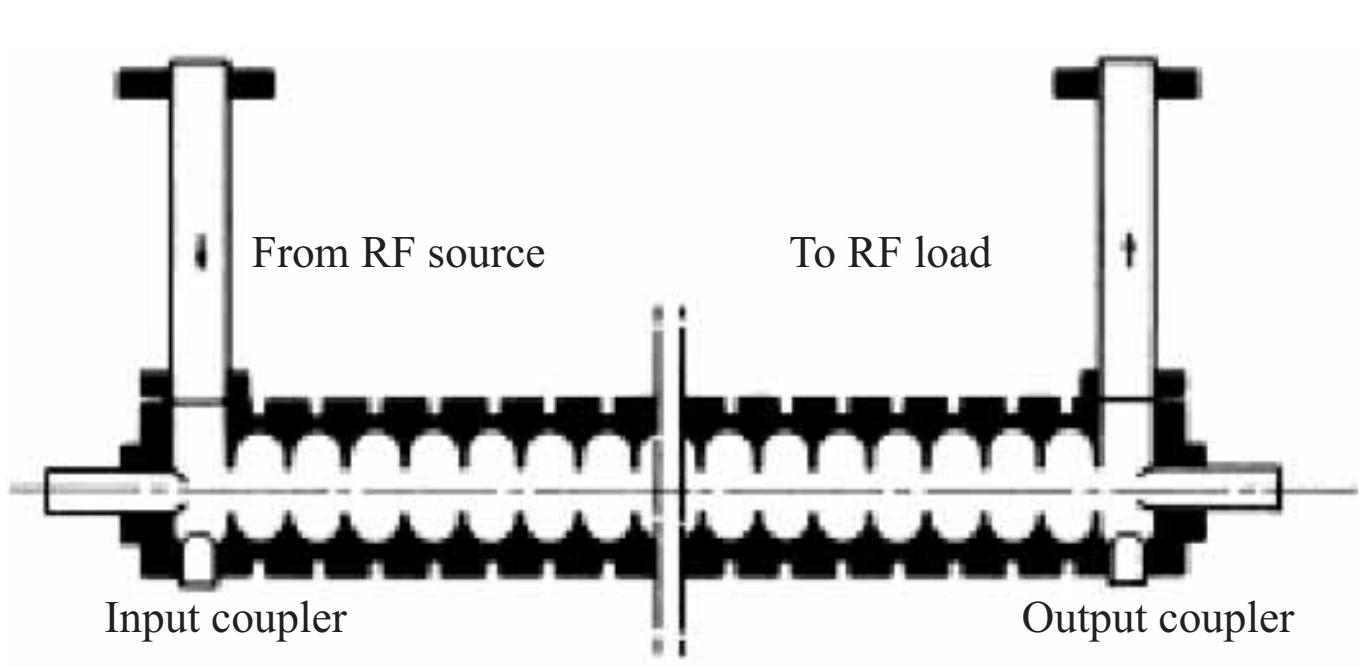

Fig. 85

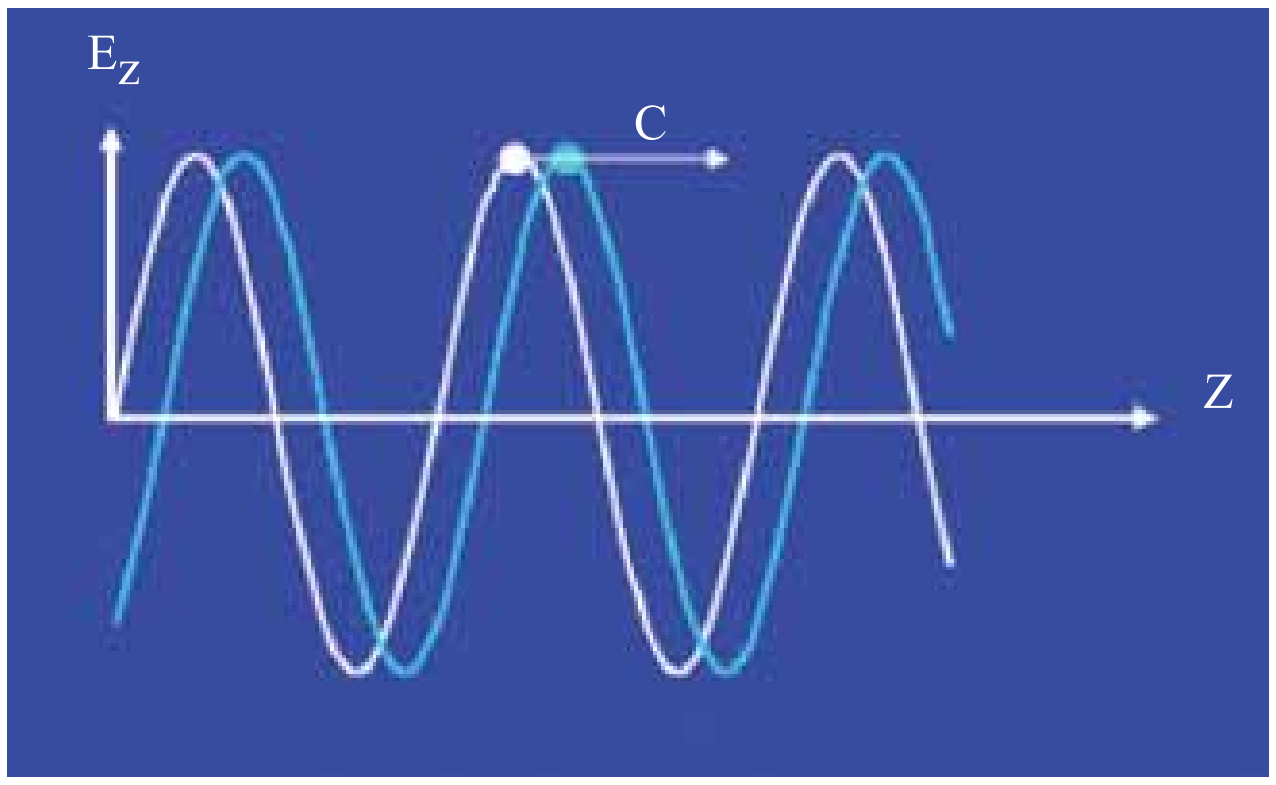

Fig. 86

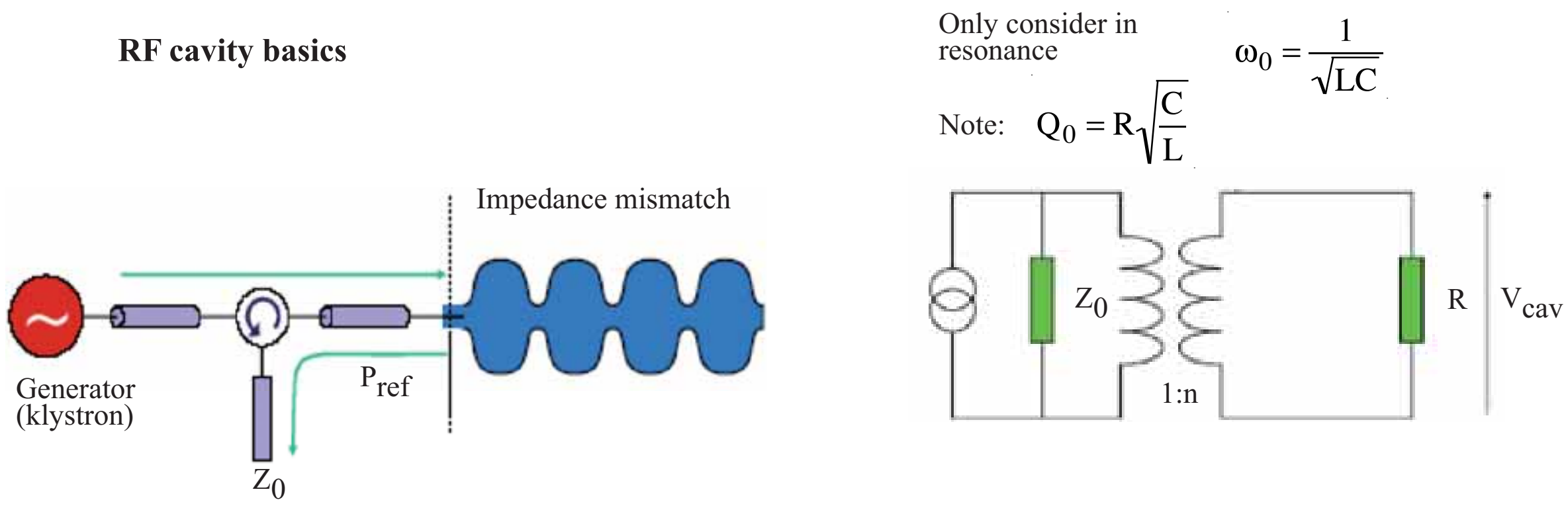

Fig. 87

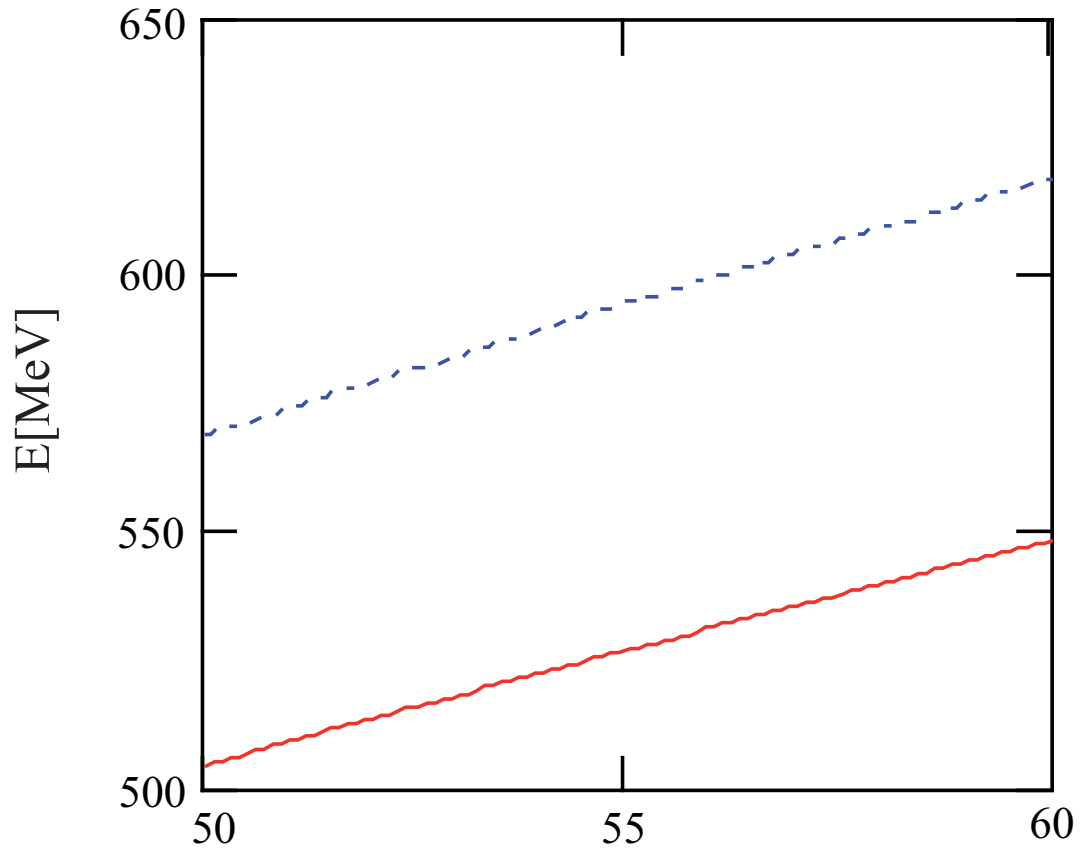

Fig. 88

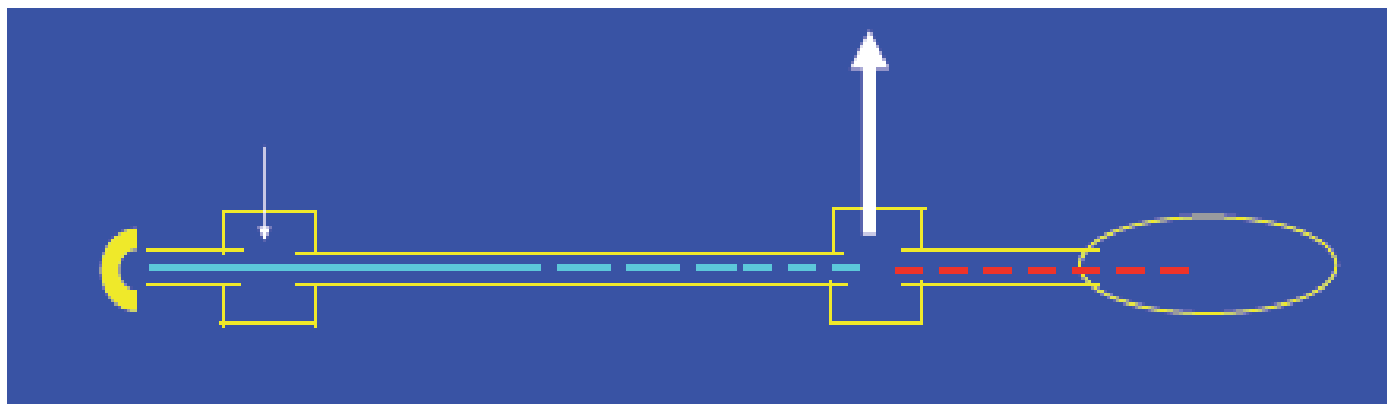

Fig. 89